\documentclass[12pt,a4paper]{article}

\usepackage{amsmath,amssymb}
\usepackage[nosort]{cite}
\usepackage[hyperref,bulletsep]{collect}
\def\gfxon{\usepackage[final]{graphicx}}

\gfxon

\makeatletter
\let\old@makecaption=\@makecaption
\def\@makecaption{\small\old@makecaption}
\makeatother

\makeatletter \@addtoreset{equation}{section} \makeatother

\makeatletter
\let\old@startsection=\@startsection
\renewcommand{\@startsection}[6]{\old@startsection{#1}{#2}{#3}{#4}{#5}{#6\mathversion{bold}}}
\makeatother

\newcommand{\fes}[4]{#2}

\let\oldPhi=\Phi
\let\oldPsi=\Psi
\let\oldGamma=\Gamma
\let\oldDelta=\Delta
\let\oldSigma=\Sigma
\let\oldTheta=\Theta
\let\oldPi=\Pi
\renewcommand{\Phi}{\mathnormal{\oldPhi}}
\renewcommand{\Psi}{\mathnormal{\oldPsi}}
\renewcommand{\Gamma}{\mathnormal{\oldGamma}}
\renewcommand{\Sigma}{\mathnormal{\oldSigma}}
\renewcommand{\Delta}{\mathnormal{\oldDelta}}
\renewcommand{\Theta}{\mathnormal{\oldTheta}}
\renewcommand{\Pi}{\mathnormal{\oldPi}}

\newcommand{\alg}[1]{\mathfrak{#1}}

\newcommand{\alSU}{\alg{su}}

\newcommand{\alSO}{\alg{so}}

\newcommand{\alPSU}{\alg{psu}}

\newcommand{\indup}[1]{_{\mathrm{#1}}}
\newcommand{\indups}[1]{_{\mathrm{\scriptscriptstyle #1}}}

\newcommand{\matr}[2]{\left(\begin{array}{#1}#2\end{array}\right)}

\newcommand{\atopfrac}[2]{\genfrac{}{}{0pt}{}{#1}{#2}}
\newcommand{\sfrac}[2]{{\textstyle\frac{#1}{#2}}}
\newcommand{\half}{\sfrac{1}{2}}
\newcommand{\quarter}{\sfrac{1}{4}}

\newcommand{\ellE}{\mathrm{E}}
\newcommand{\ellK}{\mathrm{K}}
\newcommand{\ellPi}{\oldPi}
\newcommand{\pint}{\makebox[0pt][l]{\hspace{3.4pt}$-$}\int}

\newcommand{\superN}{\mathcal{N}}
\newcommand{\gym}{g\indups{YM}}
\newcommand{\fldZ}{\mathcal{Z}}

\newcommand{\Op}{\mathcal{O}}
\newcommand{\Tr}{\mathop{\mathrm{Tr}}}

\newcommand{\gaugepar}{\mathfrak{p}}
\newcommand{\fldperm}{\mathcal{P}}

\newcommand{\lambdabmn}{\lambda'}
\newcommand{\coup}{g}

\newcommand{\ham}{\mathbf{H}}

\newcommand{\eng}{\mathbf{D}}
\newcommand{\dil}{\mathbf{D}}
\newcommand{\charge}{\mathbf{Q}}
\newcommand{\excharge}{\mathbf{q}}
\newcommand{\transfer}{\mathbf{T}}
\newcommand{\shift}{\mathbf{U}}
\newcommand{\contourgauge}{\mathbf{C}}
\newcommand{\resolv}{\mathbf{G}}

\newcommand{\engs}{\mathcal{E}}
\newcommand{\charges}{\mathcal{Q}}

\newcommand{\contourstring}{\mathcal{C}}
\newcommand{\resolvs}{\mathcal{G}}
\newcommand{\eival}{\psi}

\newcommand{\rap}{\varphi}
\newcommand{\rapi}{\phi}

\newcommand{\order}[1]{\mathcal{O}(#1)}

\newcommand{\trans}{{\scriptscriptstyle\mathsf{T}}}

\newcommand{\PTerm}[1]{\{#1\}}

\newcommand{\lrbrk}[1]{\left(#1\right)}
\newcommand{\bigbrk}[1]{\bigl(#1\bigr)}
\newcommand{\brk}[1]{(#1)}

\newcommand{\nln}{\nonumber\\}
\newcommand{\nl}{\nonumber\\&&\mathord{}}
\newcommand{\nlnum}{\\&&\mathord{}}
\newcommand{\earel}[1]{\mathrel{}&#1&\mathrel{}}
\newcommand{\eq}{\earel{=}}
\newenvironment{myeqnarray}{\arraycolsep0pt\begin{eqnarray}}{\end{eqnarray}\ignorespacesafterend}
\newenvironment{myeqnarray*}{\arraycolsep0pt\begin{eqnarray*}}{\end{eqnarray*}\ignorespacesafterend}

\def\[{\begin{equation}}
\def\]{\end{equation}}
\def\<{\begin{myeqnarray}}
\def\>{\end{myeqnarray}}

\makeatletter
\def\mr@ignsp#1 {\ifx\:#1\@empty\else #1\expandafter\mr@ignsp\fi}%
\newcommand{\multiref}[1]{\begingroup
\xdef\mr@no@sparg{\expandafter\mr@ignsp#1 \: }%
\def\mr@comma{}%
\@for\mr@refs:=\mr@no@sparg\do{\mr@comma\def\mr@comma{,}\ref{\mr@refs}}%
\endgroup}
\makeatother

\newcommand{\hypref}[2]{\ifx\href\asklfhas #2\else\href{#1}{#2}\fi}

\newcommand{\secref}[1]{Sec.~\multiref{#1}}

\newcommand{\appref}[1]{App.~\multiref{#1}}

\newcommand{\tabref}[1]{Tab.~\multiref{#1}}

\newcommand{\figref}[1]{Fig.~\multiref{#1}}
\renewcommand{\eqref}[1]{(\multiref{#1})}

\ifx\href\asklfhas\newcommand{\href}[2]{#2}\fi
\newcommand{\arxivno}[1]{\href{http://arxiv.org/abs/#1}{#1}}

\begin{document}

\setcounter{page}{0}\thispagestyle{empty}
\begin{flushright}\footnotesize
\texttt{\arxivno{hep-th/0405001}}\\
\texttt{AEI 2004-036}
\end{flushright}
\vspace{2cm}

\begin{center}
{\Large\textbf{\mathversion{bold} A Novel Long-Range Spin
Chain\\and Planar $\mathcal{N}=4$ Super Yang-Mills }\par}
\vspace{2cm}

\textsc{N.~Beisert, V.~Dippel and M.~Staudacher} \vspace{5mm}

\textit{Max-Planck-Institut f\"ur Gravitationsphysik\\
Albert-Einstein-Institut\\
Am M\"uhlenberg 1, 14476 Potsdam, Germany} \vspace{3mm}

\texttt{nbeisert,dipvi,matthias@aei.mpg.de}\par\vspace*{2cm}\vspace*{\fill}

\textbf{Abstract}\vspace{7mm}

\begin{minipage}{12.7cm}\small
We probe the long-range spin chain approach to planar $\superN=4$
gauge theory at high loop order. A recently employed hyperbolic
spin chain invented by Inozemtsev is suitable for the
$\alSU(2)$ subsector of the state space up to three loops, but
ceases to exhibit the conjectured thermodynamic scaling properties
at higher orders. We indicate how this may be bypassed while
nevertheless preserving integrability, and suggest the
corresponding all-loop asymptotic Bethe ansatz. We also propose
the local part of the all-loop gauge transfer matrix, leading to
conjectures for the asymptotically exact formulae for all local
commuting charges.
The ansatz is finally shown to be related to a standard
inhomogeneous spin chain.
A comparison of our ansatz to semi-classical
string theory uncovers a detailed, non-perturbative agreement
between the corresponding expressions for the infinite tower of
local charge densities. However, the respective Bethe equations
differ slightly, and we end by refining and elaborating a
previously proposed possible explanation for this disagreement.
\end{minipage}\vspace*{\fill}

\end{center}

\newpage

\section{Introduction}
\label{sec:Intro}

\subsection{Spins \ldots}

The calculation of anomalous dimensions of local composite operators in
a conformal quantum field theory such as $\superN=4$
gauge theory in four dimensions is difficult even in perturbation theory.
At one loop the relevant Feynman diagrams are easily computed, but
in general a formidable mixing problem for fields of equal classical dimension
has to be resolved. At higher loops the mixing problem worsens, and, in
addition, the Feynman diagram technique rapidly becomes prohibitive in
complexity. Recently much progress was achieved in dealing with both
problems. It was understood, initially for a scalar subsector of the fields,
that the computation can be quite generally reformulated in a combinatorial fashion
\cite{Kristjansen:2002bb,Constable:2002hw,Beisert:2002bb,Constable:2002vq},
and that this combinatorics may be efficiently treated by
Hamiltonian methods \cite{Minahan:2002ve,Beisert:2002ff}.

At one loop, Minahan and Zarembo recognized that this Hamiltonian
is, in the \emph{planar} limit, identical to the one of an
\emph{integrable} quantum spin chain \cite{Minahan:2002ve}. This
picture was successfully extended to all $\superN=4$ fields in
\cite{Beisert:2003yb}, exploiting the planar structure of the
complete non-planar one-loop dilatation operator obtained in
\cite{Beisert:2003jj}. The resulting (non-compact) $\alPSU(2,2|4)$
super spin chain does not only extend the integrable structures of
\cite{Minahan:2002ve}, it also unifies them with the ones observed
earlier for different types of operators in the context of QCD
\cite{Braun:1998id,Braun:1999te,Belitsky:1999bf}.

The second development was the realization that Hamiltonian methods are
also applicable at higher loops \cite{Beisert:2003tq}.%
\footnote{One should not confuse the Hamiltonian approach with the
integrable spin chain approach. E.g.~in \cite{Beisert:2003tq} the
full non-planar one- and two-loop dilatation operator was derived
in the $\alSU(2)$ sector. It acts on a grand-canonical ensemble of
disconnected spin chains. The $\frac{1}{N}$ corrections lead to
spin chain splitting and joining, as in \cite{Beisert:2002ff}, and
break integrability.} Most importantly, it was shown that planar
\emph{integrability} extends to at least two loops, and it was
conjectured that the full non-perturbative planar dilatation
operator of $\superN=4$ theory is identical to the Hamiltonian of
some integrable \emph{long-range} spin chain. This was achieved by
studying the two-loop deformations of the hidden commuting charges
responsible for the one-loop integrability. Based on this (and
certain further assumptions, see below) the planar three-loop
dilatation operator for the $\alSU(2)$ subsector of the state
space was derived. One of its predictions was the previously
unknown three-loop anomalous dimension of the Konishi field.
Very recently, entirely independent arguments
resulted in a conjecture for the three-loop anomalous dimension
of a twist-two Konishi descendant \cite{Kotikov:2004er}.%
\footnote{This conjecture has been confirmed by an explicit calculation
in $\superN=4$ at the two-loop level.}
It is based on extracting the $\superN=4$ anomalous dimensions
of twist operators from the exact QCD result by truncating
to contributions of the `highest transcendentality'.
The three-loop QCD result recently became available after an impressive,
full-fledged and rigorous field theoretic computation
by Moch, Vermaseren and Vogt \cite{Moch:2004pa}.
The conjecture of \cite{Kotikov:2004er} \emph{agrees} with the
prediction of \cite{Beisert:2003tq} in a spectacular fashion.

What is the evidence that the full $\superN=4$ planar dilatation
operator is indeed described by an integrable long-range spin
chain? In \cite{Beisert:2003ys} three-loop integrability was
\emph{proven} for the maximally compact closed $\alSU(2|3)$
subsector of $\alPSU(2,2|4)$. Independent confirmation comes from
a related study of dimensionally reduced
$\superN=4$ theory at
three loops \cite{Klose:2003qc}. Furthermore, the procedure of
\cite{Beisert:2003tq} of deriving the $\alSU(2)$ dilatation
operator by assuming integrability was pushed to four loops,
leading to a unique result after imposition of two further
assumptions \cite{Beisert:2003jb}. The first of these is suggested
by rather firmly established structural constraints derived from
inspection (as opposed to calculation) of Feynman diagrams. The
second, somewhat less compelling, assumption postulates a certain
\emph{thermodynamic scaling behavior}, i.e.~the $L$ dependence of
the anomalous dimension in the limit of large spin chain length $L$.

What is the relevance of the observed integrability? Apart from its
ill-understood conceptual importance for planar $\superN=4$ theory,
it allows for very efficient
calculations of anomalous dimensions by means of the Bethe ansatz,
as first derived at one loop in \cite{Minahan:2002ve,Beisert:2003yb}.
For ``long{}'' composite operators where $L \gg 1$ this computational tool is not only
useful, but indispensable.
Beyond one loop, a Bethe ansatz is also expected to exist, as the latter
is based on the principle of factorized scattering. This means that
the problem of diagonalizing a spin chain with $M$ excitations (magnons)
may be reduced to the consideration of a sequence of pairwise interactions,
i.e.~the two-body problem. This principle is one of several possible ways to
characterize integrability. And indeed a Bethe ansatz technique was
derived in \cite{Serban:2004jf} for the $\alSU(2)$ sector up to
three-loops. This involved embedding the three-loop dilatation operator
of \cite{Beisert:2003tq} into an integrable long-range spin chain invented by Inozemtsev
\cite{Inozemtsev:1989yq,Inozemtsev:2002vb}.

At four loops, however, the Inozemtsev chain clashes with the
postulate of thermodynamic scaling in an irreparable fashion
\cite{Serban:2004jf}. It thus also contradicts the four-loop
integrable structure found in \cite{Beisert:2003jb}. This is
somewhat surprising, as Inozemtsev presented arguments which
suggest that his integrable long-range chain should be the most
general one possible. One might wonder whether a spin chain with
``good{}'' thermodynamic behavior could lead to inconsistencies at
even higher loop levels. In \secref{sec:Ansatz} we will see that
this does not happen up to \emph{five} loops. In fact, it appears
that the principles of integrability, field theory structure, and
thermodynamic behavior result in a \emph{unique}, novel long-range
spin chain. As a crucial second test of this claim, we will show,
by working out a large number of rather non-trivial examples, that
the scattering of our new chain is still factorized up to five
loops. A byproduct is the successful test of the validity of the
three-loop ansatz of \cite{Serban:2004jf} for a larger set of
multi-magnon states. Our study allows us to find the Bethe ansatz
corresponding to the new chain. What is more, our findings suggest
a general pattern for the scattering which appears to be
applicable at arbitrary order in perturbation theory. Stated
differently, we propose an integrability-based
non-perturbative procedure for calculating anomalous dimensions in
the $\superN=4$ model without the need of knowing the precise
all-loop structure of its dilatation operator!

Of course the reader should keep in mind that our model is
merely an \emph{ansatz} for the treatment of the gauge theory, and
proving (or disproving) it will require new insights.%
\footnote{Accordingly, when working out the potential consequences
of our ansatz for the gauge theory below, we will for simplicity
mostly write ``gauge theory result{}'' instead of ``gauge theory
result under the assumption of the validity of the novel
long-range spin chain ansatz{}'', etc.} Furthermore, the validity
of the all-loop Bethe ansatz we are proposing is still subject to
one serious restriction. It is, as in
\cite{Inozemtsev:2002vb,Serban:2004jf}, \emph{asymptotic} in the
sense that the length of the chain (and thus of the operator) is
assumed to exceed the range of the interaction (and thus the order
in perturbation theory). We hope that this restriction will be
overcome in the future, as it might then allow to find the
spectrum of planar $\superN=4$ of finite length operators. In
addition, we suspect the restriction to be at the heart of a
recently discovered vexing discrepancy between the anomalous
dimensions of certain long operators and the energies of the
related IIB super string states in the $AdS_5 \times S^5$
background.

\subsection{\ldots and Strings}

While anomalous dimensions are of intrinsic interest to the gauge field theory,
further strong motivation for their study comes from a conjectured relation
to energies of superstrings on curved backgrounds.
Spin chains in their ferromagnetic ground state describe $\superN=4$ half-BPS
operators. Long spin chains near the ground state, with a finite number of
excitations, correspond then to near-BPS operators. These are dual, via
the AdS/CFT correspondence, to certain string states carrying large charges,
whose spectrum can be computed exactly as they are effectively propagating
in a near-flat metric \cite{Berenstein:2002jq}. This yields an all-loop
prediction of the anomalous dimensions of the near-BPS operators.
The prediction is interesting even qualitatively, as it leads to the above
mentioned thermodynamic ``BMN'' scaling behavior of the spin chain. One then finds,
combining this behavior with integrability, that the BMN prediction is also
quantitatively reproduced, as will be argued below up to five loops. This
feature is a strong argument in favor of our new chain, and thus against (beyond
three loops) the Inozemtsev model.
The all-loop extension of the five-loop spin chain assumes the repetition of the observed
pattern ad infinitum, and therefore reproduces the quantitative BMN formula
to all orders by construction. One should nevertheless stress that
BMN scaling is very hard to \emph{prove} on the gauge side (much harder than
e.g.~integrability!).
So far this has not been done rigorously beyond two loops.
Therefore the Inozemtsev model has not yet been completely ruled out
either.%
\footnote{The result of \cite{Moch:2004pa,Kotikov:2004er} fixes one of
the two remaining \cite{Beisert:2003ys} parameters of the three-loop
dilatation operator of \cite{Beisert:2003tq} which are left undetermined
unless one assumes BMN scaling behavior. Likewise, one further
three-loop dimension for a different field in the $\alSU(2)$ sector
would complete the proof of scaling at this order.}

The BMN limit is not the only situation where a quantitative comparison between
long gauge operator dimensions and large charge string energies was
successful. The large charge limit may be interpreted as a semi-classical
approximation to the string sigma model \cite{Gubser:2002tv}.
Following an idea of Frolov and Tseytlin,
this allowed to perform explicit calculations of the energies of strings rapidly
spinning in two directions on the five sphere
\cite{Frolov:2003qc,Frolov:2003xy,Arutyunov:2003uj} and successfully
comparing them, at one-loop, to $\alSU(2)$ Bethe ansatz computations
\cite{Beisert:2003xu,Beisert:2003ea}.
Using the mentioned higher-loop Inozemtsev-Bethe ansatz, it was possible to
also confirm the matching at two loops \cite{Serban:2004jf}.
The string sigma model is classically integrable and this makes explicit
computations of the energies and charges feasible
\cite{Arutyunov:2003uj,Arutyunov:2003za}. The agreement
accordingly also extends to the tower of one- and two-loop
commuting charges \cite{Arutyunov:2003rg,Arutyunov:2004xy}.
A more intuitive understanding of this agreement in sigma model
language was achieved in \cite{Kruczenski:2003gt,Kruczenski:2004kw}.

Unfortunately this encouraging pattern breaks down at \emph{three}
loops \cite{Serban:2004jf}. A similar three-loop disagreement
appeared earlier in the string analysis of the near BMN limit
presented in \cite{Callan:2003xr,Callan:2004uv}. This case will be
discussed in detail in \secref{sec:nearBMN}. One might wonder
whether the trouble is either due to a faulty embedding of the
three-loop dilatation operator into the Inozemtsev long-range spin
chain, or else, due to problems with the three-loop asymptotic
Inozemtsev-Bethe ansatz. In this paper an extended and detailed
study of the multi-magnon diagonalization using this ansatz
definitely rules out these potential pitfalls. Let us stress once
more that the three-loop disagreement of the Inozemtsev-Bethe
ansatz with string theory is \emph{unrelated} to the four-loop
breakdown of thermodynamic scaling in the Inozemtsev model. In
this paper we are definitely bypassing the second problem, and
suggesting a potential explanation of the first problem in chapter
\secref{sec:resolution}. There we will refine the conjecture,
first made in \cite{Serban:2004jf}, that the disagreement might be
explained as an order of limits problem. In particular, we
shall argue in the last chapter of this paper
that if we were to implement the same scaling procedure as in
string theory, we should include \emph{wrapping interactions}
into the gauge theory computations.
These are precisely excluded by the weak-coupling asymptotic Bethe ansatz.
The refinement also allows to gain a qualitative understanding why
the strict BMN limit might indeed agree at all orders, while
subtleties arise in the near BMN and the Frolov-Tseytlin
situations.

Further confirmation for this picture, as well as for the validity
of our novel long-range spin chain, comes from a detailed all-loop
comparison of the long-range asymptotic Bethe ansatz with the
classical Bethe equation for the string sigma model. The latter
was recently derived for the $\alSU(2)$ sector in an important
work by Kazakov, Marshakov, Minahan and Zarembo
\cite{Kazakov:2004qf}. It allowed to successfully compare the
string and gauge integrable structures%
\footnote{This was shown earlier on two specific examples by using
the B\"acklund transformation \cite{Arutyunov:2003rg} (see also
\cite{Arutyunov:2004xy}). One major advantage of the systematic
approach of \cite{Kazakov:2004qf} is its generality. It would be
interesting to also treat the B\"acklund approach in a general
fashion.} by showing that the classical equation may be mapped to
the thermodynamic limit of the quantum spin chain Bethe and
Inozemtsev-Bethe equations at, respectively, one and two-loops
\cite{Kazakov:2004qf}. Here we will extend this comparison in
chapter \secref{sec:Trouble} to all-loops. We find the intriguing
result that the \emph{local excitations} of string and gauge
theory agree to all orders in perturbation theory! The Bethe
equations describing the dynamics of the excitations however
differ, leading to differences in the expectation values of the
global charges. We suspect that this is due to the global effect
of wrapping interactions, as will be discussed in the final
\secref{sec:resolution}. These disagreements between the energy
eigenvalues of the string sigma model on the one hand and those of
the long-range spin chain (in the thermodynamic limit) on the
other hand will be illustrated by way of example in
\appref{sec:elliptic}. There we explicitly derive the energies
of the folded and the circular string using the Bethe ans\"atze
for both the novel spin chain and the string sigma model as
introduced in \cite{Kazakov:2004qf}.

Let us end by mentioning a crucial issue which is not addressed in
this paper, namely the extension to subsectors larger than the
closed $\alSU(2)$ spin $\half$ chain. In fact, essentially nothing
is known about the Bethe ansatz for larger sectors, except at one
loop \cite{Minahan:2002ve,Beisert:2003yb}. At this loop order,
there is much evidence that the triality between string theory,
gauge theory and spin chains extends to other sectors, see
e.g.~the string
\cite{Russo:2002sr,Minahan:2002rc,Tseytlin:2002ny,Frolov:2003tu,Dimov:2004qv,Hernandez:2004uw,Stefanskijr.:2004cw,Mikhailov:2004xw,Mikhailov:2004qf}
and Bethe
\cite{Engquist:2003rn,Kristjansen:2004ei,Engquist:2004bx}
computations, or even to other superconformal models containing
open strings \cite{Stefanski:2003qr,Chen:2004yf}. See also the
review paper \cite{Tseytlin:2003ii}. Subsectors containing
covariant derivatives or field strengths should be very important
in the QCD context
\cite{Braun:1998id,Braun:1999te,Belitsky:1999bf}, cf.~also the
very recent paper \cite{Ferretti:2004ba}. It would be extremely
interesting to find the analog of our asymptotic Bethe ansatz for
some larger closed sector, and ideally for the full
$\alPSU(2,2|4)$ super spin chain. Given that the full super spin
chain is certainly \emph{dynamic} \cite{Beisert:2003ys}, i.e.~the
length of the chain becomes itself a quantum variable beyond one
loop, such an ansatz, if it exists at all, will presumably contain
novel features not yet encountered in traditional exactly solvable
spin chains. In particular it should be fascinating to see how
such an ansatz might reconcile two seemingly contradictory
features of the interactions of the elementary excitations of such
chains. These features are, for one, the principle of elastic
scattering, as usually required by integrability, and, secondly,
the occurrence of particle production in dynamic, long-range spin
chains, where fermionic and bosonic degrees of freedom are not
separately conserved. Could it be that supersymmetry will lead to
a generalization of the traditional notion of a factorized
S-matrix?

\section{The Long-Range Spin Chain}
\label{sec:Ansatz}

The $\alSU(2)$ sector of $\superN=4$ SYM consists of local operators
composed from two charged scalar fields $\fldZ,\phi$.
In the planar limit, local operators are single trace, and of the form
\[\label{eq:ops}
\Tr \left(\fldZ^{L-M} \phi^M \right) + \ldots \, ,
\]
where the dots indicate linear mixing of the elementary fields $\fldZ,\phi$
inside the trace in order to form eigenstates of the dilatation generator.
These can be represented by eigenstates of a cyclic $\alSU(2)$ quantum spin chain
of length $L$ with
elementary spin $\frac{1}{2}$ \cite{Minahan:2002ve}.
In this picture $M$ is the number of ``down spins{}'' or
``magnons{}'', and the dilatation operator, which closes on this
subsector\cite{Beisert:2003tq},
corresponds to the spin chain Hamiltonian. For future use let us also
introduce the letter $J$ for the number of ``up{}'' spins
\[\label{eq:charge}
J=L-M \, ,
\]
which is the standard notation \cite{Berenstein:2002jq} for the
total $\alSO(2) \subset \alSO(6)$ charge of the chiral scalar
fields $\fldZ$. Minahan and Zarembo have discovered that the
dilatation operator at the one-loop level is in fact integrable
\cite{Minahan:2002ve} and thus isomorphic to the Heisenberg
XXX$_{1/2}$ spin chain. In \cite{Beisert:2003tq} it was shown that
integrability extends to two-loops and conjectured that it might
hold at all orders in perturbation theory or even for finite
't~Hooft coupling constant.

In this section we will investigate a possible
extension of the dilatation operator to higher loop orders.
We will make use of three key assumptions:
\begin{list}{.}{\leftmargin2.5em\itemsep0pt}
\item[($i$)] Integrability,
\item[($ii$)] field theoretic considerations, and
\item[($iii$)] BMN scaling behavior.
\end{list}
These are not firm facts from gauge theory, but there are reasons
to believe in them. For example, at three-loops integrability
follows from (firm) field theoretic constraints and superconformal
symmetry \cite{Beisert:2003ys}.
One could argue that BMN scaling is an analog of the
`highest transcendentality' conjecture
(see \cite{Kotikov:2004er} and references therein):
In $\superN=4$ SYM all contributions
scale with the maximum allowed power of $1/J$.
What is more, a
conjecture for a three-loop anomalous dimension
in $\superN=4$ SYM \cite{Kotikov:2004er}, which
rests on entirely unrelated assumptions while being based
on a rigorous \emph{tour-de-force} computation for QCD
\cite{Moch:2004pa},
agrees with the prediction of the spin chain and thus confirms our
premises to some extent. Whether or not the assumptions are
fully justified in (perturbative) $\superN=4$ SYM will not be the
subject of this chapter, but we believe that the model shares
several features with higher-loop gauge theory and therefore
deserves an investigation. Intriguingly, it will turn out to be
\emph{unique} up to (at least) five-loops and agree with the
excitation energy formula in the BMN limit! At any rate, this
makes it a very interesting model to consider in its own right.
For a related, very recent study see \cite{Ryzhov:2004nz}.

There are two approaches to integrable quantum spin chain models.
One uses the Hamiltonian and the corresponding commuting charges.
The other employs factorized scattering and the Bethe ansatz
technique. We will discuss these two approaches in the following
two subsections. By means of example we shall demonstrate the
equivalence of both models in \secref{sec:Results} and
\appref{sec:Spec}.

\subsection{Commuting Charges}
\label{sec:CommCharge}

Let us start by describing the set of charges as operators
acting on the spin chain.
Introducing a coupling constant $g$ by
\[
g^2=\frac{\gym^2N}{\fes{4}{8}{16}{8}\pi^2}=\frac{\lambda}{\fes{4}{8}{16}{8}\pi^2}\,,
\]
we expand the charges in a Taylor series
\[\charge_{r}(g)=\sum_{\ell=1}^\infty \charge_{r,2\ell-2}\,g^{2\ell-2}.
\]
The dilatation operator $\dil$ can be expressed in terms of the
spin chain Hamiltonian $\ham$ which is defined as the second
charge $\charge_2$
\[
\dil(g)=L+g^2\,\ham(g),\qquad \ham(g)=\charge_2(g).
\]
Any $\alSU(2)$ invariant interaction can be written
as a permutation of spins.
These can in turn be represented in terms
of elementary permutations $\fldperm_{p,p+1}$ of
adjacent fields. A generic term will thus be written as%
\footnote{This notation was introduced in \cite{Beisert:2003tq}
where one can find a set of rules for the simplification of involved
expressions.}
\[\label{eq:SU2.Perm}
\PTerm{p_1,p_2,\ldots}=
\sum_{p=1}^L
\fldperm_{p+p_1,p+p_1+1}
\fldperm_{p+p_2,p+p_2+1}\ldots\,.
\]
For example, in this notation the one-loop dilatation generator
\cite{Minahan:2002ve} is given by
\[\label{eq:SU2.PermH0}
\ham_0=\charge_{2,0}=\fes{\half}{}{2}{}\bigbrk{\PTerm{}-\PTerm{1}}.
\]
This notation is useful due to the nature of maximal scalar
diagrams as discussed in \cite{Beisert:2003tq}: An interaction of
scalars at $\ell$-loops with the maximal number of $2+2\ell$ legs
can be composed of $\ell$ crossings of scalar lines. In the planar
limit, the crossings correspond to the elementary permutations and
at $\ell$-loops there should be no more than $\ell$ permutations.
In field theory this is a feature of maximal diagrams, but here we
will assume the pattern to hold in general. Furthermore, a general
feature of ordinary (one-loop) spin chains is that the $r$-th
charge at leading order can be constructed from $r-1$ copies of
the Hamiltonian density which, in this case \eqref{eq:SU2.PermH0},
is essentially an elementary permutation. We will therefore assume
the contributions to the charges to be of the form
\[\label{eq:SU2.ChargePerm}
\charge_{r,2\ell-2}\sim\PTerm{p_1,\ldots,p_m}\quad
\mbox{with }m\leq r+\ell-2\mbox{ and }1\leq p_i\leq r+\ell-2.
\]
Finally, the even (odd) charges should be
parity even (odd) and (anti)symmetric.%
\footnote{In fact, the Hamiltonian $\ham(g)$ and charges
$\charge_r(g)$ should be hermitian. The coefficients
of the interaction structures should therefore be real (imaginary)
for even (odd) $r$. Reality of the Hamiltonian follows from the
equivalence of the Hamiltonian for the $\alSU(2)$ sector and its
conjugate.}
Parity~$\gaugepar$ acts on the interactions as
\[\label{eq:SU2.PermParity}
\gaugepar\,\PTerm{p_1,\ldots,p_m}\,\gaugepar^{-1}
=\PTerm{-p_1,\ldots,-p_m},
\]
whereas symmetry acts as
\[\label{eq:SU2.PermSymmetry}
\PTerm{p_1,\ldots,p_m}^{\trans}=\PTerm{p_m,\ldots,p_1}.
\]
Symmetry will ensure that the eigenvalues of the
charges are real.

We can now write the dilatation operator up to three loops
as given in \cite{Beisert:2003tq}
\<\label{eq:SU2.SU23}
\ham_0\eq
\fes{\half}{}{2}{} \PTerm{}
-\fes{\half}{}{2}{}\PTerm{1},
\nln
\ham_2\eq
-\fes{\sfrac{1}{2}}{2}{8}{2}\PTerm{}
+\fes{\sfrac{3}{4}}{3}{12}{3}\PTerm{1}
-\fes{\sfrac{1}{8}}{\half}{2}{\half}\bigbrk{\PTerm{1,2}+\PTerm{2,1}},
\nln
\ham_4\eq
\fes{\sfrac{15}{16}}{\sfrac{15}{2}}{60}{\sfrac{15}{2}}\PTerm{}
-\fes{\sfrac{13}{8}}{13}{104}{13}\PTerm{1}
+\fes{\sfrac{1}{16}}{\half}{4}{\half}\PTerm{1,3}
\nl
+\fes{\sfrac{3}{8}}{3}{24}{3}\bigbrk{\PTerm{1,2}+\PTerm{2,1}}
-\fes{\sfrac{1}{16}}{\half}{4}{\half}\bigbrk{\PTerm{1,2,3}+\PTerm{3,2,1}}.
\>
The one-loop contribution has been computed explicitly in field theory
\cite{Minahan:2002ve}.
To obtain the higher-loop contributions
it is useful to rely on the (quantitative) BMN limit,
this suffices for the two-loop contribution \cite{Beisert:2003tq}.
At three-loops the BMN limit fixes
all but a single coefficient.
It can be uniquely fixed if,
in addition, integrability is imposed \cite{Beisert:2003tq}.%
\footnote{Alternatively one may use superconformal invariance
\cite{Beisert:2003ys} or the input from the conjectured
result of \cite{Kotikov:2004er}, based on the rigorous computation of
\cite{Moch:2004pa}, all of which are compatible with
\eqref{eq:SU2.SU23}.}
The same is true at four-loops
\cite{Beisert:2003jb} and five-loops \cite{Beisert:2004ry}, the
BMN limit and integrability \emph{uniquely} fix the Hamiltonian.
We present the five-loop Hamiltonian
in \tabref{tab:SU2.FourFive} in
\appref{sec:Spec.Low}.
Expressions for the higher charges can be found in
\cite{Beisert:2004ry} along with a set of \texttt{Mathematica}
routines to compute scaling dimensions explicitly and commutators
of charges in an abstract way.

There are some interesting points to be mentioned regarding this
solution. First of all, integrability and the thermodynamic limit
fix exactly the right number of coefficients for a unique solution
(up to five loops). Moreover we can give up on the quantitative
BMN limit and only require proper scaling behavior. This merely
allows for two additional degrees of freedom and the most general
Hamiltonian would be given by $\ham'(g)=c_1 \ham (c_2g)$. The
constants $c_1,c_2$ correspond to symmetries of the defining
equations, they can therefore not be fixed by algebraic arguments,
but the quantitative BMN limit requires $c_1=c_2=1$. For this
solution, the contribution $\delta \dil_n$ of one excitation of
mode $n$ to the scaling dimension in the BMN limit is given by
\[\label{eq:SU2.BMN}
\delta\dil_n=c_1
\lrbrk{\sqrt{1+c_2^2\,\lambdabmn\, n^2 }-1}
+\order{g^{12}}.
\]
where the BMN coupling constant $\lambda'$ is defined as
\[\label{eq:bmncoupling}
\lambdabmn=\fes{4}{8}{16}{8}\pi^2\frac{g^2}{J^2}
\qquad \mbox{i.e.} \qquad \lambdabmn=\frac{\lambda}{J^2},
\]
and $J$ has been defined in \eqref{eq:charge}.
It is interesting to observe that the BMN square-root formula
\eqref{eq:BMN.Energy} for the energy of one
excitation is predicted correctly.
It would be important to better understand this intriguing
interplay between integrability, and (qualitative and quantitative)
thermodynamic scaling behavior.
See also \cite{Ryzhov:2004nz}.

There is however one feature of the dilatation operator which
cannot be accounted for properly. For increasing loop order $\ell$
the length of the interaction, $\ell+1$, grows. For a fixed length
$L$ of a state, the interaction is longer than the state when
$\ell\geq L$. In this case the above Hamiltonian does not apply,
it needs to be extended by \emph{wrapping interactions} which
couple only to operators of a fixed length. In planar field theory
these terms exist, they correspond to Feynman diagrams which fully
encircle the state. We will comment on this kind of interactions
in \secref{sec:wrapping}. Here we only emphasize that $\charge_r$
is reliable only up to and including $\order{g^{2L-2r}}$.

\subsection{Long-Range Bethe Ansatz}
\label{sec:LongRange}

Minahan and Zarembo have demonstrated the equivalence of the
one-loop, planar dilatation operator in the $\alSU(2)$ subsector
with the XXX$_{1/2}$ Heisenberg spin chain \cite{Minahan:2002ve}.
The discovery of integrability opens up an alternative way to
compute energies, namely by means of the algebraic Bethe ansatz.
Serban and one of us have recently shown how to extend the Bethe
ansatz to account for up to three-loop contributions
\cite{Serban:2004jf}. This ansatz is based on the Inozemtsev spin
chain \cite{Inozemtsev:1989yq,Inozemtsev:2002vb}
after a redefinition of the coupling constant and the charges.
For the Inozemtsev spin chain there exists an
asymptotic%
\footnote{\emph{Asymptotic} refers to the fact that wrapping
interactions are probably not taken into account correctly, see
also \secref{sec:wrapping}.} 
Bethe ansatz. It makes use of the
Bethe equations
\[\label{eq:Ansatz}
\exp(iLp_k)=\prod_{\textstyle\atopfrac{j=1}{j\neq k}}^M
\frac{\rap(p_k)-\rap(p_j)+\fes{2i}{i}{\sfrac{i}{2}}{i}}
{\rap(p_k)-\rap(p_j)-\fes{2i}{i}{\sfrac{i}{2}}{i}}\, .
\]
The left hand side is a free plane wave phase factor for the
$k$-th magnon, with momentum $p_k$, going around the chain.
The right hand side is ``almost{}'' one, except for a sequence
of \emph{pairwise, elastic} interactions with the $M-1$ other magnons,
leading to a small phase shift. Without this phase shift,
the equation simply leads to the standard momentum quantization
condition for a free particle on a circle. The details of the
exchange interactions are encoded into the functions $\rap(p_k)$,
and definitely change from model to model, but the two-body
nature of the scattering is the universal feature leading to
integrability. It allows the reduction of an $M$-body problem to
a sequence of two-body problems.
The energy and higher charge eigenvalues%
\footnote{Note that here and in the rest of the paper we will, somewhat
loosely, avoid to notionally or, at times, semantically distinguish
between operators and their eigenvalues, as it should always be
clear from the context what is meant. If it is not, the statement
should be true in both interpretations.}
are then given by the linear sum of contributions
from the individual magnons
\[\label{eq:partonsum}
\charge_r=\sum^M_{k=1} \excharge_r(p_k),\qquad \ham=\charge_2.
\]
Again, this additive feature is due to the nearly complete independence
of the individual excitations. However, the details of the
contribution of an individual excitation to the $r$-th
charge, $\excharge_r(p_k)$, depend once more on the precise integrable model.
For example, the XXX$_{1/2}$ Bethe ansatz is obtained by setting
\[\label{eq:oneloopansatz}
\rap(p)=\fes{}{\half}{\quarter}{\half}\cot(\half p),\qquad
\excharge_r(p)=
\frac{\fes{2}{2^r}{2^{2r-1}}{2^r}}{r-1}\,\sin\bigbrk{\half(r-1)p}\sin^{r-1}(\half p).
\]
The Bethe roots or \emph{rapidities} $\rap_k$,
defined as $\rap_k=\rap(p_k)$ are
also denoted by $\lambda_k$ or $u_k$ in the literature.
The inversion of \eqref{eq:oneloopansatz}
\[
\exp(ip)=
\frac{\rap+\fes{i}{\sfrac{i}{2}}{\sfrac{i}{4}}{\sfrac{i}{2}}}
{\rap-\fes{i}{\sfrac{i}{2}}{\sfrac{i}{4}}{\sfrac{i}{2}}}\,,
\qquad
\excharge_r(\rap)=\frac{i}{r-1}
\lrbrk{\frac{1}{(\rap+\fes{i}{\sfrac{i}{2}}{\sfrac{i}{4}}{\sfrac{i}{2}})^{r-1}}-
       \frac{1}{(\rap-\fes{i}{\sfrac{i}{2}}{\sfrac{i}{4}}{\sfrac{i}{2}})^{r-1}}}
\]
leads to the common and fully algebraic description in terms of
rapidities $\rap_k$ instead of momenta $p_k$.

The phase relation
$\rap(p)=\rap(p,g)$ of the modified Inozemtsev spin chain is given by
\cite{Serban:2004jf}
\[\label{eq:inophase}
\rap(p)=\fes{}{\half}{\quarter}{\half} \cot(\half p)
\bigbrk{
1+\fes{2}{4}{8}{4}g^2\sin^2(\half p)
-\fes{2}{8}{32}{8}g^4\sin^4(\half p)
+\fes{}{8}{64}{8}g^6\sin^4(\half p)
+\fes{2}{16}{128}{16}g^6\sin^6(\half p)
+\ldots
}
\]
and the single-excitation energy is
\[\label{eq:inoeng}
\excharge_2(p)=
\fes{2}{4}{8}{4}\sin^2(\half p)
-\fes{2}{8}{32}{8}g^2\sin^4(\half p)
+\fes{4}{32}{256}{32}g^4\sin^6(\half p)
-\fes{10}{160}{2560}{160}g^6\sin^8(\half p)
+\ldots
\]
This reproduces scaling dimensions in gauge theory up to three
loops, $\order{g^6}$, when the dilatation operator $\dil$ is
identified with the spin chain Hamiltonian $\ham$ as follows
\[\label{eq:Energy}
\dil(g)=L+g^2 \ham(g),\qquad \textnormal{with }
\ham=\charge_2=\sum_{k=1}^M \excharge_2(p_k).
\]
Starting at four loops%
\footnote{Order $\order{g^8}$ in $\dil(g)=L+g^2 \ham(g)$
correspond to $\order{g^6}$ in $\ham(g)$.} it was noticed that the
scaling dimensions do not obey BMN scaling. As this was an
essential input for the construction of the model in
\secref{sec:CommCharge}, the energies of the modified Inozemtsev
spin chain cannot agree with the model. The breakdown of BMN
scaling can be traced back to the term
proportional to $g^6\sin^4(\half p)$ in \eqref{eq:inophase}. While
all the other terms are of $\order{J^0}$ in the BMN limit, this
one is of order $J^2$.

Our aim is to find a Bethe ansatz for the spin chain model described
in \secref{sec:CommCharge}, therefore we shall make an ansatz for
$\rap(p),\excharge_2(p)$ which is similar to \eqref{eq:inophase}, but
which manifestly obeys BMN scaling
\<\label{eq:phaseengexp}
\rap(p)\eq\fes{}{\half}{\quarter}{\half}\cot(\half p)
\sum_{\ell=1}^\infty \alpha_\ell \,\sin^{2\ell-2}(\half p)\,g^{2\ell-2}\,,
\nln
\excharge_2(p)\eq
\sum_{\ell=1}^\infty \beta_\ell \,\sin^{2\ell}(\half p)\,g^{2\ell-2}\,.
\>
By comparison to \eqref{eq:inophase,eq:inoeng} we find
$\alpha_{1,2,3}=+1,+\fes{2}{4}{8}{4},-\fes{2}{8}{32}{8}$ and
$\beta_{1,2,3}=+\fes{2}{4}{8}{4},-\fes{2}{8}{32}{8},+\fes{4}{32}{256}{32}$.
Interestingly, a comparison of energies
at four and five loops
with our spin chain model shows that we can
indeed achieve agreement
(see \secref{sec:Results} and \appref{sec:Spec}).
The correct coefficients are $\alpha_{4,5}=+\fes{4}{32}{256}{32},-\fes{10}{160}{2560}{160}$ and
$\beta_{4,5}=-\fes{10}{160}{2560}{160},+\fes{28}{896}{28672}{896}$.
Now it is not hard to guess, by ``physicist's induction{}'',
analytic expressions for the phase relation
\[\label{eq:Phase}
\rap(p)=\fes{}{\half}{\quarter}{\half} \cot(\half p)\sqrt{1+\fes{4}{8}{16}{8}g^2\sin^2(\half p)}
\]
and the magnon energy
\[\label{eq:magnoneng}
\excharge_2(p)=
\frac{1}{g^2}\lrbrk{\sqrt{1+\fes{4}{8}{16}{8}g^2\sin^2(\half p)}-1}
\]
which agree up to $\ell=5$ in \eqref{eq:phaseengexp}.
Furthermore, we have also found a generalization of the
higher charges \eqref{eq:oneloopansatz} to higher loops
\[\label{eq:Charges}
\charge_r=\sum^M_{k=1} \excharge_r(p_k),\qquad
\excharge_r (p)=
\frac{2\sin\bigbrk{\half (r-1)p}}{r-1}
\lrbrk{\frac{\sqrt{1+\fes{4}{8}{16}{8}g^2\sin^2(\half p)}-1}
{2g^2\sin(\half p)}}^{r-1}.
\]
A non-trivial consistency check of the conjectured all-order expressions
\eqref{eq:Phase,eq:magnoneng,eq:Charges} will be
performed in \secref{sec:Trouble} where we will find
a remarkable link to the predictions of semi-classical string theory.
Note that the one-particle momentum does not depend on the coupling $g$ and
is given by $\excharge_1(p,g)=p$.
The charges can be summed up into the
``local{}'' part of the transfer matrix
\[\label{eq:translocal}
\transfer(x)=\exp i \sum_{r=1}^{\infty} \, x^{r-1}\,\charge_r + \dots\, ,
\]
and we find its eigenvalue from \eqref{eq:Charges} to be
\[\label{eq:Transfer}
\transfer(x)=\prod_{k=1}^M \frac{\,\,\,\displaystyle
x-\frac{2g^2\exp(+\sfrac{i}{2}p_k)\sin(\half p_k)}
                                  {\sqrt{1+\fes{4}{8}{16}{8}g^2\sin^2(\half p_k)}-1}\,\,\,}
{\,\,\,\displaystyle
x-\frac{2g^2\exp(-\sfrac{i}{2}p_k)\sin(\half p_k)}
                            {\sqrt{1+\fes{4}{8}{16}{8}g^2\sin^2(\half p_k)}-1}\,\,\,}+\ldots\,.
\]

The transfer matrix at $x=0$ gives the total phase shift along the chain
\[\label{eq:Shift}
\shift=\transfer(0)=\prod_{k=1}^M \exp(ip_k),
\]
which should equal $\shift=1$ for gauge theory states with cyclic symmetry.
The dots in \eqref{eq:translocal,eq:Transfer}
indicate further possible terms like $x^L$ or
$g^{2L}$ which cannot be seen for the lower charges or at lower
loop orders. It is possible that finding these terms will allow for a
modification of the asymptotic Bethe equations
\eqref{eq:Ansatz} so as to make them exact,
i.e.~they would then correctly take into account the gauge theoretic
wrapping interactions.
In chapter \secref{sec:resolution} we will suggest that the latter are
responsible for the recently observed gauge-string disagreements
for long operators \cite{Callan:2003xr,Callan:2004uv,Serban:2004jf}.

Our long-range Bethe equations \eqref{eq:Ansatz,eq:Phase},
along with the expressions \eqref{eq:Charges} for the charge densities,
look somewhat involved. We will now show in \secref{sec:Rapid}
that they may be significantly simplified by fully eliminating
the momentum variables $p_k$ and expressing them through
rapidity variables $\rap_k$. In particular,  this replaces all
trigonometric expressions by rational or algebraic ones.
Furthermore, it uncovers the remarkable analytic structure of the ansatz,
which, interestingly, largely survives in the thermodynamic
limit, cf.~\secref{sec:Trouble}. What is more, it allows
us to find an intriguing link to \emph{inhomogeneous} spin chains,
as we shall elaborate in \secref{sec:Inhomo}. Apart from its
potential conceptual importance, the last observation allows one to
also correctly treat certain ``singular{}'' solutions of the Bethe ansatz,
as will be explained in some detail in \appref{sec:Spec.Singular}.
After these conceptual elaborations we will proceed in
\secref{sec:Results} to actually \emph{test} that our long-range
Bethe ansatz indeed properly diagonalizes the Hamiltonian
proposed in \secref{sec:CommCharge}. An application to
the near-BMN limit is presented in \secref{sec:nearBMN}.

\subsection{The Rapidity Plane}
\label{sec:Rapid}

The Bethe equation \eqref{eq:Ansatz} involves momenta $p_k$ on
the left hand side and rapidities $\rap_k$ on
the right hand side. The relation
$\rap(p)$ defined in \eqref{eq:Phase}
specifies the precise nature of the model.
In the previous section we have used the momenta $p_k$
as the fundamental variables and $\rap_k=\rap(p_k)$
as derived variables. Here we would like to take the opposite
point of view and consider $\rap_k$ as fundamental.
For that purpose we need to invert the relation \eqref{eq:Phase},
there turns out to be a remarkably simple form%
\[\label{eq:pofphi}
\exp(ip)=\frac{x(\rap+\fes{i}{\sfrac{i}{2}}{\sfrac{i}{4}}{\sfrac{i}{2}})}
{x(\rap-\fes{i}{\sfrac{i}{2}}{\sfrac{i}{4}}{\sfrac{i}{2}})}
\]
where%
\footnote{Note that $x(\rap)$ is odd under $\rap\mapsto -\rap$.
This property is more manifest if we replace the square root by
$\rap\sqrt{1-\fes{4}{2}{}{}\coup^2/\rap^2}$
which also straightforwardly yields the correct perturbative expansion
for small $g$.}
\[\label{eq:xofphi}
x(\rap)
=\sfrac{1}{2}\rap+
\sfrac{1}{2}\sqrt{\rap^2-\fes{4}{2}{}{2}g^2}.
\]
The Bethe equations \eqref{eq:Ansatz} can now be conveniently written without
trigonometric functions as
\[\label{eq:BethePhi}
\frac{x(\rap_k+\fes{i}{\sfrac{i}{2}}{\sfrac{i}{4}}{\sfrac{i}{2}})^L}
     {x(\rap_k-\fes{i}{\sfrac{i}{2}}{\sfrac{i}{4}}{\sfrac{i}{2}})^L}=
\prod_{\textstyle\atopfrac{j=1}{j\neq k}}^M
\frac{\rap_k-\rap_j+\fes{2i}{i}{\sfrac{i}{2}}{i}}
     {\rap_k-\rap_j-\fes{2i}{i}{\sfrac{i}{2}}{i}}\,
\]
in great similarity with the Bethe ansatz for the Heisenberg model.
In the new variables $\transfer(x)$ simplifies drastically to
\[\label{eq:TransferPhi}
\transfer(x)=
\prod_{k=1}^M\frac{x-x(\rap_k+\fes{i}{\sfrac{i}{2}}{\sfrac{i}{4}}{\sfrac{i}{2}})}
                  {x-x(\rap_k-\fes{i}{\sfrac{i}{2}}{\sfrac{i}{4}}{\sfrac{i}{2}})}+\ldots\,.
\]
The local charges $\charge_r$ follow from the expansion
\eqref{eq:translocal}
\[\label{eq:ChargesPhi}
\charge_r=
\sum_{k=1}^M
\excharge_r(\rap_k),\qquad
\excharge_r(\rap)=\frac{i}{r-1}
\lrbrk{\frac{1}{x(\rap+\fes{i}{\sfrac{i}{2}}{\sfrac{i}{4}}{\sfrac{i}{2}})^{r-1}}-
       \frac{1}{x(\rap-\fes{i}{\sfrac{i}{2}}{\sfrac{i}{4}}{\sfrac{i}{2}})^{r-1}}}.
\]

It is interesting to see that the transfer matrix $\transfer(x)$
is not the obvious guess related to the Bethe equation
\eqref{eq:BethePhi}.
In analogy to the Heisenberg model, see e.g.~\cite{Faddeev:1996iy},
the immediate guess $\bar\transfer(\rap)$ would be%
\[\label{eq:BarTransfer}
\bar\transfer(\rap)=
x(\rap+\fes{i}{\sfrac{i}{2}}{\sfrac{i}{4}}{\sfrac{i}{2}})^L
\prod_{k=1}^M
\frac{\rap-\rap_k-\fes{2i}{i}{\sfrac{i}{2}}{i}}
     {\rap-\rap_k}
+x(\rap-\fes{i}{\sfrac{i}{2}}{\sfrac{i}{4}}{\sfrac{i}{2}})^L
\prod_{k=1}^M
\frac{\rap-\rap_k+\fes{2i}{i}{\sfrac{i}{2}}{i}}
     {\rap-\rap_k}\,.
\]
The Bethe equations \eqref{eq:BethePhi} follow from the
cancellation of poles at $\rap=\rap_j$.
The charges $\bar\charge_r$ obtained from
$\bar\transfer(\rap)$ as
$\bar\transfer(\rap+\frac{i}{2})/x(\rap+i)^L=\exp i\sum_{r=1}^\infty \rap^{r-1}\bar\charge_r+\ldots$
\[\label{eq:BarCharge}
\bar\charge_r=
\sum_{k=1}^M \frac{i}{r-1}
\lrbrk{\frac{1}{(\rap_k+\fes{i}{\sfrac{i}{2}}{\sfrac{i}{4}}{\sfrac{i}{2}})^{r-1}}-
       \frac{1}{(\rap_k-\fes{i}{\sfrac{i}{2}}{\sfrac{i}{4}}{\sfrac{i}{2}})^{r-1}}}.
\]
In perturbation theory we can relate
these charges to the physical charges $\charge_r$ by
\[\label{eq:ChargesTranslate}
\charge_r=\bar\charge_r
+\fes{}{\sfrac{1}{2}}{\sfrac{1}{4}}{\sfrac{1}{2}}(r+1)g^2\bar\charge_{r+2}
+\fes{\sfrac{1}{2}}{\sfrac{1}{8}}{\sfrac{1}{32}}{\sfrac{1}{8}}(r+2)(r+3)g^4\bar\charge_{r+4}
+\ldots
\]
The form of these charges on an operatorial level is not clear to us.
At one-loop they agree with the physical charges, but
at higher loops their range grows twice as fast with the loop order
as for $\charge_r$. Therefore $\bar\charge_2$ is clearly
not a suitable candidate for the dilatation operator $\dil$.
In the next section we will find a
natural explanation using, however, a different basis.

Let us comment on \eqref{eq:xofphi} and its inverse
\[\label{eq:uofx}
\rap(x)=x+\frac{g^2}{\fes{}{2}{4}{2}x}\,.
\]
The map between $x$ and $\rap$ is a double covering map.
For every value of $\rap$ there are two corresponding values of $x$,
namely
\[
\rap\longleftrightarrow
\left\{
x,
\frac{g^2}{\fes{}{2}{4}{2}x}
\right\}.
\]
For small values of $g$, where the Bethe ansatz
describes the long-range spin chain, we will always assume that
$x\approx \rap$.
When $g$ is taken to be large
(if this makes sense at all is a different question), however,
special care is needed in selecting the appropriate
branch.
The double covering map for $x$ and $\rap$ has an analog
for the transfer matrices $\transfer(x)$ and $\bar\transfer(\rap)$.
We find the relation
\[
\frac{\transfer(x)\,
\transfer(g^2/\fes{}{2}{4}{2}x)}{\transfer(0)}
\approx\frac{\bar\transfer(\rap(x)+\fes{i}{\sfrac{i}{2}}{\sfrac{i}{4}}{\sfrac{i}{2}})}
      {x(\rap(x)+\fes{2i}{i}{\sfrac{i}{2}}{i})^L}\,.
\]
which holds if the second term in \eqref{eq:BarTransfer} is
dropped. It can be proved by using the double covering relation
\[
(x-x')\lrbrk{1-\frac{g^2}{\fes{}{2}{4}{2}xx'}}
=
\lrbrk{x+\frac{g^2}{\fes{}{2}{4}{2}x}}-
\lrbrk{x'+\frac{g^2}{\fes{}{2}{4}{2}x'}}
=
\rap-\rap'.
\]
We believe it is important to further study the implications of the double
covering maps. This might lead to insight into the definition of our
model, possibly even beyond wrapping order.

\subsection{The Inhomogeneous Bethe Ansatz}
\label{sec:Inhomo}

The equations \eqref{eq:BethePhi,eq:BarTransfer} are very similar
to the Bethe ansatz for an inhomogeneous spin chain,
see e.g.~\cite{Faddeev:1996iy}.%
\footnote{See \appref{sec:InhomoChains} for a reformulation
of more general long-range spin chains,
in particular the Inozemtsev spin chain, in terms
of an inhomogeneous spin chain.}
The only difference is that
the inhomogeneous Bethe ansatz requires a polynomial of degree $L$ in $\rap$ whereas
the function $x(\rap)^L$ also contains negative powers of $\rap$
(when expanded for small $g$).
In fact, for the function $x(\rap)$ as defined in \eqref{eq:xofphi},
the negative powers only start at $g^{2L}$ and are
irrelevant for the desired accuracy of our asymptotic model.
In order to relate our equations to a well-known model
we could truncate the expansion of $x(\rap)^L$ at $\order{g^{2L}}$.
Remarkably this truncation can be achieved analytically:
The inverse
\eqref{eq:uofx} can be used to show that
\[
P_L(\rap)=x(\rap)^L+\lrbrk{\frac{g^2}{\fes{}{2}{4}{2}x(\rap)}}^L
\]
is a polynomial
of degree $L$ in $\rap$,
which is exactly the proposed truncation at
$\order{g^{2L}}$.
In fact, the polynomial is easily factorized as follows
\[
P_L(\rap)
=\prod_{p=1}^{L}(\rap-\rapi_p)\quad
\mbox{with}\quad
\rapi_p=\fes{2}{\sqrt{2}\,}{}{\sqrt{2}\,}g\cos\frac{\pi(2p-1)}{2L}\,.
\]
We now replace $x(\rap)^L$ by
$P_L(\rap)$ in the Bethe equation \eqref{eq:BethePhi}
\[\label{eq:BethePhiNew}
\frac{P_L(\rap_k+\fes{i}{\sfrac{i}{2}}{\sfrac{i}{4}}{\sfrac{i}{2}})}
     {P_L(\rap_k-\fes{i}{\sfrac{i}{2}}{\sfrac{i}{4}}{\sfrac{i}{2}})}=
\prod_{\textstyle\atopfrac{j=1}{j\neq k}}^M
\frac{\rap_k-\rap_j+\fes{2i}{i}{\sfrac{i}{2}}{i}}
     {\rap_k-\rap_j-\fes{2i}{i}{\sfrac{i}{2}}{i}}
\]
and transfer matrix \eqref{eq:BarTransfer}
\[\label{eq:BarTransferNew}
\bar\transfer(\rap)=
P_L(\rap+\fes{i}{\sfrac{i}{2}}{\sfrac{i}{4}}{\sfrac{i}{2}})
\prod_{k=1}^M
\frac{\rap-\rap_k-\fes{2i}{i}{\sfrac{i}{2}}{i}}
     {\rap-\rap_k}
+P_L(\rap-\fes{i}{\sfrac{i}{2}}{\sfrac{i}{4}}{\sfrac{i}{2}})
\prod_{k=1}^M
\frac{\rap-\rap_k+\fes{2i}{i}{\sfrac{i}{2}}{i}}
     {\rap-\rap_k}\,,
\]
and obtain the Bethe ansatz for an inhomogeneous spin chain with
shifts $\rapi_p$.

Let us first of all comment on the inhomogeneities.
Our spin chain is homogeneous, how can the Bethe ansatz of
an inhomogeneous spin chain describe our model?
The equation \eqref{eq:ChargesTranslate}
relates a homogeneous charge $\charge_r$ on the left hand side
with inhomogeneous charges $\bar\charge_s$ on the right hand side.
A possible resolution may be found by investigating the
inhomogeneous spin chain itself:
On the one hand, the order of the
inhomogeneities $\rapi_p$ does not matter for the Bethe ansatz and thus
for the eigenvalues of the charges.
On the other hand, it should certainly influence the eigenstates.
Consequently, the eigenstates should be related by a similarity
transformation%
\footnote{The inhomogeneities $\rapi_p$ and $\rapi_{p+1}$ can
be interchanged by conjugation with $R_{p,p+1}(\rapi_p-\rapi_{p+1})$,
we thank K.~Zarembo for a discussion on this point.}
and \eqref{eq:ChargesTranslate} is merely an equation
for the eigenvalues of the involved operators.
Alternatively, the $\bar\charge_s$ in \eqref{eq:ChargesTranslate}
can be interpreted as homogenized charges which are related to the
naive, inhomogeneous charges $\bar\charge_s$
by a change of basis.
To understand our model better, it would be essential
to investigate this point further and find the map
between our homogeneous spin chain model and the
common inhomogeneous spin chain.

Now we have totally self-consistent Bethe equations
\eqref{eq:BethePhiNew} but the physical transfer matrix
$\transfer(x)$ as defined in \eqref{eq:TransferPhi}
(consequently the energy $\dil$ and charges $\charge_r$)
does not agree with the inhomogeneous transfer matrix $\bar\transfer(\rap)$.
The physical transfer matrix involves the function $x(\rap)$.
When the coupling constant $g$ is arbitrary,
this function is ambiguous due to the two branches of the square root.
This is not a problem in perturbation theory, however,
even there inconsistencies are observed at higher order in $g$,
see \appref{sec:Spec.Singular}.
Remarkably, these appear precisely at the order where
wrapping interactions start to contribute
and our asymptotic Bethe ansatz is fully consistent
to the desired accuracy.
Conversely, there are signs of the missing of
wrapping terms.
We hope that finding a cure for the
problems beyond wrapping order
might help to find a generalization of the
Bethe equations which include wrapping interactions.
Presumably these equations will have a substantially different form.

\subsection{Comparison}
\label{sec:Results}

It is obviously important to check that our asymptotic
Bethe ansatz as developed in \secref{sec:LongRange,sec:Rapid,sec:Inhomo}
indeed properly diagonalizes the original, rather involved five-loop
Hamiltonian \eqref{eq:SU2.SU23}, \tabref{tab:SU2.FourFive} which
is currently known up to five loops. This is particularly important
as asymptotic Bethe ans\"atze such as ours are usually
not easy to derive rigorously. However, we can certainly check
its exactness by comparing its predictions to the results
obtained by brute-force diagonalization of the Hamiltonian
for specific states.
Here we summarize the results of the comparisons we have performed,
and refer the reader to \appref{sec:Spec} and \tabref{tab:Spec.Low}
for further details.

\paragraph{Two excitations.}

The perturbative Bethe ansatz gives results for two-excitation
states of arbitrary length away from the near BMN limit (see
\appref{sec:Spec.TwoEx} for details). The structure of the
energy agrees with the conjectured formula
\eqref{eq:Spec.All}, see also \cite{Beisert:2003jb}, and
the coefficients agree at five-loop accuracy.

\paragraph{Three excitations.}

For three excitations, there exist paired and unpaired solutions.
The unpaired three-excitation states are singular and
a direct computation requires the Bethe ansatz
of \secref{sec:Inhomo}. Alternatively, their
energies can be computed via \emph{mirror solutions}.
The fact that both methods (see \appref{sec:Spec.Singular})
lead to the same result hints at the
consistency of our equations.
The paired solutions for $L=7$ and $L=8$ agree with the
perturbative gauge theory results
(c.f.~\appref{sec:Spec.ThreeExPair} for details).
Finding the Bethe roots for longer spin chains becomes
more and more involved.

\paragraph{More excitations.}

All states with up to length $L\leq 8$ and a few examples
up to length $10$ have been computed in the Bethe ansatz.
Their energies and charges agree with the eigenvalues of
the spin chain operators.

\paragraph{Higher charges.}

For the afore mentioned states we have also computed the first few
orders of the higher charges $\charge_{3,4,5,6}$, using our Bethe
ansatz, and compared them to the direct diagonalization of the
long-range spin chain model. We find agreement for all instances
of states with low excitation numbers.

\paragraph{BMN limit.}

The general BMN energy formula
\[\label{eq:BMN.Energy}
\dil(\lambdabmn)-J=\sum^M_{k=1} \sqrt{1+\lambdabmn n^2_k}
\]
is easily confirmed. Regarding the single excitation energy
formula \eqref{eq:magnoneng} this is not a miracle. However, it is
fascinating to have found an integrable model where the proper BMN
behavior is more or less \emph{implemented}. This indicates that
there might be a deeper connection between integrability and the
BMN/planar limit.

\paragraph{Conclusion.}

In conclusion, we can say that for all considered examples
(including all states of length $L$ up to 8)
the Bethe ansatz yields precisely the same spectrum as the Hamiltonian
approach described in \secref{sec:CommCharge}. It shows that an
integrable spin chain of infinitely long-range, and with a
well-defined thermodynamic limit, is very likely to exist, largely
putting to rest the concerns expressed in \cite{Serban:2004jf}.
These were based on arguments \cite{Inozemtsev:2002vb} that the
elliptic Inozemtsev chain should be the most general integrable
model, paralleling analogous results for the Calogero-Moser
multi-particle system. However, the proof seems to implicitly
assume that the lowest charge only contains two-spin interactions,
whereas our new chain definitely is not of this type (see again
\cite{Serban:2004jf}). In terms of the Bethe ansatz there may seem to
be many such models. These would be obtained by appropriately
modifying the coefficients in \eqref{eq:phaseengexp}. 
If we however demand that the model is related to
an inhomogeneous spin chain as in \secref{sec:Inhomo}
we find a unique model with thermodynamic scaling behavior,
see \appref{sec:InhomoChains}.

The upshot for the integrable spin chain model is similar: 
In its construction we have assumed a very specific form of
interactions and the obtained Hamiltonian has turned out to be
unique (at five loops). In other words, the very relations
\eqref{eq:Phase,eq:magnoneng} are special and
correspond to the assumed form of interactions ($ii$).%
\footnote{This picture is rather similar to the Inozemtsev spin
chain where the requirement of pairwise interactions of spins at a
distance was shown to lead to the phase relation of the Inozemtsev
Bethe ansatz.} 
At any rate, these relations are very suggestive in
view of a correspondence to string theory on plane waves. It is
therefore not inconceivable that our Bethe ansatz does indeed
asymptotically describe planar $\superN=4$ gauge theory in the
$\alSU(2)$ subsector at higher-loops.

\subsection{The Near-BMN Limit}
\label{sec:nearBMN}

Let us now use our novel ansatz to obtain all-orders predictions
for the $1/J$ corrections to the anomalous dimensions of
BMN type operators, i.e.~let us consider the so-called
near-BMN limit. In fact, in the first non-trivial
case of states with two excitations an all-loop gauge
theory expression for this correction has been guessed
in \cite{Beisert:2003jb}. Excitingly, we shall find that
our ansatz precisely reproduces this conjecture!
The expression in question, which
agrees with \eqref{eq:Spec.All} and with
\tabref{tab:Spec.TwoEx} at five-loops, is
\[\label{eq:Spec.TwoNearBMN} \dil (J,n,\lambdabmn)=
J+2\sqrt{1+\lambdabmn\, n^2}
-\frac{4\lambdabmn\,n^2}{J\sqrt{1+\lambdabmn\,n^2}}
+\frac{2\lambdabmn\,n^2}{J(1+\lambdabmn\,n^2)} +\order{1/J^2}
\]
where $J$ and $\lambdabmn$ have been defined in, respectively,
\eqref{eq:charge} and \eqref{eq:bmncoupling}. The first $1/J$
term can be regarded as a renormalization of the term $\lambdabmn
n^2$ in the first square root. For instance, we might replace $J$
in the definition of $\lambdabmn$ by $J+4$ to absorb the second
term into the leading order energy. Unfortunately, as has already
been pointed out in \cite{Beisert:2003jb}, this formula does not
agree with the expression for the near plane-wave limit on the
string side derived in \cite{Callan:2003xr,Callan:2004uv}
\[\label{eq:Spec.TwoNearPlane}
\dil (J,n,\lambdabmn)=J+ 2\sqrt{1+\lambdabmn\, n^2}
-\frac{2\lambdabmn\,n^2}{J} +\order{1/J^2}.
\]

Now let us compute the momenta $p_k$ for the case $M=2$ and in the
near BMN limit. That is the momenta are expanded around
$J=\infty$:
\[\label{eq:BMN.TwoExc.CorrP}
p=\frac{p^{(0)}}{J}+\frac{p^{(2)}}{J^2}+\frac{p^{(4)}}{J^3}+\ldots\, .
\]
Then the Bethe equations \eqref{eq:Ansatz} are solved order by
order in $1/J$. In general, we would have to start with two
distinct Bethe equations, which determine the two roots $\rap_1,\rap_2$.
However, the momentum constraint $\shift=1$
\eqref{eq:Shift} implies in this case a symmetric distribution of
the roots in the complex plane \cite{Minahan:2002ve}: 
$\rap\equiv\rap_1=-\rap_2$, i.e.~$p\equiv p_1=-p_2$. We are thus left with
one root determined by
\[\label{eq:BMN.Class}
\exp\bigbrk{ip(J+2)}=\frac{\cot (\half p)
\,\sqrt{1+\lambdabmn J^2\pi^{-2} \sin^2 (\half p)}+i}{\cot
(\half p)\, \sqrt{1+\lambdabmn J^2\pi^{-2}
\sin^2 (\half p)}-i}\,.
\]
The general solution of this equation at leading order in $1/J$ is
then given by $p^{(0)}=2 \pi n$ where $n$ is an integer. Substituting
this back into \eqref{eq:BMN.TwoExc.CorrP}
and expanding up to $\order {1/J^2}$ yields the first correction
to the momentum
\[\label{eq:BMN.TwoExc.P1}
p^{(2)}=-\frac{2 n \pi (2 \sqrt{1+\lambdabmn
n^2}-1)}{\sqrt{1+\lambdabmn n^2}} \, .
\]
The conjectured near-BMN energy formula \eqref{eq:Spec.TwoNearBMN}
is then indeed obtained by inserting the perturbed momentum into
\eqref{eq:Energy} with \eqref{eq:magnoneng} and
expanding the result in $1/J$.

In view of recent results which might soon also allow the
computation of the $1/J^2$ corrections on the string side
\cite{Swanson:2004mk},
one easily obtains in much the same way the next order gauge
correction to the momentum
\[
p^{(4)}=
\frac{2 \pi n ( -4-6\lambdabmn n^2-2 \lambdabmn^2 n^4 )}
{(1+\lambdabmn n^2)^{5/2}}
+
\frac{2 \pi n (5+8\lambdabmn n^2+4 \lambdabmn^2 n^4)}
{(1+\lambdabmn n^2)^{2}}
\]
and the energy now reads
\<
\dil (J,n,\lambdabmn)
\eq
 J+2\sqrt{1+\lambdabmn\, n^2}
-\frac{4\lambdabmn\,n^2}{J\sqrt{1+\lambdabmn\,n^2}}
+\frac{2\lambdabmn\,n^2}{J(1+\lambdabmn\,n^2)}
\nl
+ \frac{15 \lambdabmn  n^2 + 20 \lambdabmn^2 n^4 + 8\lambdabmn^3 n^6}
{J^2 (1+\lambdabmn n^2)^{5/2}}
- \frac{n^2\pi^2(\lambdabmn n^2 +2\lambdabmn^2 n^4+\lambdabmn^3 n^6)}
{3 J^2 (1+\lambdabmn n^2)^{5/2}}
\nl
- \frac{12 \lambdabmn n^2+16 \lambdabmn^2 n^4+4\lambdabmn^3 n^6}
{J^2 (1+\lambdabmn n^2)^3} +\order{1/J^3}.
\>
Of course, agreement with string theory is not expected beyond,
at most, two loops.

\section{Stringing Spins and Spinning Strings at All Loops}
\label{sec:Trouble}

\subsection{Perturbative Gauge Theory: The Thermodynamic Limit}
\label{sec:Stringing}

The thermodynamic limit is the limit in which the length of the
spin chain $L$ as well as the number of excitations $M$ is taken
to infinity while focusing on the the low-energy spectrum. In this
limit, it was observed that the $r$-th charge $\excharge_{r,0}$
of one magnon
\eqref{eq:Charges}
at one-loop scales as $L^{-r}$ \cite{Arutyunov:2003rg}.
Here, we would like to generalize the thermodynamic limit to higher-loops.
{}From the investigation of the closely related BMN limit as well as
from classical spinning strings, we infer that each power of the
coupling constant $g$ should be accompanied by one power of $1/L$.
We thus replace $g$ according to
\[
\fes{g^2}{g^2}{g^2}{\frac{g^2}{2}}=\frac{\lambda}{\fes{4}{8}{16}{16}\pi^2}
\mapsto L^2\,\coup^2 \, ,
\]
where we have used the same symbol $\coup$ for the rescaled,
effective coupling constant in thermodynamic limit.
It is common belief that this scaling
behavior holds for perturbative gauge theory, but it is clearly
not a firm fact. We shall assume its validity for several reasons:
Firstly, it was not only confirmed at one-loop, but also at
two-loops \cite{Gross:2002su,Beisert:2003tq}. It is a nice
structure and conceptually it would be somewhat disappointing if
broken at some higher loop order. Secondly, the AdS/CFT
correspondence seems to suggest it. Thirdly, it will allow us to
define charges uniquely, see \cite{Beisert:2003ys,Arutyunov:2003rg}.

In conclusion, we expect that the scaling of charges in the
thermodynamic limit is given by
\[\label{eq:Higher.Thermo}
\excharge_{r}(g)\mapsto L^{-r}\excharge_r(\coup) \, ,
\]
Due to the large number of excitations $M=\order{L}$,
the total charge scales as
$\charge_{r}(g)\mapsto L^{-r+1}\charge_r(\coup)$.%
\footnote{In the BMN limit the total charges scale as in
\eqref{eq:Higher.Thermo}. We recall that the difference
between the BMN and the thermodynamic limit is that in the former
the magnon number $M$ stays finite.}
In particular the scaling dimension is
\[\label{eq:gaugeenergy}
\dil(g)\mapsto L\, \eng(\coup)
\qquad\mbox{with}\qquad \eng(\coup)=
1+\fes{}{}{}{2} \coup^2\,\charge_2(\coup) \, .
\]
The relevant quantities of the Bethe ansatz should behave as follows:
\[\label{eq:Scaling}
\rap_k \mapsto L \rap_k \, , \qquad x(\rap_k) \mapsto L x(\rap_k)
, \qquad p(\rap_k) \mapsto p(\rap_k)/L
\, .
\]
where%
\footnote{The correct sign for the square root
$\sqrt{\rap^2-\fes{4}{2}{}{}\coup^2}$ in perturbation theory is
the same as of $\rap$. More accurately we should write
$\rap\sqrt{1-\fes{4}{2}{}{}\coup^2/\rap^2}$\,.}
\[
x(\rap)=\half\rap+\half\sqrt{\rap^2-\fes{4}{2}{}{4} \coup^2}
\qquad\mbox{and}\qquad
p(\rap)=\frac{\fes{2}{1}{1}{1}}{\fes{}{}{2}{}\sqrt{\rap^2-\fes{4}{2}{}{4}\coup^2}}
 \,.
\]
In the scaling limit the charges \eqref{eq:ChargesPhi} become%
\[\label{eq:gaugechargedensity}
\excharge_r(\rap)=
\frac{\fes{2}{1}{1}{1}}{\fes{}{}{2}{}\sqrt{\rap^2-\fes{4}{2}{}{}\coup^2}}\,
\frac{1}{\lrbrk{\half\rap+\half\sqrt{\rap^2-\fes{4}{2}{}{4} \coup^2}}^{r-1}}=
\frac{p(\rap)}{x(\rap)^{r-1}}\,,
\]
where the momentum and energy densities are, respectively,
\[\label{eq:scaledgaugeenergy}
\excharge_1(\rap)=p(\rap)
=\frac{\fes{2}{1}{1}{1}}{\fes{}{}{2}{}\sqrt{\rap^2-\fes{4}{2}{}{4}\coup^2}}
\qquad \mbox{and} \qquad \excharge_2(\rap)=
\frac{\fes{2}{1}{1}{1}}{\fes{}{}{2}{}\sqrt{\rap^2-\fes{4}{2}{}{4}\coup^2}}\,
\frac{1}{\half\rap+\half\sqrt{\rap^2-\fes{4}{2}{}{4} \coup^2}}\,.
\]
%
%
%
%
%
The discrete sums \eqref{eq:partonsum} over excitation charges turn
into integrals over the distribution density
\[\label{eq:Spinning.Folded.Density}
\rho (\rap) = \frac{1}{L} \sum^M_{k=1} \delta \brk{ \rap -\rap_k }
\qquad\mbox{or}\qquad \frac{1}{L}\sum_{k=1}^M f(\rap_k)\mapsto
\int_{\contourgauge}d\rap\,\rho(\rap)\, f(\rap) \, ,
\]
with support on a discrete union of $K$ smooth contours
$\contourgauge=\contourgauge_1 \cup \ldots \cup \contourgauge_K$,
i.e.~the charges are given by
\[\label{eq:gaugecharge}
\charge_r=\int_{\contourgauge} d\rap\,\rho(\rap) \,\excharge_r(\rap) \, .
\]
Note the normalization of the density:
\[\label{eq:norm}
\int_{\contourgauge} d \rap\, \rho (\rap) =
\alpha \qquad \mbox{with} \qquad \alpha=\frac{M}{L}\, ,
\]
where $\alpha$ is termed filling fraction.
%
%

The distribution density, in addition to the normalization
condition \eqref{eq:norm}, is subject to the momentum constraint
\[
\charge_1=\int_{\contourgauge} d\rap\,\rho(\rap)\,p(\rap)=2 \pi m \, ,
\]
where $m$ is an integer mode number as required by cyclic
symmetry. Finally, the continuum Bethe equations derived from
\eqref{eq:Ansatz} lead to a system of singular integral equations,
determining the distribution density $\rho(\rap)$,
\[\label{eq:gaugebethe}
2\pint_{\contourgauge} \frac{d\rap'\,\rho(\rap')}{\rap-\rap'}=
 \frac{1}{\sqrt{\rap^2-\fes{4}{2}{}{4} \coup^2}}
+\fes{}{2}{4}{2}\pi n_{\nu}
\qquad \mbox{with} \qquad
\rap \in \contourgauge_\nu \, ,
\]
which are to hold on all $K$ cuts $\contourgauge_1, \ldots ,
\contourgauge_K$, i.e.~$\nu=1, \ldots, K$. These equations are
solved explicitly in \appref{sec:gaugefolded} for the folded
string and in \appref{sec:gaugecircular} for the circular string.
The charges are then determined from the solution
$\rho(\rap)$ by computing the integrals \eqref{eq:gaugecharge},
using \eqref{eq:gaugechargedensity}. The parameters $n_\nu$ are
integer mode numbers obtained from taking the logarithm of
\eqref{eq:Ansatz}. In fact, the right hand side of
\eqref{eq:gaugebethe} is just
$\fes{\half(p(\rap)+2\pi n_\nu)}{p(\rap)+2\pi n_\nu}{2(p(\rap)+2\pi n_\nu)}{p(\rap)+2\pi n_\nu}$,
while the
left hand side describes the factorized scattering of the
excitations in the thermodynamic limit.%
\footnote{For more
information on the mode numbers $n_\nu$ and $m$, as well as on
certain subtleties involving root ``condensates{}'', we refer to the
detailed one and two-loop discussion in \cite{Kazakov:2004qf},
which largely generalizes to our all-loop equations. However, note
in \eqref{eq:gaugebethe} the appearance of an additional square
root cut in the potential $p(\rap)$.
}
One easily checks that the perturbative expansion of the Bethe
equations \eqref{eq:gaugebethe} and of the expressions for the
energy \eqref{eq:scaledgaugeenergy}
reproduces, by construction, the three-loop thermodynamic
Inozemtsev-Bethe ansatz in \cite{Serban:2004jf}.

Note that the Bethe equation
\eqref{eq:gaugebethe}
can alternatively be obtained as a consistency
condition on the transfer matrix $\bar\transfer(u)$.
In the thermodynamic limit, the transfer matrix
becomes%
\[
\frac{\bar\transfer(\rap)}{x(\rap)^L},
\frac{\bar\transfer(\rap)}{P_L(\rap)}
\to 2\cos \bar\resolv\indup{sing}(\rap)
\qquad\mbox{with}\qquad
\bar\resolv\indup{sing}(\rap)=\frac{1}{\fes{}{2}{4}{2}\sqrt{\rap^2-\fes{4}{2}{}{4}\coup^2}}
+\bar\resolv(\rap)
\]
where $\bar\resolv(\rap)$ is the singularity-free $\rap$ resolvent.
The resolvent has many sheets, but the
transfer matrix $2\cos \bar\resolv\indup{sing}(\rap)$ must be single-valued on
the complex $\rap$ plane. This requires
\[
\bar\resolv\indup{sing}(\rap+i\epsilon)+\bar\resolv\indup{sing}(\rap-i\epsilon)=2\pi n
\]
across a branch cut of $\bar\resolv$ at $\rap$,
which is an equivalent to the
Bethe equation \eqref{eq:gaugebethe}.
At this point,
it is however not clear how the physical transfer matrix
$\transfer(x)$ is related to the physical
resolvent $\resolv(x)$ and if there is also a
consistency requirement which leads to the
Bethe equations.
This is largely related to mirror cuts
in $\transfer(\coup^2/\fes{}{2}{4}{}x)$ which
are due to the double covering map
$x(\rap)$.

\subsection{Semi-classical String Theory: The Bethe Equation}
\label{sec:Spinning}

In \cite{Kazakov:2004qf} Kazakov, Marshakov, Minahan and Zarembo developed
a general approach for finding the semi-classical solutions of the string sigma model
in the large charge limit. They astutely exploited the
(classical) integrability of the sigma model and derived the
equations determining the monodromy matrix of the system.
These singular integral equations were termed ``classical{}'' Bethe equations in
\cite{Kazakov:2004qf}, and it was shown that the monodromy matrix
may be interpreted as their resolvent.
Not surprisingly, the derivation conceptionally differs from the one of
the algebraic Bethe ansatz for quantum spin chains. In particular, one may introduce
a pseudodensity $\sigma(x)$ as the imaginary part of the resolvent
$\resolvs(x)$ along the discontinuities
$\contourstring=\contourstring_1 \cup \ldots \cup \contourstring_K$
in the complex plane of the spectral parameter $x$:
\[\label{eq:stringresolv}
\resolvs(x)=\int_{\contourstring} dx'\, \frac{\sigma(x')}{x'-x}\, ,
\]
i.e.~$\resolvs(x)$ is manifestly analytic on the physical sheet of the complex
$x$-plane, except for the set of cuts $\contourstring$.
The normalization condition of $\sigma(x)$ was found to be
\[\label{eq:stringnorm}
\int_{\contourstring}
dx\,\sigma(x)\,\fes{\half}{}{2}{}\lrbrk{1-\frac{\coup^2}{\fes{}{2}{4}{}x^2} }=
\alpha \, ,
\]
where the filling fraction $\alpha=M/L$ is defined in the same
fashion as in \secref{sec:Stringing}. Written in this form, we see
that the function $\sigma(x)$ should not be interpreted as a
distribution density of the local excitations in the spectral $x$
plane (see, however, the next \secref{sec:Matching}). The local
charges are then neatly expressed as
\[\label{eq:stringlocal}
\charges_r=\int_{\contourstring} dx\,\frac{\sigma(x)}{x^r} \, .
\]
This means that the Taylor expansion of the string resolvent around $x=0$
generates these charges:
\[\label{eq:stringtaylor}
\resolvs(x)=\sum_{r=1}^{\infty}\, \charges_r \, x^{r-1} \,.
\]
As in gauge theory the first charge is the
momentum $\charges_1$. It is subject to the integer constraint
\[
\charges_1=2 \pi m \, ,
\]
while the rescaled string energy is also precisely given by
\[\label{eq:stringenergy}
\engs=1+\fes{}{}{}{2} \coup^2  \charges_2(\coup) \, .
\]
Finally the Bethe equation, which is solved in
\appref{sec:stringfolded} and \appref{sec:stringcircular} for the
folded and the circular string, respectively, reads
\[\label{eq:stringbethe}
2\pint_{\contourstring} dx'\,\frac{\sigma(x')}{x-x'}=
\fes{2\engs}{\engs}{\frac{\engs}{2}}{\engs}\,
\frac{x}{x^2-\fes{}{\half}{\quarter}{}\coup^2}
+2\pi n_\nu
\qquad \mbox{with} \qquad
x \in \contourstring_\nu \, .
\]
This ends our brief summary of the
classical Bethe equations of \cite{Kazakov:2004qf}.
In \appref{sec:density} we present an equivalent
set of equations using the true density of excitations $\rho$ instead
of the pseudodensity $\sigma$.

\subsection{Structural Matching of Gauge and String Theory}
\label{sec:Matching}

Let us structurally compare the gauge and string ans\"atze.%
The normalization condition \eqref{eq:stringnorm} of string theory
appears to be incompatible with the one in gauge theory \eqref{eq:norm}
This may be fixed by relating the string and gauge spectral
measures through
\[
d\rap =
\left(1-\frac{\coup^2}{\fes{}{2}{4}{}x^2} \right)\,
dx \, .
\]
Upon integration we recover the relation \eqref{eq:xofphi,eq:uofx}
between the two spectral parameters $\rap$ and $x$
from the study of the discrete system
\[\label{eq:map}
\rap=x+\frac{\coup^2}{\fes{}{2}{4}{}x} \, ,
\qquad
x=\half \rap+\half\sqrt{\rap^2-\fes{4}{2}{}{4} \coup^2}\,.
\]
Interestingly, this is the same change of variables employed in
\cite{Kazakov:2004qf} in order to show the two-loop agreement
between string and gauge theory.%
\footnote{A very similar change of
variables also relates gauge theory and the generating function of
B\"acklund charges, see \cite{Arutyunov:2004xy}.}
Here we see
that the relationship should actually hold to \emph{all} loops if
we are to compare the two structures. One now easily checks the
elegant formula, c.f.~\eqref{eq:gaugechargedensity},
\[\label{eq:thepoint}
\excharge_r(\rap)\,d\rap=\fes{2}{}{\frac{1}{2}\,}{}\frac{dx}{x^r} \, ,
\]
i.e.~the scaled gauge charge densities $\excharge_r(\rap)$ in \eqref{eq:gaugechargedensity}
and the string charge densities $x^{-r}$ \emph{precisely} agree for all $r$!
Equation \eqref{eq:thepoint} is one of the key results of this paper, as it demonstrates
the structural equivalence of the elementary excitations in string and
gauge theory. We will have more to say about this at the end of \secref{sec:resolution}.
The all-loop agreement between the infinite set of gauge and string theory charges could be
established if one could show that the gauge theory distribution
density $\rho\indup{g}(\rap)$
and the function $\sigma\indup{s}(x(\rap))$ coincide:
\[
\rho\indup{g}(\rap)
\stackrel{?}{=}\fes{\half}{}{2}{}
\sigma\indup{s}(x) \, .
\]
This is however not the case, as was observed previously \cite{Serban:2004jf}.
Indeed, as a first sign of trouble we observe that only at one loop the expansion
point $\rap=0$ gets mapped to $x=0$. For $\coup \neq 0$ the map \eqref{eq:map}
is singular and actually represents the $\rap$-plane as a double
cover of the $x$-plane.
Under this map the gauge Bethe equation \eqref{eq:gaugebethe}
turns into%
\footnote{The second integral can be interpreted as the effect of
mirror cuts $\contourstring'=\coup^2/\fes{}{2}{4}{}\contourstring$
due to the double covering map $x(\rap)$.}
\[\label{eq:gaugebethe2}
2\pint_{\contourstring}  dx'\,\frac{\sigma\indup{g}(x')}{x-x'}=
\frac{\fes{2}{1}{1}{1}}{\fes{}{}{2}{}x}\,\frac{1}{1-\frac{\coup^2}{\fes{}{2}{4}{}x^2}}
+\frac{\fes{2}{}{}{2}\coup^2}{\fes{}{}{2}{}x}\,\int_{\contourstring}\,
dx'\,\frac{\sigma\indup{g}(x')}{x'^2}\,
\frac{1}{1-\frac{\coup^2}{\fes{}{2}{4}{}x x'}}
+2\pi n_\nu \, .
\]
Here the gauge pseudodensity is $\sigma\indup{g}(x):=\rho(\rap)$,
while the string Bethe equation \eqref{eq:stringbethe} may be rewritten as
\[\label{eq:stringbethe2}
2\pint_{\contourstring} dx'\,\frac{\sigma\indup{s}(x')}{x-x'}=
\frac{\fes{2}{1}{1}{1}}{\fes{}{}{2}{}x}\,\frac{1}{1-\frac{\coup^2}{\fes{}{2}{4}{}x^2}}
+\frac{\fes{2}{}{}{2}\coup^2}{\fes{}{}{2}{}x}\,\int_{\contourstring}\,
dx'\,\frac{\sigma\indup{s}(x')}{x'^2}\,
\frac{1}{1-\frac{\coup^2}{\fes{}{2}{4}{}x^2}}
+2\pi n_\nu \, ,
\]
where we used \eqref{eq:stringenergy}. The only distinction between
\eqref{eq:gaugebethe2,eq:stringbethe2} is the slightly different integrand
of the integral on the right hand side of each equation. In this form the agreement of
all charges up to exactly two loops is manifest.

The reader may find it instructive to also contemplate the form of the string Bethe
equation \eqref{eq:stringbethe} or \eqref{eq:stringbethe2}
on the spectral $\rap$-plane:
\<\label{eq:stringbethe3}
2\pint_{\contourgauge}
\frac{d\rap'\,\rho\indup{s}(\rap')}{\rap-\rap'}\eq
\frac{1}{\sqrt{\rap^2-\fes{4}{2}{}{4}\coup^2}}+
\fes{}{2}{4}{2}\pi n_{\nu}+
\fes{4}{2}{}{4} \coup^2
\int_{\contourgauge} \frac{d\rap' \rho\indup{s}(\rap')}
{\sqrt{\rap^2-\fes{4}{2}{}{4}\coup^2}
\sqrt{\rap'^2-\fes{4}{2}{}{4}\coup^2}} \times
\nlnum\nonumber\qquad\qquad\qquad\qquad
\times\frac{\rap-\sqrt{\rap^2-\fes{4}{2}{}{4}\coup^2}
-\rap'+\sqrt{\rap'^2-\fes{4}{2}{}{4}\coup^2}}
{\lrbrk{\rap+\sqrt{\rap^2-\fes{4}{2}{}{4}\coup^2}}
\lrbrk{\rap'+\sqrt{\rap'^2-\fes{4}{2}{}{4}\coup^2}}
-\fes{4}{2}{}{4}\coup^2}  \, , \>
where $\rho\indup{s}(\rap):= \fes{\half}{}{2}{}\sigma\indup{s}(x)$.
Comparing to the perturbative gauge Bethe equation
\eqref{eq:gaugebethe} we notice that the difference is generated
at $\order{\coup^4}$ (corresponding to three loop order) by the
last term on the right hand side of \eqref{eq:stringbethe3}. In
this form it becomes very easy (for details, see appendix
\secref{sec:curious}) to prove in generality, for even (i.e.~the
odd charges are zero) solutions, the ``curious observation{}'' of
\cite{Serban:2004jf,Arutyunov:2004xy} which involved a conjecture
about the structure of the three-loop disagreement between
perturbative gauge and semi-classical string theory.

\section{Resolution of the Puzzle: A Proposal}
\label{sec:resolution}

It is by now rather well established that there is a disagreement
between gauge theory and string theory at three-loop order, unless
the three-loop dilatation operator proposed in
\cite{Beisert:2003tq} and confirmed in \cite{Beisert:2003ys} is
incorrect. However, note the recent strong support for its
validity from the conjecture \cite{Kotikov:2004er} based on the
three-loop computation of \cite{Moch:2004pa}.
The disagreement first showed up in the
near BMN limit \cite{Callan:2003xr,Callan:2004uv}, and
subsequently in the similar but different case of Frolov-Tseytlin
(FT) spinning strings \cite{Serban:2004jf}. Assuming that the
AdS/CFT correspondence is indeed correct and exact, a possible
reason for the mismatch was first pointed out in
\cite{Serban:2004jf}. Indeed, gauge and string theory employ
slightly different scaling procedures. Here we will elaborate and
significantly refine this possible explanation. Our discussion
should apply equally well to the near BMN and FT situations. To be
specific, we will use the BMN notation $\lambdabmn$ in order to
explain our argument.%
\footnote{Note the following subtlety: For the
near BMN limit it is convenient to define this parameter as in
\eqref{eq:charge,eq:bmncoupling}. In the spinning string
discussion we should instead define $\lambdabmn=\fes{4}{8}{16}{8}\pi^2g^2/L^2$. In
the strict BMN limit the difference is of course irrelevant.}

\subsection{Order of Limits}

\begin{figure}\centering
\includegraphics[scale=1.2]{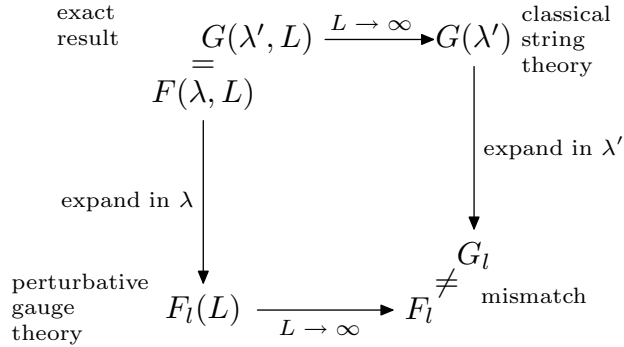}

\caption{A possible explanation for both the near BMN and the FT spinning strings
disagreement. $F_\ell$ excludes gauge theory wrapping effects,
while $G_\ell$ is expected to include them.}
\label{fig:explain}
\end{figure}

The comparison takes place in the thermodynamic limit
$L\to\infty$ and in an expansion around $\lambdabmn=0$.
However starting with an exact function $F(\lambda,L)$,
we must decide which limit is taken first.
It turns out that for classical string theory,
the thermodynamic limit $L\to\infty$ is a basic assumption.
The resulting energy may then be expanded in powers of $\lambdabmn$.
In contrast, gauge theory takes the other path.
The computations are based on perturbation theory around $\lambda=0$.
This expansion happens to coincide with the
expansion in $\lambdabmn$ and for the thermodynamic limit
one may drop subleading terms in $1/L$.
The claim has been that the order of limits does potentially matter \cite{Serban:2004jf}.
This is best illustrated in the noncommutative diagram \figref{fig:explain}.
Semi-classical string theory corresponds to the upper right corner
of the diagram, i.e.~it requires the large spin limit.
Conversely, perturbative gauge theory is situated at
the lower left corner, where the length $L$ is finite, but
only the first few orders in $\lambda$ are known. (However, we recall that
the number of known terms grows with $L$, if our spin
chain ansatz is correct.)

The BMN and FT proposals are both based on the assumption that the
diagram in \figref{fig:explain} does commute. In other words one
should be able to compare, order by order, the gauge theory loop
expansion with the string theory expansion in $\lambdabmn$. That
this might in fact not be true was first hinted at in
\cite{Klebanov:2002mp}, where also an example was given. Another,
more closely related, instance where the different limiting
procedures lead to different results can be found in
\cite{Serban:2004jf}. For the hyperbolic Inozemtsev spin chain it
was shown that the order of limits does matter.
In the ``gauge theory{}'' order, this spin chain appears to have no proper
thermodynamic limit. For the ``string theory{}'' order, i.e.~when
the thermodynamic limit is taken right from the start, it is
meaningful! (However, the resulting asymptotic Inozemtsev-Bethe ansatz also fails to
reproduce the three-loop string results, cf.~(59) of\cite{Serban:2004jf}).

In order to make contact with string theory we propose (in
agreement with \cite{Klebanov:2002mp}) to sum up the perturbation
series in $\lambda$ before taking the thermodynamic limit. With
the all-loop spin chain at hand this may indeed be feasible. In
contrast to the Inozemtsev chain, there appears to be no
difference between the two orders of limits (essentially because
the thermodynamic limit is well-behaved in perturbation theory).
However one has to take into account wrapping interactions. These
arise at higher loop orders $\ell$ when the interaction stretches
all around the state, i.e.~when $\ell\geq L$. We will discuss them
after the following example, which illustrates the potential
importance of these interactions.

\subsection{Example}

Here we present an example where one can see the importance
of the order of limits.
We choose a function
\[
F(\lambda,L)=\frac{\lambda^L}{(c+\lambda)^L}
=\lrbrk{1+\frac{c}{\lambdabmn L^2}}^{-L}=G(\lambdabmn,L).
\]
In perturbation theory around $\lambda=0$ we find that
the function vanishes at $L$ leading loop orders
\[\label{eq:ExampleExpand}
F(\lambda,L)=\sum_{\ell=0}^\infty F_\ell\, \lambda^\ell=
\frac{\lambda^L}{c^L}
-\frac{\lambda^{L+1}}{c^{L+1}}
+\frac{\lambda^{L+2}}{c^{L+2}}
+\ldots,
\qquad
\mbox{i.e.}\quad
F_\ell(L)=0\quad
\mbox{for}\quad \ell<L
\]
The factor $\lambda^L$ mimics the effect
of wrapping interactions in gauge theory as
explained below.
When we now go to the thermodynamic limit $L\to\infty$,
we see that all coefficients $F_\ell$ are zero.

Now let us take the thermodynamic limit first. The large $L$ limit
of $G(\lambdabmn,L)=\sum_\ell G_\ell\,\lambdabmn^\ell$
yields $G(\lambdabmn)=1$ in a straightforward
fashion. This result depends crucially on the function $F(\lambda,L)$.
Currently, we do not know how to incorporate wrapping
interactions, but $\lambda^L$ alone would not have a sensible
thermodynamic limit. To compensate this, we have introduced some
function ${1/(c+\lambda)^L}$. Clearly we cannot currently prove
that gauge theory produces a function like this, but it appears to be
a definite possibility. In our toy example, the expansion in
$\lambdabmn$ gives $G_0=1$ and $G_\ell=0$ otherwise.

In conclusion we find $G_\ell=\delta_{\ell,0}$
while $F_0=0$ which demonstrates the noncommutativity
of the diagram in \figref{fig:explain} in an example potentially relevant to our context.
It is not hard to construct a function $F(\lambda,L)$ which yields
arbitrary coefficients $G_\ell$ while all $F_\ell$ remain zero.

Note however that there is a sign of the non-commutativity
in \eqref{eq:ExampleExpand}:
A correct scaling behavior would require the coefficient
$F_\ell$ to scale as $L^{-2\ell}$. In particular for
$\ell=L$, the coefficient should scale as $L^{-2L}$ instead of
$c^{-L}$. Therefore one can say that the function $F$ violates
the scaling law at weak coupling, but in a mild way that is easily
overlooked.
This parallels observations made for the Inozemtsev spin chain
for which scaling is manifestly violated at weak coupling.
Nevertheless, when one does not expand for small
$\lambda$ scaling might be recovered \cite{Serban:2004jf}.

\subsection{Wrapping Interactions}
\label{sec:wrapping}

In perturbative field theory, the contributions to the dilatation
operator are derived from Feynman diagrams.
Let us consider a local operator with finitely many fields.
At lower loop orders $\ell$ a planar Feynman diagram attaches to
a number of neighboring sites along the spin chain.
When the loop order increases, this region stretches until
it wraps completely around the trace. At this point,
when $\ell\geq L$, our methods cease to work: We know there are
further contributions which couple only to states of a fixed length,
but we currently have no information about their structure.
Furthermore, it is not quite clear how to achieve a BMN limit
or integrability.

Something very similar is true for long-range spin chains, and
their solution by the \emph{asymptotic} Bethe ansatz. It is again
useful to take inspiration from the Inozemtsev model
\cite{Inozemtsev:1989yq,Inozemtsev:2002vb}. There the second
charge (i.e.~the Hamiltonian) is $L$ dependent, and the strength
of the spin-spin interactions is governed by a doubly-periodic
elliptic Weierstrass function. Its imaginary period is related to
the coupling constant, and its real period to $L$. In the
asymptotic limit $L \rightarrow \infty$ the real period disappears
and the interaction strength turns into a hyperbolic function. It
is precisely this reduced model which is properly described by the
asymptotic Bethe ansatz, as in \eqref{eq:Ansatz}. This ansatz
actually works even better than one might expect at first sight:
It does not strictly require $L=\infty$, but only that $\ell< L$,
where $\ell$ measures the interaction range, as in field theory.

Our novel long-range spin chain is clearly very closely related to
the Inozemtsev model. We expect that its Hamiltonian, along with
all other charges, also has a ``periodized{}'' extension, in full
analogy with going from the hyperbolic to the elliptic Inozemtsev
interaction. Hopefully this extension will still be consistent
with our construction principles spelled out at the beginning of
\secref{sec:Ansatz}, namely ($i$) integrability, ($ii$)
compatibility with field theory, and ($iii$) BMN scaling. If we
are lucky, the full model will still be \emph{unique}, and should
then correspond to the non-perturbative planar dilatation operator
of $\superN=4$ theory.

While it is very reasonable to assume that this Hamiltonian exists,
it is currently unclear whether it may be explicitly written down in any useful
fashion. Luckily, our results above suggest that this might not be
necessary or even desirable, if we succeed in properly including the
effect of wrappings into the Bethe ansatz. This is however not an easy problem,
which has not yet been solved for the Inozemtsev model either \cite{Inozemtsev:2002vb}.

We cannot currently offer a quantitative theory of wrapping effects,
and are thus unable to explain why they only modify the thermodynamic
limit starting at $\order\lambdabmn^3$.
Nevertheless, on a qualitative level much can be said in favor of the proposal
that their inclusion will lead to a reconciliation of the current disagreements
between string and gauge theory. First of all, it is reasonable to expect
that wrappings should not affect the energy formula of the
strict BMN limit \cite{Berenstein:2002jq}. This formula is obtained, from
the point of view of our long-range Bethe ansatz, in a rather trivial fashion by
neglecting magnon scattering altogether!
In this ``dilute gas{}'' approximation
one sets the right hand side of the Bethe equations
\eqref{eq:Ansatz} to 1. In contrast, the near BMN limit
takes into account finite size $1/J$ corrections, which
\emph{are} including the scattering effects, as may be seen by
inspecting the calculations of \secref{sec:nearBMN}. Clearly
wrapping effects should be included into such finite size corrections
when we scale according to the north-east corner of \figref{fig:explain},
but not when we scale as in the south-west corner of that diagram.
As for spinning strings and the FT proposal, it is clear that
magnon scattering is heavily influencing the computation of the
energy (and all other charges). In fact, the Hilbert kernel
of the relevant singular integral equations is precisely
describing the local, pairwise interaction of a macroscopically
large number of excitations. Therefore, as for near BMN,
the two different scaling procedures of \figref{fig:explain},
which in- or exclude wrappings, are expected to influence
the scattering phase shifts, and therefore the
distribution density of magnon momenta. Wrappings are \emph{not}
expected to change the \emph{form} of the contributions (i.e.~their
functional dependence on the magnon momentum and the coupling constant)
of individual magnons to the overall expectation values of charges, which
is precisely what we have been finding in
\secref{sec:Trouble}!

On a more technical level, we suspect that the unknown
terms in the transfer matrix eigenvalues \eqref{eq:Transfer}
will contribute when we scale according to the north-east corner
of \figref{fig:explain}. It may be expected that a non-asymptotic
Bethe ansatz (if it exists at all) will follow from the full analytic
structure of the complete transfer matrix, just as in the one-loop case.
Likewise, one would hope that the latter will also generate the
correct expressions for the global charges as functions of the
individual magnon contributions.

\subsection*{Acknowledgments}

We are very grateful to G.~Arutyunov, D.~Serban and K.~Zarembo
for useful discussions.
N.B.~dankt der \emph{Studienstiftung des
deutschen Volkes} f\"ur die Unterst\"utzung durch ein
Promotions\-f\"orderungsstipendium.

\appendix

\section{The Planar SYM Spectrum}
\label{sec:Spec}

In this appendix we compare the planar gauge theory
spectrum, which was obtained using the Hamiltonian approach
described in \secref{sec:CommCharge}, to the results of the
long-range Bethe ansatz for small excitation numbers.

\subsection{Lowest-Lying States}
\label{sec:Spec.Low}

In the following we will describe how to obtain results
using the spin chain Hamiltonian as well as
the long-range Bethe ansatz in general.
In the subsequent sections will be go into details
for certain classes of states.

\begin{table}\small
\fes{wrong}{}{wrong}{}
\<
\ham_0\eq
\fes{\half}{}{2}{} \PTerm{}
-\fes{\half}{}{2}{}\PTerm{1},
\nonumber
\\[12pt]
\ham_2\eq
-\fes{\sfrac{1}{2}}{2}{8}{2}\PTerm{}
+\fes{\sfrac{3}{4}}{3}{12}{3}\PTerm{1}
-\fes{\sfrac{1}{8}}{\half}{2}{\half}\bigbrk{\PTerm{1,2}+\PTerm{2,1}},
\nonumber
\\[12pt]
\ham_4\eq
\fes{\sfrac{15}{16}}{\sfrac{15}{2}}{60}{\sfrac{15}{2}}\PTerm{}
-\fes{\sfrac{13}{8}}{13}{104}{13}\PTerm{1}
+\fes{\sfrac{1}{16}}{\half}{4}{\half}\PTerm{1,3}
\nl
+\fes{\sfrac{3}{8}}{3}{24}{3}\bigbrk{\PTerm{1,2}+\PTerm{2,1}}
-\fes{\sfrac{1}{16}}{\half}{4}{\half}\bigbrk{\PTerm{1,2,3}+\PTerm{3,2,1}}.
\nonumber
\\[12pt]
\ham_{6}\eq
-35 \PTerm{}
+\bigbrk{67+4\alpha}\PTerm{1}
+\bigbrk{-\sfrac{21}{4}-2\alpha} \PTerm{1,3}
-\sfrac{1}{4} \PTerm{1, 4}
\nl
+\bigbrk{-\sfrac{151}{8}-4\alpha} \bigbrk{\PTerm{1,2}+\PTerm{2,1}} 
+2\alpha \bigbrk{\PTerm{1,3,2}+\PTerm{2,1,3}} 
\nl
+\sfrac{1}{4}\bigbrk{\PTerm{1,2,4}+\PTerm{1,3,4}+\PTerm{1,4,3}+\PTerm{2,1,4}}
+\bigbrk{6+2\alpha} \bigbrk{\PTerm{1,2,3}+\PTerm{3,2,1}}
\nl
+\bigbrk{-\sfrac{3}{4}-2\alpha} \PTerm{2, 1, 3, 2}
+\bigbrk{\sfrac{9}{8}+2\alpha} \bigbrk{\PTerm{1,3,2,4}+\PTerm{2,1,4,3}} 
\nl
+\bigbrk{-\sfrac{1}{2}-\alpha} \bigbrk{\PTerm{1,2,4,3}+\PTerm{1,4,3,2}+\PTerm{2,1,3,4}+\PTerm{3,2,1,4}}  
\nl
-\sfrac{5}{8}\bigbrk{\PTerm{1, 2, 3, 4} + \PTerm{4, 3, 2, 1}},
\nonumber
\\[12pt]
\ham_8\eq
+\sfrac{1479}{8}\PTerm{}
+\bigbrk{-\sfrac{1043}{4}-12\alpha+4\beta_1}\PTerm{1}
+\bigbrk{-19+8\alpha-2\beta_1-4\beta_2}\PTerm{1,3}
\nl
+\bigbrk{5+2\alpha+4\beta_2+4\beta_3}\PTerm{1,4}
+\sfrac{1}{8}\PTerm{1,5}
+\bigbrk{11\alpha-4\beta_1+2\beta_3}\bigbrk{\PTerm{1,2}+\PTerm{2,1}}
\nl
-\sfrac{1}{4}\PTerm{1,3,5}
+\bigbrk{\sfrac{251}{4}-5\alpha+2\beta_1-2\beta_3}\bigbrk{\PTerm{1,3,2}+\PTerm{2,1,3}}
\nl
+\bigbrk{-3-\alpha-2\beta_3}\bigbrk{\PTerm{1,2,4}+\PTerm{1,3,4}+\PTerm{1,4,3}+\PTerm{2,1,4}}
\nl
-\sfrac{1}{8}\bigbrk{\PTerm{1,2,5}+\PTerm{1,4,5}+\PTerm{1,5,4}+\PTerm{2,1,5}}
\nl
+\bigbrk{\sfrac{41}{4}-6\alpha+2\beta_1-4\beta_3}\bigbrk{\PTerm{1,2,3}+\PTerm{3,2,1}}
+\bigbrk{-\sfrac{107}{2}+4\alpha-2\beta_1}\PTerm{2,1,3,2}
\nl
+\bigbrk{\sfrac{1}{4}+\beta_2}\bigbrk{\PTerm{1,3,2,5}+\PTerm{1,3,5,4}+\PTerm{1,4,3,5}+\PTerm{2,1,3,5}}
\nl
+\bigbrk{\sfrac{183}{4}-6\alpha+2\beta_1-2\beta_2}\bigbrk{\PTerm{1,3,2,4}+\PTerm{2,1,4,3}}
\nl
+\bigbrk{-\sfrac{3}{4}-2\beta_2}\bigbrk{\PTerm{1,2,5,4}+\PTerm{2,1,4,5}}
+\bigbrk{1+2\beta_2}\bigbrk{\PTerm{1,2,4,5}+\PTerm{2,1,5,4}}
\nl
+\bigbrk{-\sfrac{51}{2}+\sfrac{5}{2}\alpha-\beta_1+\beta_2+3\beta_3}\bigbrk{\PTerm{1,2,4,3}+\PTerm{1,4,3,2}+\PTerm{2,1,3,4}+\PTerm{3,2,1,4}}
\nl
-\beta_2\bigbrk{\PTerm{1,2,3,5}+\PTerm{1,3,4,5}+\PTerm{1,5,4,3}+\PTerm{3,2,1,5}}
\nl
+\bigbrk{\sfrac{35}{4}+\alpha+2\beta_3}\bigbrk{\PTerm{1,2,3,4}+\PTerm{4,3,2,1}}
\nl
+\bigbrk{-\sfrac{7}{8}-\alpha+2\beta_3}\bigbrk{\PTerm{1,4,3,2,5}+\PTerm{2,1,3,5,4}}
\nl
+\bigbrk{\sfrac{1}{2}+\alpha}\bigbrk{\PTerm{1,3,2,5,4}+\PTerm{2,1,4,3,5}}
\nl
+\bigbrk{\sfrac{5}{8}+\half\alpha-\beta_3}\bigbrk{\PTerm{1,3,2,4,3}+\PTerm{2,1,3,2,4}+\PTerm{2,1,4,3,2}+\PTerm{3,2,1,4,3}}
\nl
+\bigbrk{\sfrac{1}{4}-2\beta_3}\bigbrk{\PTerm{1,2,5,4,3}+\PTerm{3,2,1,4,5}}
\nl
+\bigbrk{\sfrac{1}{4}+\half\alpha+\beta_3}\bigbrk{\PTerm{1,2,4,3,5}+\PTerm{1,3,2,4,5}+\PTerm{2,1,5,4,3}+\PTerm{3,2,1,5,4}}
\nl
+\bigbrk{-\half\alpha-\beta_3}\bigbrk{\PTerm{1,2,3,5,4}+\PTerm{1,5,4,3,2}+\PTerm{2,1,3,4,5}+\PTerm{4,3,2,1,5}}
\nl
-\sfrac{7}{8}\bigbrk{\PTerm{1,2,3,4,5}+\PTerm{5,4,3,2,1}}
\nonumber
\>

\caption{The spin chain Hamiltonian up to five-loops, $\order{g^8}$.
The constants $\alpha,\beta_{1,2,3}$ do not influence the
spectrum.}
\label{tab:SU2.FourFive}
\end{table}

We start by computing the eigenvalues of the first few commuting
charges in perturbative gauge theory.
To obtain a matrix representation for the operators, we
have applied the Hamiltonian $\ham\equiv\charge_2$
(up to five loops, see \tabref{tab:SU2.FourFive}
\footnote{The four-loop and five-loop contributions contain
some undetermined constants $\alpha,\beta_{1,2,3}$.
These are unphysical and correspond to perturbative rotations of the
space of states. They change the (unphysical) eigenstates,
but not the (physical) eigenvalues and thus cannot be fixed
(unless one finds a canonical way to write the higher order
Hamiltonians).})
as well as the charges $\charge_{3},\charge_{4}$ (up to four loops,
see \cite{Beisert:2004ry})
and $\charge_{5},\charge_{6}$ (up to two loops)
to all states with a
given length $L$ and number of excitations $M$. The computations
were performed using a set of \texttt{Mathematica} routines
which will be given in \cite{Beisert:2004ry}. Then, the
leading order energy matrix was diagonalized in order to obtain
the leading order energy eigenvalues. Next, the off-diagonal terms
at higher-loops were removed iteratively by a sequence of
similarity transformations. Afterwards, the Hamiltonian is
diagonal and we can read off the energy eigenvalues. The same
similarity transformations which were used to make $\charge_2$
diagonal also diagonalize $\charge_{3,4,5,6}$ and we may read off
their eigenvalues.

\begin{table}\centering\small\arraycolsep3pt
\fes{
$\begin{array}{|ccc|lllll|llll|}\hline L&M&\gaugepar
  & g^0x^0                & g^2x^0               & g^4x^0                     & g^6x^0                    &  g^8x^0
  & g^0x^2                & g^2x^2               & g^4x^2                     & g^6x^2
\\\hline\hline\vphantom{A}
4&2& +
  & +3                    & -3                   & +\frac{21}{4}              & \ast                      & \ast
  & +0                    & \ast                 & \ast                       & \ast
\\[1pt]\hline\vphantom{\hat A}
5&2& -
  & +2                    & -\frac{3}{2}         & +\frac{17}{8}              & -\frac{115}{32}           & \ast
  & +\frac{1}{3}          & -\frac{1}{2}         & \ast                       & \ast
\\[1pt]\hline\vphantom{\hat A}
6&2& +
  & +5\eival              & -\frac{17}{4}\eival  &  +\frac{117}{16}\eival     & -\frac{1037}{64}\eival    & +\frac{10525}{256}\eival
  & -\frac{5}{12}\eival   & +\frac{15}{8}\eival  & -\frac{381}{64}\eival      & \ast
\\[2pt]
& &
  & -5                    & +\frac{15}{2}        & -\frac{115}{8}             & +\frac{1025}{32}          & -\frac{10165}{128}
  & +0                    & -\frac{35}{24}       & +\frac{105}{16}            & \ast
\\[1pt]\hline\vphantom{\hat A}
6&3& -
  & +3                    & -\frac{9}{4}         & +\frac{63}{16}             & -\frac{621}{64}           & +\frac{7047}{256}
  & -\frac{3}{4}          & +\frac{9}{4}         & -\frac{405}{64}            & \ast
\\[1pt]\hline\vphantom{\hat A}
7&2& -
  & +1                    & -\frac{3}{8}         & +\frac{37}{128}            & -\frac{283}{1024}         & +\frac{9597}{32768}
  & +\frac{1}{6}          & -\frac{5}{32}        & +\frac{81}{512}            & -\frac{707}{4096}
\\[1pt]\hline\vphantom{\hat A}
7&2& -
  & +3                    & -\frac{21}{8}        & +\frac{555}{128}           & -\frac{8997}{1024}        & +\frac{651651}{32768}
  & +0                    & +\frac{9}{32}        & -\frac{513}{512}           & +\frac{11907}{4096}
\\[1pt]\hline\vphantom{\hat A}
7&3& \pm
  & +5\eival              & -\frac{15}{4}\eival  & +\frac{25}{4}\eival        & -\frac{875}{64}\eival     & +\frac{4365}{128}\eival
  &-\frac{5}{12}\eival    & +\frac{25}{16}\eival & -\frac{285}{64}\eival      & +\frac{1615}{128}\eival
\\[2pt]
&&
  & -\frac{25}{4}         & +\frac{75}{8}        & -\frac{1225}{64}           & +\frac{5875}{128}         & -\frac{61775}{512}
  & +\frac{245}{192}      & -\frac{45}{8}        & +\frac{28145}{1536}        & -\frac{86875}{1536}
\\[1pt]\hline\vphantom{\hat A}
8&2& +
  & +7\eival^2            & -\frac{23}{4}\eival^2& +\frac{79}{8}\eival^2      & -\frac{349}{16}\eival^2   & +\frac{3527}{64}\eival^2
  & -\frac{7}{12}\eival^2 &+\frac{39}{16}\eival^2& -\frac{125}{16}\eival^2    & +\frac{4691}{192}\eival^2
\\[2pt]
&&
  & -14\eival             & +\frac{43}{2}\eival  & -\frac{695}{16}\eival      & +\frac{1627}{16}\eival    & -\frac{8373}{32}\eival
  & +\frac{7}{6}\eival    & -\frac{175}{24}\eival& +\frac{2629}{96}\eival    & -\frac{5911}{64}\eival
\\[2pt]
&&
  & +7                    & -14                  & +\frac{483}{16}            & -\frac{4585}{64}         & +\frac{23555}{128}
  &+0                     & +\frac{21}{8}        & -\frac{2527}{192}          & +\frac{38269}{768}
\\[1pt]\hline\vphantom{\hat A}
8&3& -
  & +3                    & -\frac{9}{4}         & +\frac{33}{8}              & -\frac{81}{8}             & +\frac{1803}{64}
  & -\frac{3}{4}          & +\frac{33}{16}       & -6                         & +\frac{1191}{64}
\\[1pt]\hline\vphantom{\hat A}
8&3 & \pm
  & +4\eival              & -\frac{5}{2}\eival   & +\frac{7}{2}\eival         & -\frac{51}{8}\eival       & +\frac{211}{16}\eival
  & +\frac{1}{6}\eival    & -\frac{1}{8}\eival   & +\frac{1}{8}\eival         & -\frac{13}{96}\eival
\\[2pt]
&&
  & -4                    & +5                   & -\frac{137}{16}            & +\frac{137}{8}            & -\frac{1197}{32}
  & -\frac{1}{12}         & -\frac{5}{12}        & +\frac{41}{24}             & -\frac{487}{96}
\\[1pt]\hline\vphantom{\hat A}
8&4& +
  & +10\eival^2           & -8\eival^2           & +14\eival^2                & -\frac{511}{16}\eival^2   & +\frac{2665}{32}\eival^2
  & -\frac{4}{3}\eival^2  & +\frac{9}{2}\eival^2 & -\frac{221}{16}\eival^2    & +\frac{1033}{24}\eival^2
\\[2pt]
&&
  & -29\eival             & +\frac{85}{2}\eival  & -\frac{175}{2}\eival       & +\frac{3469}{16}\eival    & -\frac{9561}{16}\eival
  & -\frac{49}{6}\eival   & -\frac{775}{24}\eival& +\frac{5177}{48}\eival     & -\frac{11337}{8}\eival
\\[2pt]
&&
  & +25                   & -50                  & +\frac{225}{2}             & -\frac{575}{2}            & +\frac{51475}{64}
  & -10                   & +\frac{175}{4}       & -\frac{1825}{12}           & +\frac{24355}{48}
\\[1pt]\hline
\end{array}$
}{
$\begin{array}{|ccc|lllll|llll|}\hline L&M&\gaugepar
  & g^0x^0                & g^2x^0               & g^4x^0                     & g^6x^0                    & g^8x^0
  & g^0x^2                & g^2x^2               & g^4x^2                     & g^6x^2
\\\hline\hline\vphantom{A}
4&2& +
  & +6                    & -12                  & +42                        & \ast                      & \ast
  & +0                    & \ast                 & \ast                       & \ast
\\[1pt]\hline\vphantom{\hat A}
5&2& -
  & +4                    & -6                   & +17                        & -\frac{115}{2}            & \ast
  & +\frac{8}{3}          & -8                   & \ast                       & \ast
\\[1pt]\hline\vphantom{\hat A}
6&2& +
  & +10\eival             & -17\eival            &  +\frac{117}{2}\eival      & -\frac{1037}{4}\eival     & +\frac{10525}{8}\eival
  & -\frac{10}{3}\eival   & +30\eival            & -\frac{381}{2}\eival       & \ast
\\[2pt]
& &
  & -20                   & +60                  & -230                       & +1025                     & -\frac{10165}{2}
  & +0                    & -\frac{140}{3}       & +420                       & \ast
\\[1pt]\hline\vphantom{\hat A}
6&3& -
  & +6                    & -9                   & +\frac{63}{2}              & -\frac{621}{4}            & +\frac{7047}{8}
  & -6                    & +36                  & -\frac{405}{2}             & \ast
\\[1pt]\hline\vphantom{\hat A}
7&2& -
  & +2                    & -\frac{3}{2}         & +\frac{37}{16}             & -\frac{283}{64}           & +\frac{9597}{1024}
  & +\frac{4}{3}          & -\frac{5}{2}         & +\frac{81}{16}             & -\frac{707}{64}
\\[1pt]\hline\vphantom{\hat A}
7&2& -
  & +6                    & -\frac{21}{2}        & +\frac{555}{16}            & -\frac{8997}{64}          & +\frac{651651}{1024}
  & +0                    & +\frac{9}{2}         & -\frac{513}{16}            & +\frac{11907}{64}
\\[1pt]\hline\vphantom{\hat A}
7&3& \pm
  & +10\eival             & -15\eival            & +50\eival                  & -\frac{875}{4}\eival      & +\frac{4365}{4}\eival
  &-\frac{10}{3}\eival    & +25\eival            & -\frac{285}{2}\eival       & +\frac{1615}{2}\eival
\\[2pt]
&&
  & -25                   & +75                  & -\frac{1225}{4}            & +\frac{5875}{4}           & -\frac{61775}{8}
  & +\frac{245}{12}       & -180                 & +\frac{28145}{24}          & -\frac{86875}{12}
\\[1pt]\hline\vphantom{\hat A}
8&2& +
  & +14\eival^2           & -23\eival^2          & +79\eival^2                & -349\eival^2              & +\frac{3527}{2}\eival^2
  & -\frac{14}{3}\eival^2 & +39\eival^2          & -250\eival^2               & +\frac{4691}{3}\eival^2
\\[2pt]
&&
  & -56\eival             & +172\eival           & -695\eival                 & +3254\eival               & -16746\eival
  & +\frac{56}{3}\eival   & -\frac{700}{3}\eival & +\frac{5258}{3}\eival      & -11822\eival
\\[2pt]
&&
  & +56                   & -224                 & +966                       & -4585                     & +23555
  & +0                    & +168                 & -\frac{5054}{3}            & +\frac{38269}{3}
\\[1pt]\hline\vphantom{\hat A}
8&3& -
  & +6                    & -9                   & +33                        & -162                      & +\frac{1803}{2}
  & -6                    & +33                  & -192                       & +1191
\\[1pt]\hline\vphantom{\hat A}
8&3 & \pm
  & +8\eival              & -10\eival            & +28\eival                  & -102\eival                & +422\eival
  & +\frac{4}{3}\eival    & -2\eival             & +4\eival                   & -\frac{26}{3}\eival
\\[2pt]
&&
  & -16                   & +40                  & -137                       & +548                      & -2394
  & -\frac{4}{3}          & -\frac{40}{3}        & +\frac{328}{3}             & -\frac{1948}{3}
\\[1pt]\hline\vphantom{\hat A}
8&4& +
  & +20\eival^2           & -32\eival^2          & +112\eival^2               & -511\eival^2              & +2665\eival^2
  & -\frac{32}{3}\eival^2 & +72\eival^2          & -442\eival^2               & +\frac{8264}{3}\eival^2
\\[2pt]
&&
  & -116\eival            & +340\eival           & -1400\eival                & +6938\eival               & -38244\eival
  & +\frac{392}{3}\eival  & -\frac{3100}{3}\eival& +\frac{20708}{3}\eival     & -45348\eival
\\[2pt]
&&
  & +200                  & -800                 & +3600                      & -18400                    & +102950
  & -320                  & +2800                & -\frac{58400}{3}           & +\frac{389680}{3}
\\[1pt]\hline
\end{array}$
}{
\remark{I give up}
}{
$\begin{array}{|ccc|lllll|llll|}\hline L&M&\gaugepar
  & g^0x^0                & g^2x^0               & g^4x^0                     & g^6x^0                    &  g^8x^0
  & g^0x^2                & g^2x^2               & g^4x^2                     & g^6x^2
\\\hline\hline\vphantom{A}
4&2& +
  & +6                    & -12                  & +42                        & \ast                      & \ast
  & +0                    & \ast                 & \ast                       & \ast
\\[1pt]\hline\vphantom{\hat A}
5&2& -
  & +4                    & -6                   & +17                        & -\frac{115}{2}            & \ast
  & +\frac{8}{3}          & -8                   & \ast                       & \ast
\\[1pt]\hline\vphantom{\hat A}
6&2& +
  & +10\eival             & -17\eival            &  +\frac{117}{2}\eival      & -\frac{1037}{4}\eival     & +\frac{10525}{8}\eival
  & -\frac{10}{3}\eival   & +30\eival            & -\frac{381}{2}\eival       & \ast
\\[2pt]
& &
  & -20                   & +60                  & -230                       & +1025                     & -\frac{10165}{2}
  & +0                    & -\frac{140}{3}       & +420                       & \ast
\\[1pt]\hline\vphantom{\hat A}
6&3& -
  & +6                    & -9                   & +\frac{63}{2}              & -\frac{621}{4}            & +\frac{7047}{8}
  & -6                    & +36                  & -\frac{405}{2}             & \ast
\\[1pt]\hline\vphantom{\hat A}
7&2& -
  & +2                    & -\frac{3}{2}         & +\frac{37}{16}             & -\frac{283}{64}           & +\frac{9597}{1024}
  & +\frac{4}{3}          & -\frac{5}{2}         & +\frac{81}{16}             & -\frac{707}{64}
\\[1pt]\hline\vphantom{\hat A}
7&2& -
  & +6                    & -\frac{21}{2}        & +\frac{555}{16}            & -\frac{8997}{64}          & +\frac{651651}{1024}
  & +0                    & +\frac{9}{2}         & -\frac{513}{16}            & +\frac{11907}{64}
\\[1pt]\hline\vphantom{\hat A}
7&3& \pm
  & +10\eival             & -15\eival            & +50\eival                  & -\frac{875}{4}\eival      & +\frac{4365}{4}\eival
  &-\frac{10}{3}\eival    & +25\eival            & -\frac{285}{2}\eival       & +\frac{1615}{2}\eival
\\[2pt]
&&
  & -25                   & +75                  & -\frac{1225}{4}            & +\frac{5875}{4}           & -\frac{61775}{8}
  & +\frac{245}{12}       & -180                 & +\frac{28145}{24}          & -\frac{86875}{12}
\\[1pt]\hline\vphantom{\hat A}
8&2& +
  & +14\eival^2           & -23\eival^2          & +79\eival^2                & -349\eival^2              & +\frac{3527}{2}\eival^2
  & -\frac{14}{3}\eival^2 & +39\eival^2          & -250\eival^2               & +\frac{4691}{3}\eival^2
\\[2pt]
&&
  & -56\eival             & +172\eival           & -695\eival                 & +3254\eival               & -16746\eival
  & +\frac{56}{3}\eival   & -\frac{700}{3}\eival & +\frac{5258}{3}\eival      & -11822\eival
\\[2pt]
&&
  & +56                   & -224                 & +966                       & -4585                     & +23555
  & +0                    & +168                 & -\frac{5054}{3}            & +\frac{38269}{3}
\\[1pt]\hline\vphantom{\hat A}
8&3& -
  & +6                    & -9                   & +33                        & -162                      & +\frac{1803}{2}
  & -6                    & +33                  & -192                       & +1191
\\[1pt]\hline\vphantom{\hat A}
8&3 & \pm
  & +8\eival              & -10\eival            & +28\eival                  & -102\eival                & +422\eival
  & +\frac{4}{3}\eival    & -2\eival             & +4\eival                   & -\frac{26}{3}\eival
\\[2pt]
&&
  & -16                   & +40                  & -137                       & +548                      & -2394
  & -\frac{4}{3}          & -\frac{40}{3}        & +\frac{328}{3}             & -\frac{1948}{3}
\\[1pt]\hline\vphantom{\hat A}
8&4& +
  & +20\eival^2           & -32\eival^2          & +112\eival^2               & -511\eival^2              & +2665\eival^2
  & -\frac{32}{3}\eival^2 & +72\eival^2          & -442\eival^2               & +\frac{8264}{3}\eival^2
\\[2pt]
&&
  & -116\eival            & +340\eival           & -1400\eival                & +6938\eival               & -38244\eival
  & +\frac{392}{3}\eival  & -\frac{3100}{3}\eival& +\frac{20708}{3}\eival     & -45348\eival
\\[2pt]
&&
  & +200                  & -800                 & +3600                      & -18400                    & +102950
  & -320                  & +2800                & -\frac{58400}{3}           & +\frac{389680}{3}
\\[1pt]\hline
\end{array}$
}

\caption{Five-loop energies and four-loop eigenvalues of the
charges $\charge_{3,4}$. Please refer to
App.~\ref{sec:Spec.Low} for an explanation.}
\label{tab:Spec.Low}
\end{table}

We present our findings for $\charge_2$, $\charge_3$ and
$\charge_4$ up to $L=8$ in \tabref{tab:Spec.Low} (we omit
the protected states with $M=0$) which is read as follows.
For each state there is a polynomial and we write down its
coefficients up to $\order{g^8}$ and $\order{x^2}$.
For single states the polynomial $P(x,g)$ equals simply
\[\label{eq:Spec.LowSingle}
P(x,g)=\charge_2(g)+x^2\,\charge_4(g).
\]
If there is more than one state transforming in the same
representation, the eigenvalues are solutions to algebraic
equations. These could be solved numerically, here we prefer to
state the exact polynomial $P(\eival,x,g)$ of degree $k-1$ in
$\eival$. The energy and charge eigenvalues are determined by
the formula
\[\label{eq:Spec.LowMulti}
\eival=\charge_2(g)+x\, \charge_3(g)+x^2\, \charge_4(g)+\dots,
\qquad \eival^k=P(\eival,g,x).
\]
At first sight the terms linear in $x$ may appear wrong and
the corresponding charge $\charge_3(g)$ would have to be zero.
For unpaired states with non-degenerate
$\charge_2(g)$ this is true, but not so for pairs of degenerate states.
Then the solution of the algebraic equation leads to terms
of the sort $\sqrt{0+x^2}=\pm x$, where the $0$ symbolizes
the degeneracy.
For some states the interaction is longer than the
state. In such a case, indicated by $\ast$ in the table, we do not
know the energy/charge eigenvalue, see \secref{sec:wrapping}.

Before we turn to comparing the Bethe ansatz to the Hamiltonian
method for a number of specific examples, let us briefly describe
how the energy/charge eigenvalues are obtained from the Bethe
equations in general. As we are interested in the higher loop
corrections to these eigenvalues,
we expand the Bethe roots $\rap_k$ in the coupling:
\[
\rap_k=\rap_{k,0}+g^2 \rap_{k,2}+g^4\rap_{k,4}+\dots.
\]
This is inserted into the Bethe equations \eqref{eq:BethePhi} where
both, the left and the right hand side, are expanded in $g^2$.
The zeroth order of this expansion is just the one-loop
Bethe equation. These are relatively easy to solve for short spin
chains with a small number of magnons. For increasing $L$ and
$M$, however, it becomes more and more difficult to find the
solutions to these equations. In \tabref{tab:Spec.Class} the
one-loop roots of the states of
\tabref{tab:Spec.Low} are listed. Note that we have omitted the
roots of the two-excitation states as the general formula
for their momenta will be given in
\secref{sec:Spec.TwoEx}.
The states marked $^\ast$ are mirror solutions and will be explained
in \secref{sec:Spec.Singular}.
Instead of writing down the
approximate numerical values of the Bethe roots, we prefer to
give the exact algebraic equations whose roots, $\rap_{k,0}$, are
exactly the one-loop Bethe roots. In the case where there is more
than one state (i.e.~more than one set of Bethe roots) for a given
$L$ and $M$, we give one polynomial for all Bethe roots in all the
different sets. It is left as an exercise for the reader to
determine which root belongs to which state (or set).%
\footnote{Here it is helpful to keep in mind some facts about how
the Bethe roots of an $\alSU(2)$ chain are distributed in the
complex plain, see for example \cite{Minahan:2002ve}.}

\begin{table}\centering
$\begin{array}{|c|c|c||l|}\hline L & M & \gaugepar &  \\
\hline \hline 6 & 4^\ast & - & \fes{3}{48}{768}{48} \rap^4_0
+\fes{18}{72}{288}{72} \rap^2_0 -1=0 \\
\hline 7 & 3 & \pm & \fes{15}{960}{61440}{960} \rap^6_0
                     +\fes{5}{80}{1280}{80}\rap^4_0
                     +\fes{45}{180}{720}{180}\rap^2_0
                     -9=0 \\ \hline
8 & 3 & \pm &  \fes{}{16}{1024}{16}\rap^6_0
              - \fes{2}{8}{128}{8}\rap^4_0
              +\fes{9}{9}{36}{9} \rap^2_0
              -\fes{4}{1}{1}{1}=0 \\
& 4 & + & \fes {135\rap_0^{12}+ 450\rap_0^{10} -
63\rap_0^8  + 1276\rap_0^6 - 2279\rap_0^4  +
162\rap_0^2-1=0} {552960\rap_0^{12}+ 460800\rap_0^{10}-
16128\rap_0^8   + 81664\rap_0^6 - 4464\rap_0^4+
648\rap_0^2  -1=0} { 2264924160\rap_0^{12}+
471859200\rap_0^{10 }- 4128768\rap_0^8 + 5226496\rap_0^6
- 71424\rap_0^4 + 2592\rap_0^2 -1=0} {552960\rap_0^{12}+
460800\rap_0^{10}- 16128\rap_0^8   + 81664\rap_0^6 -
4464\rap_0^4+ 648\rap_0^2  -1=0}
\\
& 6^\ast & - &
           \fes{}{64}{4096}{64} \rap^6_0
           -\fes{13}{208}{3328}{208} \rap^4_0
           -\fes{77}{308}{4928}{308} \rap^2_0
           +1
           =0
 \\\hline
\end{array}$

\caption{One-loop Bethe roots} \label{tab:Spec.Class}
\end{table}

After having obtained the one-loop Bethe roots, solving the expanded
Bethe equations order by order in $g^2$ for their higher-loop
corrections becomes a purely algebraic exercise.
Using these, the energy/charge eigenvalues are computed
and subsequently compared to \tabref{tab:Spec.Low}.

In this context let us point out the importance of paired and unpaired states.
The unpaired states correspond to symmetric distributions
of the Bethe roots, $\{\rap_k\}=\{-\rap_k\}$, which in
turn implies the vanishing of all odd charges.
The momentum constraint $\shift=1$ \eqref{eq:Shift} is almost automatically
satisfied. It merely implies that for odd $M$, in addition to
one root at the origin $\rap_1=0$, two of the roots must be
at the singular points
$\rap_{2,3}\approx\pm\fes{i}{\frac{i}{2}}{\frac{i}{4}}{\frac{i}{2}}$
(cf.~\cite{Beisert:2003xu}).
For even $M$ there can be no such roots.
This symmetry vastly simplifies the computation of the
Bethe roots; we need to solve only half as many equations!
For paired states this simplification does not apply and
finding the roots is a formidable problem even for smaller
values of $(L,M)$.

\subsection{Two Excitations}
\label{sec:Spec.TwoEx}

Now that both the Hamiltonian approach and the perturbative Bethe
ansatz have been described in detail, we may compare the results
that are obtained using these two procedures. Let us start by
analyzing the states with two magnons \cite{Beisert:2002tn}.
On the perturbative gauge theory side of
our discussion one can extend the conjectured all-loop formula (6)
in \cite{Beisert:2003jb}
\[\label{eq:Spec.All}
\dil (J,n,g) = J + 2 + \sum^\infty_{\ell=1}
\lrbrk{ \fes{4}{8}{16}{8}g^2 \sin^2
\frac{\pi n}{J+1} }^\ell
\lrbrk{ c_\ell +
\sum^{\ell-1}_{k,h=1} c_{\ell,k,h}
\frac{\cos^{2h} \frac{\pi n}{J+1}}{(J+1)^k} }
\]
by matching more and more coefficients $c_\ell$, $c_{\ell,k,h}$ to
sufficiently many two-excitation states. We present a summary of
findings in \tabref{tab:Spec.TwoEx}.

\begin{table}\centering
$\begin{array}{lll}
\displaystyle c_1=+1,&\quad&\\[3pt]
\displaystyle c_2=-\frac{1}{4},&\quad&
c_{2,1,1}=-1,\\[0.25cm]
\displaystyle c_3=+\frac{1}{8},&\quad&
c_{3,k,h}=\matr{ll}{+\sfrac{3}{4}&+\sfrac{1}{2}\\[2pt]-\sfrac{3}{4}&+\sfrac{5}{2}
},\\[0.4cm]
\displaystyle c_4=-\frac{5}{64},&\quad& c_{4,k,h}=\matr{lll}{
-\sfrac{5}{8}&-\sfrac{5}{12}&-\sfrac{1}{3}\\[2pt]
+\sfrac{3}{4}&- \sfrac{7}{4}&-\sfrac{7}{2}\\[2pt]
-\sfrac{1}{2}&+\sfrac{59}{12}&-\sfrac{49}{6}
},\\[0.8cm]
\displaystyle c_5=+\frac{7}{128},&\quad& c_{5,k,h}=\matr{llll}{
+\frac{35}{64}&+\frac{35}{96}&+\frac{7}{24}&+\frac{1}{4}\\[2pt]
-\frac{45}{64}&+\frac{185}{96}&+\frac{131}{48}&+\frac{33}{8}\\[2pt]
+\frac{5}{8}&-\frac{125}{24}&-\frac{13}{24}&+\frac{81}{4}\\[2pt]
-\frac{5}{16}&+\frac{305}{48}&-\frac{1319}{48}&+\frac{243}{8} }.
\end{array}$
\caption{Coefficients for the two-excitation formula (\ref{eq:Spec.All}).}
\label{tab:Spec.TwoEx}
\end{table}

When the coefficients have been determined, we may compare the
formula to the results of the Bethe equations. As mentioned above,
the states with two magnons turn out to be unpaired due to
momentum conservation, i.e.~we only have to solve one Bethe
equation \eqref{eq:BethePhi} for $L=J+2$ in
$\rap\equiv \rap_1=-\rap_2$:
\[\label{eq:TwoExc.ClassBetheEq}
\left(
\frac{\rap-\fes{i}{\sfrac{i}{2}}{\sfrac{i}{4}}{\sfrac{i}{2}}}
{\rap+\fes{i}{\sfrac{i}{2}}{\sfrac{i}{4}}{\sfrac{i}{2}}}
\right)^{J+1}
=\left(
\frac{1
+\sqrt{1-\fes{4}{2}{}{2}g^2/(\rap+\fes{i}{\sfrac{i}{2}}{\sfrac{i}{4}}{\sfrac{i}{2}})^2}}
{1
+\sqrt{1-\fes{4}{2}{}{2}g^2/(\rap-\fes{i}{\sfrac{i}{2}}{\sfrac{i}{4}}{\sfrac{i}{2}})^2}}
\right)^{J+2}.
\]
Solving this Bethe equation order by order in $g^2$ in the way
described above leads to
\<\label{eq:TwoExc.SolP} \rap \eq
\fes{}{\half}{\quarter}{\half}\cot \frac{\pi n}{J+1} \bigg[1
+\fes{2}{4}{8}{4}g^2\,\frac{J+2}{J+1}\,\sin^2 \frac{\pi n}{J+1}
\nl \qquad\qquad\qquad-2g^4\frac{(J+2)(J-1+6\cos^2\frac{\pi
n}{J+1})}{(J+1)^2}\, \sin^4\frac{n\pi}{J+1} + \order{g^6} \bigg]
\>
The higher order terms are rather lengthy which is
why we do not explicitly write them down here. After plugging this
into the energy formula \eqref{eq:Energy} together
with \eqref{eq:magnoneng}, we obtain \eqref{eq:Spec.All}
for the first few loop-orders.

At this point let us also say a few words about the inhomogeneous
Bethe equations \eqref{eq:BethePhiNew}. The procedure of computing
the Bethe roots is exactly the same as before, only the left hand
side of \eqref{eq:BethePhiNew} differs from \eqref{eq:BethePhi}.
It is possible to make this replacement since both equations agree
up to $\order{g^{2L}}$. However, this is also the order at which
the contributions of wrapping interactions have to be taken into
account. Since we do not know how to do this, both types of Bethe
equations are equivalent at the desired accuracy.
The benefit of the inhomogeneous Bethe equations will be demonstrated
in the following section.
As an example let us calculate the Bethe
roots of the Konishi descendant $(L,M)=(4,2)$.
These can be solved for exactly, we find
\[\label{eq:KonRap}
\rap_{1,2}=\pm
\sqrt{-\fes{\sfrac{1}{3}}{\sfrac{1}{12}}{\sfrac{1}{48}}{\sfrac{1}{12}}
      +\fes{\sfrac{2}{3}}{\sfrac{1}{3}}{\sfrac{1}{6}}{\sfrac{1}{3}}g^2
      +\fes{\sfrac{2}{3}}{\sfrac{1}{6}}{\sfrac{1}{24}}{\sfrac{1}{6}}
         \sqrt{1 + \fes{2}{4}{8}{4}g^2
                 + \fes{\sfrac{5}{2}}{10}{40}{10}g^4}}\,.
\]
The corresponding exact inhomogeneous transfer matrix is
\[\label{eq:KonTbar}
\bar\transfer(\rap)=
\sfrac{5}{8} -\fes{\half}{}{2}{}g^2
             +\fes{\quarter}{}{4}{} g^4
             + \sqrt{1 + \fes{2}{4}{8}{4}g^2+\fes{\sfrac{5}{2}}{10}{40}{10}g^4}
+ \fes{\sfrac{3}{4}}{3}{12}{3}\rap^2
- \fes{\half}{4}{32}{4}g^2\rap^2
+ \fes{\sfrac{1}{8}}{2}{16}{2}\rap^4.
\]
As expected,
the resulting energy eigenvalue agrees with \eqref{eq:Spec.All}
up to and including $\order{g^6}$.

\subsection{Singular Solutions}
\label{sec:Spec.Singular}

Next, let us analyze unpaired three-excitation states
\cite{Beisert:2003tq}
\[\label{eq:threeunpaired}
\sum_{p=1}^{L-4}(-1)^p\Tr \phi \fldZ^p \phi \fldZ^{L-3-p}\phi+\order{g^2}
\]
at higher loops using perturbative gauge theory techniques.
Note that this exact one-loop form of
the eigenstates is corrected at higher-loops.
We find for the scaling dimensions
\<\label{eq:Spec.ThreeEng} \dil\eq \phantom{0}2,\nln
\dil\eq \phantom{0}4
            +\fes{3}{6}{12}{6}g^2
            -\fes{\sfrac{12}{4}}{12}{48}{12}g^4
            +\fes{\sfrac{84}{16}}{\sfrac{84}{2}}{336}{\sfrac{84}{2}}g^6
            +\ldots\,,\nln
\dil\eq
\phantom{0}6+\fes{3}{6}{12}{6}g^2
            -\fes{\phantom{\scriptstyle 9}\sfrac{9}{4}}{\phantom{1}9}{36}{\phantom{1}9}g^4
            +\fes{\sfrac{63}{16}}{\sfrac{63}{2}}{252}{\sfrac{63}{2}}g^6
            -\fes{\sfrac{621}{64}}{\sfrac{621}{4}}{2484}{\sfrac{621}{4}}g^8
            +\fes{\sfrac{7047}{256}}{\sfrac{7047}{8}}{28188}{\sfrac{7047}{8}}g^{10}
            +\ldots\,,\nln
\dil\eq
\phantom{0}8+\fes{3}{6}{12}{6}g^2
            -\fes{\phantom{\scriptstyle 9}\sfrac{9}{4}}{\phantom{1}9}{36}{\phantom{1}9}g^4
            +\fes{\sfrac{66}{16}}{\sfrac{66}{2}}{264}{\sfrac{66}{2}}g^6
            -\fes{\sfrac{648}{64}}{\sfrac{648}{4}}{2592}{\sfrac{648}{4}}g^8
            +\fes{\sfrac{7212}{256}}{\sfrac{7212}{8}}{28848}{\sfrac{7212}{8}}g^{10}
            +\ldots\,,\nln
\dil\eq
10          +\fes{3}{6}{12}{6}g^2
            -\fes{\phantom{\scriptstyle 9}\sfrac{9}{4}}{\phantom{1}9}{36}{\phantom{1}9}g^4
            +\fes{\sfrac{66}{16}}{\sfrac{66}{2}}{264}{\sfrac{66}{2}}g^6
            -\fes{\sfrac{645}{64}}{\sfrac{645}{4}}{2580}{\sfrac{645}{4}}g^8
            +\fes{\sfrac{7179}{256}}{\sfrac{7179}{8}}{28716}{\sfrac{7179}{8}}g^{10}
            +\ldots\,,\nln
\dil\eq
12          +\fes{3}{6}{12}{6}g^2
            -\fes{\phantom{\scriptstyle 9}\sfrac{9}{4}}{\phantom{1}9}{36}{\phantom{1}9}g^4
            +\fes{\sfrac{66}{16}}{\sfrac{66}{2}}{264}{\sfrac{66}{2}}g^6
            -\fes{\sfrac{645}{64}}{\sfrac{645}{4}}{2580}{\sfrac{645}{4}}g^8
            +\fes{\sfrac{7182}{256}}{\sfrac{7182}{8}}{28728}{\sfrac{7182}{8}}g^{10}
            +\ldots\,,\nln
\dil\eq
14          +\fes{3}{6}{12}{6}g^2
            -\fes{\phantom{\scriptstyle 9}\sfrac{9}{4}}{\phantom{1}9}{36}{\phantom{1}9}g^4
            +\fes{\sfrac{66}{16}}{\sfrac{66}{2}}{264}{\sfrac{66}{2}}g^6
            -\fes{\sfrac{645}{64}}{\sfrac{645}{4}}{2580}{\sfrac{645}{4}}g^8
            +\fes{\sfrac{7182}{256}}{\sfrac{7182}{8}}{28728}{\sfrac{7182}{8}}g^{10}
            +\ldots\,,\nln
&&\ldots\,, \>
where we have added the dimension-two half-BPS state and the above
Konishi descendant which appear to be the natural first two
elements of this sequence.

We observe that all corrections $\dil_k$ to the scaling dimensions
below the ``diagonal{}'' $k\leq L-2$, are equal. Incidentally these
coefficients agree with the formula
\[\label{eq:Spec.ThreeAllLoop}
\dil(g)=L+\bigbrk{\sqrt{1+\fes{4}{8}{16}{8}g^2}-1}
+\bigbrk{\sqrt{1+\fes{}{2}{4}{2}g^2}-1}
+\bigbrk{\sqrt{1+\fes{}{2}{4}{2}g^2}-1}.
\]
We may interpret the three terms in parentheses as the energies of
the three excitations. Then this form can be taken as a clear
confirmation of an integrable system with elastic scattering of
excitations.

Only if the loop order is at least half the classical dimension,
the pattern breaks down. Interestingly, if the loop order is
exactly half the classical dimension, the coefficient is decreased by
$\fes{3\cdot 2^{2-2\ell}}{3\cdot 2^{2-\ell}}{12}{3\cdot 2^{2-\ell}}$.
It would be of great importance to understand the
changes further away from the diagonal.
This might provide us with
clues about wrapping interactions, which, in the above example,
obscure the scaling dimension of the Konishi state beyond
three-loops.

For completeness, we state an analogous all-loop conjecture for
the higher charges
\[\label{eq:Spec.ThreeCharges}
\charge_r=
\frac{(+i)^{r-2}+(-i)^{r-2}}{(r-1)g^{2r-2}}
\lrbrk{2^{1-r}\bigbrk{\sqrt{1+\fes{4}{8}{16}{8}g^2}-1}^{r-1}
+\bigbrk{\sqrt{1+\fes{}{2}{4}{2}g^2}-1}^{r-1}}.
\]
Alternatively, in terms of a transfer matrix:
\[\label{eq:Spec.ThreeTransfer}
\transfer(x)=
\frac{x-\sfrac{i}{4}\brk{1+\sqrt{1+\fes{4}{8}{16}{8}g^2}}}
     {x+\sfrac{i}{4}\brk{1+\sqrt{1+\fes{4}{8}{16}{8}g^2}}}\,\,
\frac{x-\sfrac{i}{2}\brk{1+\sqrt{1+\fes{}{2}{4}{2}g^2}}}
     {x+\sfrac{i}{2}\brk{1+\sqrt{1+\fes{}{2}{4}{2}g^2}}}
\]

Now let us discuss how this result can be reproduced using the
Bethe ansatz. The unpaired three-excitation states are singular
solutions of the Bethe equations.%
\footnote{This is related to the fact that
in the state \eqref{eq:threeunpaired}
two of the fields $\phi$
are \emph{always} next to each other.}
At leading order, the Bethe roots are at $\rap_1=0$ and the singular
points $\rap_{2,3}=\pm\fes{i}{\frac{i}{2}}{\frac{i}{4}}{\frac{i}{2}}$,
see e.g.~\cite{Beisert:2003xu}.
The singular roots lead to divergencies in the Bethe equations
which would have to be regularized. While it is not clear to us
how the regularization can be done at higher loops, there is an
alternative way to solve the equations without the need to
regularize.
The Bethe equations follow from the
requirement that $\bar\transfer(\rap)$ must not have poles at
$\rap=\rap_k$. Here we are forced to use the transfer matrix of
the inhomogeneous Bethe ansatz in \secref{sec:Inhomo} and not
the one of \secref{sec:Rapid}. The reason is that
the function $x(\rap\pm\fes{i}{\frac{i}{2}}{\frac{i}{4}}{\frac{i}{2}})^L$ introduces additional
overlapping singularities at $\rap=\pm\fes{i}{\frac{i}{2}}{\frac{i}{4}}{\frac{i}{2}}$, while
the polynomial $P_L(\rap\pm\fes{i}{\frac{i}{2}}{\frac{i}{4}}{\frac{i}{2}})$
certainly does not.
Therefore, only the
inhomogeneous Bethe equations can be used to find the
quantum corrections to the singular Bethe roots.
Interestingly, one finds their
positions not to be modified up to $\order{g^L}$
and \eqref{eq:Spec.ThreeTransfer}
follows straightforwardly from
\eqref{eq:TransferPhi}
\[
\transfer(x)=
\frac{x-x(0+\fes{i}{\frac{i}{2}}{\frac{i}{4}}{\frac{i}{2}})}
{x-x(0-\fes{i}{\frac{i}{2}}{\frac{i}{4}}{\frac{i}{2}})}
\,
\frac{x-x(+\fes{i}{\frac{i}{2}}{\frac{i}{4}}{\frac{i}{2}}+\fes{i}{\frac{i}{2}}{\frac{i}{4}}{\frac{i}{2}})}
{x-x(+\fes{i}{\frac{i}{2}}{\frac{i}{4}}{\frac{i}{2}}-\fes{i}{\frac{i}{2}}{\frac{i}{4}}{\frac{i}{2}})}
\,
\frac{x-x(-\fes{i}{\frac{i}{2}}{\frac{i}{4}}{\frac{i}{2}}+\fes{i}{\frac{i}{2}}{\frac{i}{4}}{\frac{i}{2}})}
{x-x(-\fes{i}{\frac{i}{2}}{\frac{i}{4}}{\frac{i}{2}}-\fes{i}{\frac{i}{2}}{\frac{i}{4}}{\frac{i}{2}})}
+\ldots
=
\frac{x-x(+\fes{i}{\frac{i}{2}}{\frac{i}{4}}{\frac{i}{2}})}
{x-x(-\fes{i}{\frac{i}{2}}{\frac{i}{4}}{\frac{i}{2}})}
\,
\frac{x-x(+\fes{2i}{i}{\frac{i}{2}}{i})}
{x-x(-\fes{2i}{i}{\frac{i}{2}}{i})}
+\ldots
\]
When the shifts of the poles at $\order{g^L}$ are
properly taken into account, one finally obtains
the corrections above the diagonal in
\eqref{eq:Spec.ThreeEng}.

As an example let us consider the state $(L,M)=(4,3)$, which is
the mirror state of the Konishi descendant we computed
in the previous subsection.
The mirror of a state $(L,M)$ is a (zero-norm) state of the type $(L,L-M+1)$
which has the precisely the same charges.
Its Bethe roots can be determined
analytically%
\[\rap_1=0,\qquad
\rap_{2,3}=\pm \sqrt{-\fes{3}{\sfrac{3}{4}}{\sfrac{3}{16}}{\sfrac{3}{4}}
-\fes{2}{}{\half}{} g^2 +
\fes{2}{\sfrac{1}{2}}{\sfrac{1}{8}}{\sfrac{1}{2}}
\sqrt{1 + \fes{2}{4}{8}{4}g^2  + \fes{\sfrac{5}{2}}{10}{40}{10}g^4}
}\,.\]
The exact transfer matrix $\bar\transfer(\rap)$ is
the same as for the $(4,2)$ state \eqref{eq:KonTbar}.
This demonstrates that the equations in \secref{sec:Inhomo}
are fully consistent when computing the unphysical
transfer matrix $\bar\transfer(\rap)$.

For the physical charges $\charge_r$
the situation is slightly different.
The charges do agree with the ones of the Konishi descendant.%
\footnote{The physical charges of the singular solutions
need to be regulated. Here, $g$ acts as a natural regulator,
when corrections to the roots are taken into account.
This leaves some spurious terms of the sort $\sqrt{g}$
in the charge eigenvalues which we shall ignore.}
As expected, this agreement
persists only for the first few orders.%
In particular, $\charge_2$ agrees up to
$\order{g^4}$ and $\charge_4$ up to $\order{g^0}$.
Remarkably, this is also precisely the accuracy at which wrappings occur,
which constitutes some evidence for the improper treatment of
wrapping interactions by our ansatz.
We therefore conclude, that the ansatz of \secref{sec:Inhomo}
yields self-consistent results for the physical charges only
for low loop orders.

In this example we considered the mirror of a regular state in
order to investigate a singular state. However, it is also
possible to reverse the line of argumentation:
When interested in a singular three-excitation state, one
can instead consider its mirror, which is a regular,
unpaired $(L-2)$-excitation state.
The procedure is the same as for ordinary unpaired states
and we will not discuss it in any more detail.
Instead we refer the reader to \tabref{tab:Spec.Class} where
the one-loop roots of some mirror states are listed
(marked by $^\ast$).
The fact that these states have the same energy eigenvalues
(up to ``wrapping order{}'') as the original, singular
states with three excitations shows that the inhomogeneous Bethe
ansatz is consistent.

\subsection{Paired Three-Excitation States}
\label{sec:Spec.ThreeExPair}

For three excitations there exist also paired states. Finding the
leading order Bethe roots for these states is more complicated as
we cannot use the symmetry argument
$\rap_{2k}=-\rap_{2k+1}$. Therefore, one has to work with
the whole set of Bethe equations. The longer the chain and the
more excitations it has, the more difficult it becomes to solve
the equations. For the states $(L,M)=(7,3)$ and $(L,M)=(8,3)$ we
have used the resultant of polynomials in several variables in
order to iteratively remove their dependence on all but a single
variable.%
\footnote{The resultant $R$ of two polynomials
$P(x)$ and $P'(x)$ is zero
if and only if they have a common root.
When the polynomials also depend on further variables $y_i$,
the vanishing of the resultant $R(y_i)=0$ gives
an algebraic constraint among the $y_i$'s alone.}
The results are given by the corresponding equations in
\tabref{tab:Spec.Class}. Each of these equations is solved by two
sets of momenta $\{\rap_k\}$ and
$\{\rap'_k\}=\{-\rap_k\}$. Note the specific distribution of
the three momenta in each set: one momentum lies on the positive
(negative) real axis while the other two momenta are related by
complex conjugation and have a negative (positive) real part.
The higher even charges of the two sets are equal
($\charge'_{2r}=\charge_{2r}$) whereas the odd ones differ by an
overall minus sign ($\charge_{2r+1}'=-\charge_{2r+1}$).
We find agreement with the eigenvalues of the spin chain charges.

\subsection{Higher Excitations}
\label{sec:Spec.HigherEx}

For all states with an even number of magnons (and $M\geq4$) the
procedure is exactly the same as for the paired and unpaired
states described above. The only complication arises when trying
to solve the one-loop Bethe ansatz.

For the unpaired states with an odd number of excitations one
knows that three of the one-loop roots are \emph{singular},
i.e.~their positions are $\rap_1=0$,
$\rap_{2,3}=\pm\fes{i}{\sfrac{i}{2}}{\sfrac{i}{4}}{\sfrac{i}{2}}$.
The remaining roots are again symmetric $\rap_{2k}=-\rap_{2k+1}$.
As in the case of the unpaired three-excitation states we find
that the singular roots do not receive corrections up to
(and including) $\order{g^{L-2}}$
whereas the roots $\rap_k$ with $k\geq4$ are corrected at every
order in $g^2$.

We have specifically checked the agreement of results
for the unpaired $(8,4)$, $(9,4)$, $(10,4)$ and $(10,5)$ states.

\section{Inhomogeneous Long-Range Spin Chains}
\label{sec:InhomoChains}

Our findings in \secref{sec:Inhomo} that the novel
spin chain may be interpreted as an \emph{inhomogeneous} spin chain
appears to be more generally true for long-range chains,
as we will show in this appendix.
In \secref{sec:Rapid} we have inverted the
relation $\rap(p)$ for our spin chain model
as
\[\label{eq:Gen.pofx}
\exp(ip)=\frac{x(\rap+\frac{i}{2})}{x(\rap-\frac{i}{2})}
\]
and found a function $x(\rap)$ so that the Bethe ansatz can be
expressed as follows%
\fes{\footnote{We use different conventions than in the remainder of the paper.}}{}
{\footnote{We use different conventions than in the remainder of the paper.}}{}
\[\label{eq:Gen.equation}
\frac{x(\rap_k+\sfrac{i}{2})^L}
     {x(\rap_k-\sfrac{i}{2})^L}=
\prod_{\textstyle\atopfrac{j=1}{j\neq k}}^M
\frac{\rap_k-\rap_j+i}
     {\rap_k-\rap_j-i}\,
\]
In \secref{sec:Inhomo} we have then truncated the
expansion of $x(\rap)^L$ to a polynomial
$P_L(\rap)$ in order to relate the model to
an inhomogeneous spin chain.
Here we would like to repeat this exercise for
more general long-range spin chains.

Let us start with the Inozemtsev spin chain
\cite{Inozemtsev:1989yq,Inozemtsev:2002vb} as treated in \cite{Serban:2004jf}.
The relation $\rap(p)$ is given by
($t$ is the coupling constant proportional to $g^2$)
\[\label{eq:Ino.uofp}
\rap(p)=
\half\cot(\half p)
+\sum_{n=1}^\infty
\frac{4t^n\sin(\half p)\cos(\half p)}{(1-t^n)^2+4t^n\sin(\half p)^2}\,.
\]
The inversion of this relation is given by
\eqref{eq:Gen.pofx} and the function%
\[\label{eq:Ino.xofu}
x(\rap)=
\rap
- \frac{t}{\rap}
- \left( \frac{1}{\rap^3} + \frac{3}{\rap} \right) t^2
- \left( \frac{2}{\rap^5} + \frac{7}{\rap^3} + \frac{4}{\rap} \right) t^3
- \left( \frac{5}{\rap^7} + \frac{22}{\rap^5} + \frac{20}{\rap^3} + \frac{7}{\rap} \right) t^4
-\ldots
\]
It is also useful to note its inverse
\[\label{eq:Ino.uofx}
\rap(x)=x
+ \frac{t}{x}
+ \frac{3t^2}{x}
+ \left( \frac{1}{x^3} + \frac{4}{x} \right)  t^3
+ \left( \frac{3}{x^3} + \frac{7}{x} \right)  t^4
+ \left( \frac{2}{x^5} + \frac{9}{x^3} + \frac{6}{x} \right)  t^5
+\ldots
\]
It would be interesting to investigate also the physical charges
or the transfer matrix.
Possibly they are also given by
\eqref{eq:TransferPhi,eq:ChargesPhi}
or similar expressions involving $x(\rap)$.
Furthermore it would be interesting to find
exact, analytic expressions for these functions.

The function \eqref{eq:Ino.xofu} has a special property that allows us to
reformulate the model as an inhomogeneous spin chain:
The expansion of $x(\rap)^L$ in powers of $t$ up
to $\order{t^{L-1}}$ is a polynomial in $\rap$.
Inverse powers of $\rap$ start contributing only at
$\order{t^L}$. At this order, however, wrapping interaction
start to contribute and the asymptotic Bethe ansatz
does not apply anymore. Thus we may truncate
$x(\rap)^L$ at $\order{t^{L-1}}$ and get a
polynomial $P_L(\rap)$ of degree $L$,
precisely what is needed for an inhomogeneous spin chain.

This property can be used to find functions $x(\rap)$
for more general long-range spin chains.
Here it is more useful to investigate the inverse $\rap(x)$.
We find that precisely the functions of
the form \eqref{eq:Ino.uofx}, but with
different coefficients, have this property.
It is interesting to note that then
the coefficients of $t^n/\rap^{2n-1}$
in $x(\rap)$
are always the Catalan numbers%
\footnote{The Catalan numbers
have also been found in $\superN=4$ SYM
as the coefficients of the
maximal shifts of a single excitation
\cite{Gross:2002su}.
If directly related, this observation
gives support to the conjecture of all-loop integrability
\cite{Beisert:2003tq}.}
from the expansion of $\half x+\half x\sqrt{1-4ct/x^2}$.
This is because these coefficients
are determined by the first two
terms in \eqref{eq:Ino.uofx}, $x+ct/x$, only.
Remarkably, the restricted form \eqref{eq:Ino.uofx}
confirms the claimed uniqueness of our model specified
by the conditions ($i$-$iii$) in \secref{sec:Ansatz}:
For a correct scaling behavior, all terms should scale
as $L$, i.e.~only terms proportional to $t^n/x^{2n-1}$.
The only allowed terms in \eqref{eq:Ino.uofx} are $x$ and $t/x$
in agreement with our function \eqref{eq:uofx}.

\section{Elliptic Solutions of the String and Gauge Bethe Equations}
\label{sec:elliptic}

In \secref{sec:Trouble} we compared the
Bethe equations for semi-classical string theory and asymptotic gauge theory
in generality. It is interesting to investigate the consequences for the
analytic structure of the respective solutions on some explicit,
solvable examples.
In this appendix we will therefore study the string and gauge
Bethe equations for the ``folded{}'' and the ``circular{}'' spinning
string.%
\footnote{The below
calculations in the case of the classical string sigma model Bethe
equation were independently performed by G.~Arutyunov
(unpublished).}
These are the simplest families of solutions
which still depend on a continuous parameter, namely the filling
fraction $\alpha=\frac{M}{L}$. As a byproduct we will verify
that the classical sigma model Bethe equation of
\cite{Kazakov:2004qf} indeed reproduces the energies of these spinning
string configurations which were previously obtained by simpler,
but less systematic methods \cite{Frolov:2003qc,Frolov:2003xy,Arutyunov:2003uj}.
We shall also find that the all-loop gauge solutions are significantly
more complicated in analytic structure.
The folded string is believed to
correspond to the ground state of the representation carrying the
charges $M$ and $L-M$. The resulting two-cut solutions may be
expressed through elliptic functions, and are closely related to
the ones describing the multicritical O$(\pm2)$ matrix model
\cite{Kostov:1992pn}. A simplifying feature is that all odd
charges are zero. These solutions were crucial in establishing for
the first time the agreement of ``long operator{}'' anomalous
dimensions and semi-classical string solutions at one-loop
\cite{Beisert:2003xu}, at two loops \cite{Serban:2004jf}, as well
as the matching of integrable structures up to two loops
\cite{Arutyunov:2003rg,Arutyunov:2004xy}. They also led to the
discovery of the three-loop disagreement \cite{Serban:2004jf}. The
calculations below are straightforward modifications of the ones
presented in
\cite{Beisert:2003xu,Beisert:2003ea,Arutyunov:2003rg,Serban:2004jf}
and we refer to these papers for further details.

Here we briefly state our conventions for the elliptic integrals appearing below.
The complete elliptic integrals of the first ($\ellK$) and second ($\ellE$)
kind are
\[\label{eq:conds2}
\ellK(q)\equiv \int_0^{\pi/2}\frac{d\rap}{\sqrt{1-q\sin^2 \rap}}
\qquad \qquad
\ellE(q)\equiv \int_0^{\pi/2} d\rap\ \sqrt{1-q\sin^2 \rap}.
\]
and the complete elliptic integral of the third kind is
\[\label{eq:third}
\ellPi(m^2,q)\equiv
\int_0^{\pi/2}\frac{d\rap}{(1-m^2 \sin^2 \rap)
\sqrt{1-q\sin^2 \rap}}\ .
\]
%

\subsection{The Folded String}
\label{sec:Spinning.Folded}

\subsubsection{Semi-classical String Solution}
\label{sec:stringfolded}

Let us write down the classical string equations of \secref{sec:Spinning}
for the case of exactly two contours $\contourstring_+$ and $\contourstring_-$
which are mutual images w.r.t.~reflection around the imaginary axis.
The Bethe equation \eqref{eq:stringbethe} becomes
\[\label{eq:stringbethesymm}
\pint_{\contourstring_+} dx'\,\frac{\sigma(x')\,x}{x^2-x'^2}=
\frac{\engs}{\fes{2}{4}{8}{4}}\,\frac{x}{x^2-\fes{}{\half}{\quarter}{}\coup^2}
+\half\pi n_+
\qquad \mbox{with} \qquad
x \in \contourstring_+ \, ,
\]
and the normalization condition \eqref{eq:stringnorm} reads
\[\label{eq:stringnormsymm}
\int_{\contourstring_+}
dx\,\sigma(x)\fes{\half}{}{2}{}\,\left( 1-\frac{\coup^2}{\fes{}{2}{4}{}x^2} \right)=
\frac{\alpha}{2} \, .
\]
The resolvent \eqref{eq:stringresolv} is a function analytic throughout
the spectral $x$-plane, except for the cuts $\contourstring_+$ and $\contourstring_-$:
\[\label{eq:stringresolvsymm}
\resolvs(x)=2 x \int_{\contourstring_+} dx'\, \frac{\sigma(x')}{x'^2-x^2}\, .
\]
For small $x$ we may expand the resolvent as a Taylor series in the local
charges
\[
\resolvs(x)=\sum_{r=1}^{\infty} \charges_{2 r} x^{2 r-1}
\qquad \mbox{with} \qquad
\charges_{2 r}=2 \int_{\contourstring_+} dx\,\frac{\sigma(x)}{x^{2 r}} \, ,
\]
cf.~\eqref{eq:stringlocal,eq:stringtaylor}.
Note that the odd charges are zero: $\charges_{2 r-1}=0$, and we recall
the relation between the scaled string energy and the second charge
\eqref{eq:stringenergy}: $\engs=1+\fes{}{}{}{2} \coup^2 \charges_2(\coup)$.
It is easily seen that the mode number $n_+$ may be absorbed,
after rescaling the spectral parameter $x \mapsto x/n_+$, into
the coupling constant $\coup \mapsto \coup/n_+$;
we nevertheless keep full $n_+=-n$ dependence.

It is technically convenient to solve this equation by analytically continuing to
a negative filling fraction $\alpha<0$, as in \cite{Beisert:2003xu}.
The complex cuts $\contourstring_+$ and $\contourstring_-$ are flipped to,
respectively, the real intervals $[a,b]$ and $[-b,-a]$.
This involves some sign changes in the above equations, which now read
\[\label{eq:stringbethefolded}
\pint_a^b dx'\,\frac{\sigma(x')\,x}{x'^2-x^2}=
\frac{\engs}{\fes{2}{4}{8}{4}}\,\frac{x}{x^2-\fes{}{\half}{\quarter}{}\coup^2}
-\frac{\pi n}{2}
\qquad \mbox{with} \qquad
\int_a^b
dx\,\sigma(x)\fes{\half}{}{2}{}\,\left( 1-\frac{\coup^2}{\fes{}{2}{4}{}x^2} \right)=
-\frac{\alpha}{2} \, ,
\]
and
\[\label{eq:stringresolvfolded}
\resolvs(x)=2 x \int_a^b dx'\, \frac{\sigma(x')}{x^2-x'^2}\, ,
\]
while \eqref{eq:stringenergy} is unchanged.
The solution of \eqref{eq:stringbethefolded} could be obtained by an inverse
Hilbert transform.
%
%
However, in line with the connotation of the word ``resolvent{}'',
it is technically easier to directly find an integral representation of
$\resolvs(x)$ with the required analytic properties and boundary conditions.
One finds (here $q=1-{a^2}/{b^2}$)
that \eqref{eq:stringresolvfolded} is explicitly given by
\<\label{eq:resolv11}
\resolvs(x)\eq
-\fes{\engs}{\frac{\engs}{2}}{\frac{\engs}{4}}{\frac{\engs}{2}}\,
\frac{x}{x^2-\fes{}{\half}{\quarter}{}\coup^2}
+\frac{2 n a^2}{x b} \sqrt{\frac{b^2-x^2}{a^2-x^2}}\,
\ellPi\left(-q \frac{x^2}{a^2-x^2},q\right)
\nl
+\frac{\engs}{\fes{}{2}{4}{2}x}\, \frac{\fes{}{\half}{\quarter}{}\coup^2}{x^2-\fes{}{\half}{\quarter}{}\coup^2}
\sqrt{\frac{(b^2-x^2)(a^2-x^2)}{(b^2-\fes{}{\half}{\quarter}{}\coup^2)(a^2-\fes{}{\half}{\quarter}{}\coup^2)}} \, ,
\>
which is the form of $\resolvs(x)$ appropriate for an expansion near
$x=0$, as needed for generating the local charges
\[\label{eq:stringresolvfolded2}
\resolvs(x)=
\sum_{r=1}^{\infty} \charges_{2 r} x^{2 r-1}
\qquad \mbox{with} \qquad
\charges_{2 r}=-2 \int_a^b dx\,\frac{\sigma(x)}{x^{2 r}} \, .
\]
We can now also read off the pseudodensity $\sigma(x)$ as
the discontinuity of $\resolvs(x)$ on the cut:
\[
\sigma(x)=\frac{1}{2 \pi x b} \sqrt{\frac{x^2-a^2}{b^2-x^2}} \left[
\fes{2}{}{\half}{}b \,\engs\, \frac{x^2}{x^2-\fes{}{\half}{\quarter}{}\coup^2}
\sqrt{\frac{b^2-\fes{}{\half}{\quarter}{}\coup^2}{a^2-\fes{}{\half}{\quarter}{}\coup^2}}-
4 n x^2\, \ellPi\left(\frac{b^2-x^2}{b^2},q\right) \right] \, .
\]
Furthermore, the known behavior of the resolvent at $x=0$, namely
$\resolvs(x)=\frac{0}{x}+\charges_2 x+ \order{x^3}$, yields two
conditions:
\[
\frac{\engs}{\sqrt{(b^2-\fes{}{\half}{\quarter}{}\coup^2)(a^2-\fes{}{\half}{\quarter}{}\coup^2)}}
=\frac{4n}{b}\, \ellK(q)
\qquad \mbox{and} \qquad
\ellK(q)=\frac{1}{\fes{2}{4}{8}{4} na}+\frac{\coup^2}{\fes{}{2}{4}{}a^2}\, \ellE(q) ,
\]
while a third condition is obtained from the behavior at $x \rightarrow \infty$,
namely $\resolvs(x) \rightarrow \frac{2}{x}\int_a^b dx' \sigma(x')$
(note that the last term in \eqref{eq:resolv11} flips sign for
large values of $x$):
\[
\alpha=\frac{1}{2}-\fes{}{2}{4}{2}n b\, \ellE(q)
+\frac{\fes{}{}{}{2}n\coup^2}{b}\, \ellK(q).
\]
These three equations determine the three unknowns $a,b$ and $\engs$
as a function of the filling fraction $\alpha$ and the coupling constant
$\coup$. One checks that they indeed reproduce the energy of the
folded string as first obtained without Bethe ansatz in \cite{Frolov:2003xy}.

\subsubsection{All-loop Asymptotic Gauge Solution}
\label{sec:gaugefolded}

In order to solve the singular integral equation \eqref{eq:gaugebethe}
for perturbative gauge theory
it is useful to introduce a $\rap$-resolvent $\bar\resolv(\rap)$ through
\[\label{eq:gaugeresolv}
\bar\resolv(\rap)=\int_{\contourgauge}
\frac{\fes{2}{}{\half}{}d\rap' \rho(\rap')}{\rap'-\rap} =
\sum_{r=1}^\infty \, \bar{\charge}_r \, \rap^{r-1}
\qquad \mbox{with} \qquad
\bar{\charge}_r=\int_{\contourgauge} \frac{\fes{2}{}{\half}{}d\rap\, \rho(\rap)}{\rap^r} \, .
\]
Here (and in similar expressions for the remainder of this appendix)
it is understood that, while $\bar\resolv(\rap)$ is defined throughout
the complex $\rap$-plane, its expansion
in local charges is only possible in a finite domain around $\rap=0$.
Note however that this resolvent does \emph{not} correspond to the
scaling limit of the (logarithm of) the transfer matrix
\eqref{eq:Transfer}, except for the one-loop
approximation. Accordingly, the proper gauge charges $\charge_r$
are not given by the moments $\bar{\charge}_r$ beyond one loop.
Instead the former are linear combinations of the latter, cf.~\cite{Serban:2004jf},
as coded into the equation \eqref{eq:gaugecharge}.

In our perturbative gauge theory ansatz,
the two-cut Bethe equation of the folded string in the last section
\secref{sec:stringfolded} is replaced by
\[\label{eq:gaugebethefolded}
\pint^b_a \frac{d \rap^\prime \, \rho (\rap^\prime) \, \rap}
{{\rap^\prime}^2-\rap^2} =\frac{1}{4} \,
\frac{1}{\sqrt{\rap^2-\fes{4}{2}{}{4}\coup^2}}
-\fes{\frac{\pi n}{4}}{\frac{\pi n}{2}}{\pi n}{\frac{\pi n}{2}}
\qquad \mbox{with} \qquad
\int^b_a d \rap \, \rho (\rap)=- \frac{\alpha}{2} \, ,
\]
where we are using the same procedure of analytical continuation to negative
filling $\alpha$. 
(For notational simplicity we will again use the interval
boundary values $a,b$ even though they functionally differ between
string and gauge theory.) The $\rap$-resolvent becomes
\[\label{eq:gaugeresolvfolded}
\bar\resolv(\rap)=\fes{4}{2}{}{2}\rap \int_a^b
\frac{d\rap'\,\rho(\rap')}{\rap^2-\rap'^2}=
\sum_{r=1}^{\infty} \bar{\charge}_{2 r} \rap^{2 r-1}
\qquad \mbox{with} \qquad
\bar{\charge}_{2 r}=-\fes{4}{2}{}{2} \int_a^b \frac{d\rap\,\rho(\rap)}{\rap^{2 r}} \, .
\]
%
%
The resolvent \eqref{eq:gaugeresolvfolded} is determined to be
\<\label{eq:gaugesolfolded}
\bar\resolv(\rap) \eq
\frac{2n a^2}{\rap b}
\sqrt{\frac{b^2-\rap^2}{a^2-\rap^2}}\,\,
\ellPi\left( -q\,\frac{\rap^2}{a^2-\rap^2}, q \right)
\nlnum
- \frac{\fes{2}{1}{1}{1}}{\fes{}{}{2}{}\pi \rap}\,
\frac{b^2}{a \sqrt{b^2-\fes{4}{2}{}{4} \coup^2}}\,
\sqrt{\frac{a^2-\rap^2}{b^2-\rap^2}}\,\,
\ellPi\left(
\frac{q}{1-q}\, \frac{\rap^2}{b^2-\rap^2},
\frac{q}{1-q}\, \frac{\fes{4}{2}{}{4} \coup^2}{b^2-\fes{4}{2}{}{4} \coup^2}
\right) \, .
\nonumber
\>
This representation is valid (without sign changes) both around
$\rap=0$ and $\rap=\infty$.
The behavior of the resolvent at $\rap=0$, namely
$\bar\resolv(\rap)=\frac{0}{\rap}+\Op(\rap)$, yields the condition
\[
\ellK(q)=\frac{1}{\fes{}{2}{4}{2} \pi n}\, \frac{b}{a \sqrt{b^2-\fes{4}{2}{}{4}\coup^2}}\,\,
\ellK \left(
\frac{q}{1-q}\, \frac{\fes{4}{2}{}{4}\coup^2}{b^2-\fes{4}{2}{}{4} \coup^2}
\right),
\]
while a second condition is obtained from the behavior at $\rap \rightarrow \infty$,
namely $\bar\resolv(\rap) \rightarrow -\fes{2}{}{\half}{}\frac{\alpha}{\rap}$:
\[
\alpha = -\fes{}{2}{4}{2}n b\, \ellE(q)+\frac{1}{\pi}\,
 \frac{b^2}{a \sqrt{b^2-\fes{4}{2}{}{4}\coup^2}}\,\,
\ellPi \left(
\frac{-q}{1-q},
\frac{q}{1-q}\, \frac{\fes{4}{2}{}{4}\coup^2}{b^2-\fes{4}{2}{}{4}\coup^2}
\right).
\]
These two equations determine the unknowns $a,b$ as a function of
the filling fraction $\alpha$ and the coupling constant $\coup$.
The density $\rho(\rap)$ is obtained as the discontinuity of the
resolvent on the cut and reads
\<\label{eq:dens12}
\rho(\rap) \eq
\frac{\fes{2}{1}{1}{1}}{\fes{}{}{2}{}\pi^2 \, \rap}\, \frac{b^2}{a \sqrt{b^2-\fes{4}{2}{}{4}\coup^2}}\,
\sqrt{\frac{\rap^2-a^2}{b^2-\rap^2}}\,\,
\ellPi \left(
\frac{b^2-\rap^2}{b^2-\fes{4}{2}{}{4}\coup^2}\,\frac{\fes{4}{2}{}{4}\coup^2}{\rap^2},
\frac{q}{1-q}\,\frac{\fes{4}{2}{}{4}\coup^2}{b^2-\fes{4}{2}{}{4}\coup^2}\right)
\nl
-\frac{2n \rap}{\pi b}\sqrt{\frac{\rap^2-a^2}{b^2-\rap^2}} \,\,\ellPi
\left(\frac{b^2-\rap^2}{b^2},q\right)  .
\>
The $\rap$-moments $\bar{\charge}_{2 r}$ are now easily
obtained explicitly by expanding \eqref{eq:gaugesolfolded}.
The proper gauge charges $\charge_{2 r}$, however, still require further,
unpleasant integrations, using \eqref{eq:gaugecharge,eq:gaugechargedensity},
which we have not been able to perform
explicitly. E.g.~the energy is obtained from the density
\eqref{eq:dens12} as the integral
\[
\eng(\coup)=1-\alpha
-2 \int_a^b d\rap\, \rho(\rap)\, \frac{\rap}{\sqrt{\rap^2-\fes{4}{2}{}{4}\coup^2}}\, .
\]
%
%
%
This integral representation is nevertheless useful for
working out the explicit perturbative expansion of the gauge energy to
any desired order.

\subsection{The Circular String}
\label{sec:Spinning.Circular}

\subsubsection{Semi-classical String Solution}
\label{sec:stringcircular}

The Bethe solution of the circular string makes the ansatz that
there is a single contour $\contourstring$ which is purely
imaginary and symmetric w.r.t.~reflection around the real axis.
The (pseudo)density $\sigma(x)$ is assumed to be a constant
$\sigma(x)=-2im$, $m$ integer, on the interval $x\in[-i c,i c]$,
but non-constant on the intervals $[i c,i d]$ and
$[-i d,-i c]$. It is convenient to rotate the spectral
$x$-plane by $\frac{\pi}{2}$ and redefine $x=iy$.
This leads to the classical Bethe equation
\[\label{eq:stringbethecircular}
\pint_c^d dy'\,\frac{i\sigma(iy')\,y}{y^2-y'^2}=
\frac{\engs}{\fes{2}{4}{8}{4}}\,\frac{y}{y^2+\fes{}{\half}{\quarter}{}\coup^2}
-m\log \frac{y+c}{y-c}
\]
with the normalization
\[
2cm \lrbrk{1-\frac{\coup^2}{\fes{}{2}{4}{}c^2}}+
\int_c^d dy\,i\sigma(iy)\,\left( 1+\frac{\coup^2}{\fes{}{2}{4}{}y^2} \right)
=
\fes{\alpha}{\frac{\alpha}{2}}{\frac{\alpha}{4}}{\frac{\alpha}{2}} \, ,
\]
and
\[\label{eq:stringresolvcircular}
i\resolvs(iy)=
2 m \log \frac{c-y}{c+y}+
2 y \int_c^d dy'\, \frac{i\sigma(iy')}{y'^2-y^2}=
i\sum_{r=1}^{\infty}  (iy)^{2 r-1} \charges_{2 r}\, ,
\]
where
\[
\charges_{2 r}=\frac{4}{2 r-1}\,
\frac{mi}{(ic)^{2 r-1}}+
2 \int_c^d dy\,\frac{i\sigma(iy)}{(iy)^{2 r}} \, .
\]
The expression for the string energy \eqref{eq:stringenergy} is
$\engs=1+\fes{}{}{}{2} \coup^2 \charges_2(\coup)$.
The solution of \eqref{eq:stringbethecircular} is, with $q'={c^2}/{d^2}$,
\<\label{eq:dens21}
i\sigma(iy)\eq
2m-
\frac{4m}{\pi y d} \sqrt{(d^2-y^2)(y^2-c^2)}\,\,
\ellPi\left(\frac{y^2}{d^2},q'\right)
\nl\qquad\qquad
+\frac{\engs}{\fes{}{2}{4}{2} \pi y }\,
\frac{\fes{}{\half}{\quarter}{}\coup^2}{y^2+\fes{}{\half}{\quarter}{}\coup^2}\,
\sqrt{\frac{(d^2-y^2)(y^2-c^2)}{(d^2+\fes{}{\half}{\quarter}{}\coup^2)(c^2+\fes{}{\half}{\quarter}{}\coup^2)}} \, ,
\>
and in a domain near the origin of the $y$-plane \eqref{eq:stringresolvcircular} is given by
\<
i\resolvs(iy)\eq-
\fes{\engs}{\frac{\engs}{2}}{\frac{\engs}{4}}{\frac{\engs}{2}}\,
\frac{y}{y^2+\fes{}{\half}{\quarter}{}\coup^2}+
\frac{4m}{y d} \sqrt{(d^2-y^2)(c^2-y^2)}\,\,
\ellPi\left(\frac{y^2}{d^2},q'\right)
\nl
\qquad\qquad-
\frac{\engs}{\fes{}{2}{4}{2} y}\,
\frac{\fes{}{\half}{\quarter}{}\coup^2}{y^2+\fes{}{\half}{\quarter}{}\coup^2}\,
\sqrt{\frac{(d^2-y^2)(c^2-y^2)}{(d^2+\fes{}{\half}{\quarter}{}\coup^2)(c^2+\fes{}{\half}{\quarter}{}\coup^2)}} \, .
\>
The known behavior of the resolvent at $y=0$, namely
$\resolvs(iy)=\frac{0}{y}+\charges_2 iy+ \order{y^3}$, yields two conditions:
\[
\frac{\engs}{\sqrt{(d^2+\fes{}{\half}{\quarter}{}\coup^2)(c^2+\fes{}{\half}{\quarter}{}\coup^2)}}
=\frac{\fes{4}{8}{16}{8}m}{d}\, \ellK(q')
\qquad \mbox{and} \qquad
\ellK(r)=\frac{1}{\fes{4}{8}{16}{8} mc}
+\frac{\coup^2}{\fes{}{2}{4}{}c^2}\, \big(\ellE(q')-\ellK(q')\big) ,
\]
while a third condition is obtained from the behavior of the resolvent
$\resolvs(iy)$ at infinity $y \rightarrow \infty$:
\[
\alpha=\frac{1}{2}+\fes{2}{4}{8}{4} md\, \big(\ellE(q')-\ellK(q')\big)
-\frac{\fes{2}{2}{2}{4}m}{d}\, \coup^2\, \ellK(q').
\]
These three equations determine the three unknowns $c,d$ and $\engs$
as a function of the filling fraction $\alpha$ and the coupling constant
$\coup$. One checks that they indeed reproduce the energy of the
circular string as first obtained without Bethe ansatz in \cite{Arutyunov:2003uj}.

There is a special ``algebraic{}'' point at half-filling $\alpha=\frac{1}{2}$
already worked out in \cite{Kazakov:2004qf}.
Here the cut extends from $c$ to $d=\infty$. Note that
$c=c_0=\frac{1}{\fes{2}{4}{8}{4} \pi m}$ becomes independent of $\coup$!
The pseudodensity simplifies to a semi-circle law
\[
i\sigma(iy)=2m-\frac{2 y m}{y^2+\fes{}{\half}{\quarter}{}\coup^2}\, \sqrt{y^2-c_0^2}
\]
while the resolvent reduces to
\[
i\resolvs(iy)= \frac{2 \pi m y}{y^2+\fes{}{\half}{\quarter}{}\coup^2}
\left(\sqrt{c_0^2-y^2}-\sqrt{c_0^2+\fes{}{\half}{\quarter}{}\coup^2}\right)
\, ,
\]
and the energy becomes
\[\label{eq:halfcircular}
\engs(\coup)
=\sqrt{1+\frac{\coup^2}{\fes{}{2}{4}{}c_0^2}}
=\sqrt{1+\fes{4}{8}{16}{16} \pi^2 m^2\coup^2}\, ,
\]
as originally found in \cite{Frolov:2003qc}.

\subsubsection{All-loop Asymptotic Gauge Solution}
\label{sec:gaugecircular}

Just as in the string theory computation we assume there to be a
\emph{condensate} of Bethe roots on the imaginary axis:
$\rho(\rap)=-\fes{}{2}{4}{2}im$, on the interval $[-ic,ic]$, where $m$ is an
integer. Outside this interval the root density is again
non-constant. Due to the condensate cut it is convenient to
perform a rotation $\rap=i\rapi$.
Thus the circular string Bethe equation of the last section
\secref{sec:stringcircular} is replaced in the perturbative gauge
theory by the two-cut singular integral equation
\[\label{eq:gaugebethecircular}
\pint^d_c \frac{d\rapi^\prime \, i\rho(i\rapi^\prime) \, \rapi}
{{\rapi}^2-\rapi'^2} =\frac{1}{4}\, \frac{1}{\sqrt{\rapi^2+\fes{4}{2}{}{4} \coup^2}}
-\fes{\half}{}{2}{}m\log \frac{\rapi+c}{\rapi-c}
\qquad \mbox{with} \qquad
\fes{}{2}{4}{2}cm+\int^d_c d \rapi \, i\rho (i\rapi)=\frac{\alpha}{2} \, .
\]
The $\rapi$-resolvent is
\[\label{eq:gaugeresolvcircular}
i\bar\resolv(i\rapi)=
2 m\log \frac{c-\rapi}{c+\rapi}+
\fes{4}{2}{}{2}\rapi \int_c^d d\rapi'\, \frac{i\rho(i\rapi')}{\rapi'^2-\rapi^2}=
i\sum_{r=1}^{\infty} (i\rapi)^{2 r-1} \bar{\charge}_{2 r}\, ,
\]
with
\[
\bar{\charge}_{2 r}=\frac{4}{2 r-1}\,
\frac{im}{(ic)^{2 r-1}}+\fes{4}{2}{}{2}\int_c^d d\rapi\,\frac{i\rho(i\rapi)}{(i\rapi)^{2 r}} \, .
\]
The form of the resolvent appropriate for the expansion in local charges is found to be
\<\label{eq:gaugesolcircular}
i\bar\resolv(i\rapi) \eq
\frac{4m}{\rapi\, d}\,
\sqrt{(d^2-\rapi^2)(c^2-\rapi^2)}\,\,
\ellPi\left(\frac{\rapi^2}{d^2},q' \right)\
\nlnum
-\frac{\fes{2}{1}{1}{1}}{\fes{}{}{2}{}\pi \rapi}\,
\frac{d^2}{c \sqrt{d^2+\fes{4}{2}{}{4} \coup^2}}\,
\sqrt{\frac{c^2-\rapi^2}{d^2-\rapi^2}}\,\,
\ellPi\left( \frac{1-q'}{q'}\,\frac{\rapi^2}{d^2-\rapi^2},
-\frac{1-q'}{q'}\, \frac{\fes{4}{2}{}{4}\coup^2}{d^2+\fes{4}{2}{}{4}\coup^2} \right)
\nonumber\>
with $q'=\frac{c^2}{d^2}$.
The behavior of the resolvent at $\rapi=0$, namely
$\bar\resolv(i\rapi)=\frac{0}{\rapi}+\order{\rapi}$, yields the condition
\[
\ellK(q')=\frac{1}{\fes{2}{4}{8}{4}\pi m}\,
\frac{d}{c \sqrt{d^2+\fes{4}{2}{}{4} \coup^2 }}\,\,
\ellK \left( \frac{q'-1}{q'}\, \frac{\fes{4}{2}{}{4} \coup^2}{d^2+\fes{4}{2}{}{4} \coup^2} \right),
\]
and, after analytic continuation of the representation \eqref{eq:gaugesolcircular}
to large values of $\rapi$, we find from the behavior of $\bar\resolv(i\rapi)$ at
infinity a second condition
\[
\alpha = \fes{2}{4}{8}{4} m d\, \bigbrk{ \ellE(q')-\ellK(q') }
+\frac{1}{\pi}\,\frac{d^2}{c \sqrt{d^2+\fes{4}{2}{}{4} \coup^2}}\,\,
\ellPi \left(
\frac{q'-1}{q'},
\frac{q'-1}{q'}\, \frac{\fes{4}{2}{}{4} \coup^2}{d^2+\fes{4}{2}{}{4} \coup^2}
\right).
\]
Finally the density is once more obtained from the behavior of
$\bar\resolv(i\rapi)$ on the cut. We found the following form
\<
i\rho(i\rapi)\eq \fes{}{2}{4}{2}m-
\frac{\fes{2}{4}{8}{4}m}{\pi \rapi d}\, \sqrt{(d^2-\rapi^2)(\rapi^2-c^2)}\,\,
\ellPi\left(\frac{\rapi^2}{d^2},q'\right)+
\frac{\fes{2}{4}{8}{4}d m}{\pi \rapi}\,\sqrt{\frac{\rapi^2-c^2}{d^2-\rapi^2}}\,\,\ellK(q')
\nlnum-
\frac{1}{\pi^2 \,\rapi}\, \frac{d^2}{c \sqrt{d^2+\fes{4}{2}{}{4} \coup^2}}\,
\sqrt{\frac{\rapi^2-c^2}{d^2-\rapi^2}}\,\,\ellPi \left(
-\frac{d^2-\rapi^2}{d^2+\fes{4}{2}{}{4} \coup^2}\,\frac{\fes{4}{2}{}{4} \coup^2}{\rapi^2},
-\frac{1-q'}{q'}\,\frac{\fes{4}{2}{}{4} \coup^2}{d^2+ \fes{4}{2}{}{4} \coup^2} \right)
\nonumber \, ,
\>
which should be compared to its much simpler string analog \eqref{eq:dens21}.
This density yields, in view of \eqref{eq:gaugecharge,eq:gaugechargedensity},
integral representations for all proper gauge charges $\charge_{2 r}$.

At the half-filling point $\alpha=\frac{1}{2}$ these expressions simplify,
but, unlike the string case \secref{sec:stringcircular},
they do \emph{not} become algebraic. The density reduces to
\[
i\rho(i\rapi)=\fes{1}{2}{4}{2}m
-\frac{1}{\pi^2\, c \,\rapi}\,
\sqrt{\rapi^2-c^2}\,\,
\ellPi \left( -\frac{\fes{4}{2}{}{4} \coup^2}{\rapi^2}, -\frac{\fes{4}{2}{}{4} \coup^2}{c^2} \right),
\]
and the resolvent becomes
\[
i\bar\resolv(i\rapi)=\frac{m}{4}-\frac{m}{4}\,
\frac{\rapi}{\sqrt{\rapi^2+\fes{4}{2}{}{4} \coup^2}}
+\frac{1}{2 \pi c}\, \sqrt{c^2-\rapi^2}\,\,
\ellPi \left( -\frac{\fes{4}{2}{}{4} \coup^2}{\rapi^2}, -\frac{\fes{4}{2}{}{4} \coup^2}{c^2} \right),
\]
while the boundary point $c$ remains coupling constant dependent at $\alpha=\frac{1}{2}$,
in contradistinction to the string theory case:
\[\label{eq:Spinning.Circular.Norm2}
\fes{\half}{}{2}{}mc=\frac{1}{2 \pi^2}\,\, \ellK\left( -\frac{\fes{4}{2}{}{4} \coup^2}{c^2} \right).
\]
In order to work out the perturbative gauge energy we still have to perform
the following integral:
\<\label{eq:messy}
\eng(\coup)\eq\frac{3}{2}-\fes{}{2}{4}{2} \pi m c+\frac{\fes{4}{2}{1}{2}}{\pi^2 m c}
\int_c^{\infty} d\rapi \,\,
\sqrt{\frac{\rapi^2-c^2}{\rapi^2+\fes{4}{2}{}{4} \coup^2}}
\nlnum\nonumber
\qquad\qquad\qquad\qquad\times
\left(
\ellPi \left( -\frac{\fes{4}{2}{}{4} \coup^2}{\rapi^2}, -\frac{\fes{4}{2}{}{4} \coup^2}{c^2} \right)
-\sqrt{1+\frac{\fes{4}{2}{}{4} \coup^2}{\rapi^2}}\,\,\ellK \left(-\frac{\fes{4}{2}{}{4} \coup^2}{c^2} \right) \right).
\>
We did not succeed in
calculating this integral in terms of algebraic or elliptic
functions. It is however straightforward to use the representation
\eqref{eq:messy} to derive the following perturbative expansion of
the energy ($c_0=\frac{1}{\fes{2}{4}{8}{4}\pi m}$):
\<\label{eq:halfgaugeenergy}
\eng(\coup)\eq 1
+\frac{1}{2}\frac{\coup^2}{\fes{}{2}{4}{}c_0^2}
-\frac{1}{8}\left(\frac{\coup^2}{\fes{}{2}{4}{}c_0^2}\right)^2
+\frac{3}{128}\left(\frac{\coup^2}{\fes{}{2}{4}{}c_0^2}\right)^4
-\frac{3}{256}\left(\frac{\coup^2}{\fes{}{2}{4}{}c_0^2}\right)^6
+\frac{267}{32768}\left(\frac{\coup^2}{\fes{}{2}{4}{}c_0^2}\right)^8
\\ \nonumber & &
-\frac{441}{65536}\left(\frac{\coup^2}{\fes{}{2}{4}{}c_0^2}\right)^{10}
+\frac{6483}{1048576}\left(\frac{\coup^2}{\fes{}{2}{4}{}c_0^2}\right)^{12}
-\frac{12813}{2097152}\left(\frac{\coup^2}{\fes{}{2}{4}{}c_0^2}\right)^{14}
+\order{\coup^{16}} .
\>
Oddly, the odd powers of $\coup^2$ are missing, except for the linear term.
This contrasts with the much simpler string result
\eqref{eq:halfcircular}, whose expansion matches \eqref{eq:halfgaugeenergy}
only up to $\order{\coup^4}$, and certainly contains all powers of $\coup^2$.

\section{A Density for the String Bethe Ansatz}
\label{sec:density}

The normalization of the string Bethe equations \eqref{eq:stringnorm}
differs from the one in gauge theory \eqref{eq:norm}.
To make the ans\"atze more similar we should transform
the pseudodensity to a true density.%
\footnote{This turns out to be the key to a possible generalization
of the Bethe ansatz to account for string sigma model quantum effects.
These $1/J$ corrections require a clean definition of individual roots.}
For that purpose we set
\[
\sigma(x)=\frac{\fes{2}{}{\half}{}\rho(x)}{1-\frac{\coup^2}{\fes{}{2}{4}{}x^2}}
\]
and obtain a proper normalization
\[\label{eq:stringnormnew}
\int_{\contourstring}dx\,\rho(x)=\alpha.
\]

The local charges are now given by
\[\label{eq:stringchargesnew}
\charges_r=\int_{\contourstring}
\frac{\fes{2}{}{\half}{}\,dx\,\rho(x)}{1-\frac{\coup^2}{\fes{}{2}{4}{}x^2}}\,
\frac{1}{x^r} \, .
\]
Assuming that $\rho$ transforms as a density,
$dx\,\rho(x)=d\rap\,\rho(\rap)$,
we see immediate agreement with the gauge theory expression
\eqref{eq:gaugechargedensity,eq:gaugecharge}
\[
\charge_r=\int_{\contourgauge} \frac{\fes{2}{}{\half}{} d\rap\,\rho(\rap)}
{\sqrt{\rap^2-\fes{4}{2}{}{}\coup^2}\,x(\rap)^{r-1}} \,,
\]
noting a relation which holds by virtue of $x=x(\rap)$ \eqref{eq:xofphi,eq:uofx}
\[
x-\frac{\coup^2}{\fes{}{2}{4}{}x}
=
\fes{\half}{}{2}{}\sqrt{\rap^2-\fes{4}{2}{}{}\coup^2}\,
.
\]

The string Bethe equation using the density reads
\[\label{eq:stringbethenew}
2\pint_{\contourstring}
\frac{dx'\,\rho\indup{s}(x')}{x-x'}
=
\frac{1}{x}\,
+\fes{}{2}{4}{2}\pi n_\nu
\lrbrk{1-\frac{\coup^2}{\fes{}{2}{4}{}x^2}}\,
-\frac{\fes{2}{}{}{2}\coup^2}{\fes{}{}{2}{}x^2}\,\charges_1.
\]
In contrast, the gauge Bethe equations read
\[\label{eq:gaugestringbethe4}
2\pint_{\contourstring}
 dx'\,\rho\indup{g}(x')
\lrbrk{\frac{1}{x-x'}+\frac{\coup^2}{\fes{}{2}{4}{}x^2x'}\,
\frac{1}{1-\frac{\coup^2}{\fes{}{2}{4}{}xx'}}}
=
\frac{1}{x}
+\fes{}{2}{4}{2}\pi n_\nu \lrbrk{1-\frac{\coup^2}{\fes{}{2}{4}{}x^2}}.
\]
The only distinction between the two
is the slightly different second part of the integrand
after substituting \eqref{eq:stringchargesnew} in
\eqref{eq:stringbethenew}.







\section{Proof of a Curious Observation}
\label{sec:curious}

Let us investigate the leading order perturbative difference between the gauge Bethe equation
and the string Bethe equation.
In the spectral $\rap$-plane the former is given by
\eqref{eq:gaugebethe} while the latter is \eqref{eq:stringbethe3}.
Expansion in $\coup$ gives for the respective equations to
three-loop order
\[\label{eq:gaugebetheexp}
2\pint_{\contourgauge}
\frac{d\rap'\,\rho(\rap')}{\rap-\rap'}=
\fes{}{2}{4}{2}\pi n_{\nu}
+\frac{1}{\rap}
+\frac{\fes{2}{}{}{2}\coup^2}{\fes{}{}{2}{}\rap^3}
+\frac{\fes{6}{3}{3}{6}\coup^4}{\fes{}{2}{8}{}\rap^5}+
\order{\coup^6}
\, ,
\]
and
\[\label{eq:stringbetheexp}
2\pint_{\contourgauge} \frac{d\rap'\,\rho\indup{s}(\rap')}{\rap-\rap'}=
\fes{}{2}{4}{2}\pi n_{\nu}
+\frac{1}{\rap}
-\frac{\fes{}{}{}{2}\coup^4}{\fes{}{2}{4}{}\rap^2}
\,\bar{\charges}_{3,0}
+\frac{\fes{2}{1}{1}{2}}{\fes{}{}{2}{}\rap^3}
\lrbrk{\coup^2
  +\fes{\half}{\half}{\half}{}\coup^4\,\bar{\charges}_{2,0}}
+\frac{\fes{6}{3}{3}{6}\coup^4}{\fes{}{2}{8}{}\rap^5}+
\order{\coup^6}
\, .
\]
Here we needed also the one-loop second and third moments
$\bar{\charges}_{2,0},\bar{\charges}_{3,0}$ which are obtained from
the loop expansion of the string theory $\rap$-moments:
\[
\bar{\charges}_r(\coup)=\sum_{\ell=1}^{\infty}
\bar{\charges}_{r,2\ell-2} \, \fes{}{}{}{2^{\ell-1}}\coup^{2 \ell-2}
\qquad \mbox{with} \qquad
\bar{\charges}_r=\int_{\contourgauge} \frac{\fes{2}{}{\half}{}d\rap\, \rho\indup{s}(\rap)}{\rap^r} \, .
\]
To two-loop order the right hand sides of equations
\eqref{eq:gaugebetheexp,eq:stringbetheexp} are identical, but for
three loops the string equation has two extra terms. The first is
proportional to $1/\rap^2$. As terms even in $\rap$
are completely absent in the gauge potential, generic solutions
will irreparably differ in structure starting from this order.
However, note that this term is multiplied by an odd expectation
value $\bar{\charges}_{3,0}$. It is therefore absent for
(unpaired) solutions symmetric in $\rap$ such as the ones
studied in appendix \secref{sec:elliptic}. Furthermore, we see
that for symmetric solutions we can introduce a shifted coupling
constant
\[\label{eq:shift}
\coup\indup{s}^2:=\coup^2+\fes{\half}{\half}{\half}{}\coup^4\,\bar{\charges}_{2,0}\, ,
\]
and rewrite the string equation \eqref{eq:stringbetheexp} as
\[\label{eq:stringbetheexp2}
2\pint_{\contourgauge} \frac{d\rap'\,\rho\indup{s}(\rap')}{\rap-\rap'}=
\fes{}{2}{4}{2}\pi n_{\nu}
+\frac{1}{\rap}
+\frac{\fes{2}{}{\half}{2}\coup\indup{s}^2}{\rap^3}
+\frac{\fes{6}{3}{3}{6}\coup\indup{s}^4}{\fes{}{2}{8}{}\rap^5}
+\order{\coup\indup{s}^6}
\qquad \mbox{if} \qquad
\bar{\charges}_{3,0}=0
\, .
\]
Therefore the equation to be solved is formally identical to the gauge
equation \eqref{eq:gaugebetheexp}, and, to this order, one will find the same
form of the density, but with the shifted coupling \eqref{eq:shift}.
The string charges can then be obtained from the gauge charges
by the simple replacement \eqref{eq:shift}.
This immediately leads to the result
\[
\bar{\charges}_{2 r}(\coup)-\bar{\charge}_{2 r}(\coup)=
\fes{\half}{\half}{\half}{2}\coup^4\,\bar{\charges}_{2,0}\,
\bar{\charges}_{2 r,2}
+ \order{\coup^6}
\, ,
\]
which, after accounting for \fes{heavily}{somewhat}{slightly}{} altered normalizations and conventions,
precisely proves in generality the finding in equation~(17) in \cite{Arutyunov:2004xy},
originally derived for two specific solutions (folded and circular string).
Likewise, using \eqref{eq:gaugeenergy,eq:stringenergy}, we find
for \emph{all} even solutions of the Bethe equations
\[
\engs(\coup)-\eng(\coup)=
\fes{\half}{\half}{\half}{4}\coup^4\,\bar{\charges}_{2,0}\,\bar{\charges}_{2,2} + \order{\coup^8}
\, ,
\]
which is, after adjusting conventions, the
general proof for the ``curious observation{}''
at the end of \cite{Serban:2004jf}.


\bibliography{bds}

\begin{thebibliography}{10}
\ifx\href\asklfhas\newcommand{\href}[2]{#2}\fi
\raggedright
\small
\parskip 0pt

\bibitem{Kristjansen:2002bb}
C.~Kristjansen, J.~Plefka, G.~W.~Semenoff and M.~Staudacher,
\textit{``A new double-scaling limit of {$\mathcal{N}=\mathord{}$4} super
  {Yang-Mills} theory and {PP}-wave strings''},
\textsf{Nucl.~Phys.~B643,~3~(2002)},
\href{http://arXiv.org/abs/hep-th/0205033}{\texttt{hep-th/0205033}}.
%
\bibitem{Constable:2002hw}
N.~R.~Constable, D.~Z.~Freedman, M.~Headrick, S.~Minwalla, L.~Motl,
  A.~Postnikov and W.~Skiba,
\textit{``{PP}-wave string interactions from perturbative {Yang-Mills}
  theory''},
\textsf{JHEP~0207,~017~(2002)},
\href{http://arXiv.org/abs/hep-th/0205089}{\texttt{hep-th/0205089}}.
%
\bibitem{Beisert:2002bb}
N.~Beisert, C.~Kristjansen, J.~Plefka, G.~W.~Semenoff and M.~Staudacher,
\textit{``BMN correlators and operator mixing in {$\mathcal{N}=\mathord{}$4}
  super Yang-Mills theory''},
\textsf{Nucl.~Phys.~B650,~125~(2003)},
\href{http://arXiv.org/abs/hep-th/0208178}{\texttt{hep-th/0208178}}.
%
\bibitem{Constable:2002vq}
N.~R.~Constable, D.~Z.~Freedman, M.~Headrick and S.~Minwalla,
\textit{``Operator mixing and the BMN correspondence''},
\textsf{JHEP~0210,~068~(2002)},
\href{http://arXiv.org/abs/hep-th/0209002}{\texttt{hep-th/0209002}}.
%
\bibitem{Minahan:2002ve}
J.~A.~Minahan and K.~Zarembo,
\textit{``The Bethe-ansatz for {$\mathcal{N}=\mathord{}$4} super Yang-Mills''},
\textsf{JHEP~0303,~013~(2003)},
\href{http://arXiv.org/abs/hep-th/0212208}{\texttt{hep-th/0212208}}.
%
\bibitem{Beisert:2002ff}
N.~Beisert, C.~Kristjansen, J.~Plefka and M.~Staudacher,
\textit{``BMN gauge theory as a quantum mechanical system''},
\textsf{Phys.~Lett.~B558,~229~(2003)},
\href{http://arXiv.org/abs/hep-th/0212269}{\texttt{hep-th/0212269}}.
%
\bibitem{Beisert:2003yb}
N.~Beisert and M.~Staudacher,
\textit{``The {$\mathcal{N}=\mathord{}$4} SYM Integrable Super Spin Chain''},
\textsf{Nucl.~Phys.~B670,~439~(2003)},
\href{http://arXiv.org/abs/hep-th/0307042}{\texttt{hep-th/0307042}}.
%
\bibitem{Beisert:2003jj}
N.~Beisert,
\textit{``The Complete One-Loop Dilatation Operator of
  {$\mathcal{N}=\mathord{}$4} Super Yang-Mills Theory''},
\textsf{Nucl.~Phys.~B676,~3~(2004)},
\href{http://arXiv.org/abs/hep-th/0307015}{\texttt{hep-th/0307015}}.
%
\bibitem{Braun:1998id}
V.~M.~Braun, S.~E.~Derkachov and A.~N.~Manashov,
\textit{``Integrability of three-particle evolution equations in {QCD}''},
\textsf{Phys.~Rev.~Lett.~81,~2020~(1998)},
\href{http://arXiv.org/abs/hep-ph/9805225}{\texttt{hep-ph/9805225}}.
%
\bibitem{Braun:1999te}
V.~M.~Braun, S.~E.~Derkachov, G.~P.~Korchemsky and A.~N.~Manashov,
\textit{``Baryon distribution amplitudes in {QCD}''},
\textsf{Nucl.~Phys.~B553,~355~(1999)},
\href{http://arXiv.org/abs/hep-ph/9902375}{\texttt{hep-ph/9902375}}.
%
\bibitem{Belitsky:1999bf}
A.~V.~Belitsky,
\textit{``Renormalization of twist-three operators and integrable lattice
  models''},
\textsf{Nucl.~Phys.~B574,~407~(2000)},
\href{http://arXiv.org/abs/hep-ph/9907420}{\texttt{hep-ph/9907420}}.
%
\bibitem{Beisert:2003tq}
N.~Beisert, C.~Kristjansen and M.~Staudacher,
\textit{``The dilatation operator of {$\mathcal{N}=\mathord{}$4} conformal
  super Yang-Mills theory''},
\textsf{Nucl.~Phys.~B664,~131~(2003)},
\href{http://arXiv.org/abs/hep-th/0303060}{\texttt{hep-th/0303060}}.
%
\bibitem{Kotikov:2004er}
A.~V.~Kotikov, L.~N.~Lipatov, A.~I.~Onishchenko and V.~N.~Velizhanin,
\textit{``Three-loop universal anomalous dimension of the Wilson operators in
  {$\mathcal{N}=\mathord{}$4} SUSY Yang-Mills model''},
\textsf{Phys.~Lett.~B595,~521~(2004)},
\href{http://arXiv.org/abs/hep-th/0404092}{\texttt{hep-th/0404092}}.
%
\bibitem{Moch:2004pa}
S.~Moch, J.~A.~M.~Vermaseren and A.~Vogt,
\textit{``The three-loop splitting functions in QCD: The non-singlet case''},
\textsf{Nucl.~Phys.~B688,~101~(2004)},
\href{http://arXiv.org/abs/hep-ph/0403192}{\texttt{hep-ph/0403192}}.
%
\bibitem{Beisert:2003ys}
N.~Beisert,
\textit{``The su(2$/$3) dynamic spin chain''},
\textsf{Nucl.~Phys.~B682,~487~(2004)},
\href{http://arXiv.org/abs/hep-th/0310252}{\texttt{hep-th/0310252}}.
%
\bibitem{Klose:2003qc}
T.~Klose and J.~Plefka,
\textit{``On the Integrability of large $N$ Plane-Wave Matrix Theory''},
\textsf{Nucl.~Phys.~B679,~127~(2004)},
\href{http://arXiv.org/abs/hep-th/0310232}{\texttt{hep-th/0310232}}.
%
\bibitem{Beisert:2003jb}
N.~Beisert,
\textit{``Higher loops, integrability and the near BMN limit''},
\textsf{JHEP~0309,~062~(2003)},
\href{http://arXiv.org/abs/hep-th/0308074}{\texttt{hep-th/0308074}}.
%
\bibitem{Serban:2004jf}
D.~Serban and M.~Staudacher,
\textit{``Planar {$\mathcal{N}=\mathord{}$4} gauge theory and the Inozemtsev
  long range spin chain''},
\textsf{JHEP~0406,~001~(2004)},
\href{http://arXiv.org/abs/hep-th/0401057}{\texttt{hep-th/0401057}}.
%
\bibitem{Inozemtsev:1989yq}
V.~I.~Inozemtsev,
\textit{``On the connection between the one-dimensional $s=1/2$ Heisenberg
  chain and Haldane Shastry model''},
\textsf{J.~Stat.~Phys.~59,~1143~(1990)}.
%
\bibitem{Inozemtsev:2002vb}
V.~I.~Inozemtsev,
\textit{``Integrable Heisenberg-van Vleck chains with variable range
  exchange''},
\textsf{Phys.~Part.~Nucl.~34,~166~(2003)},
\href{http://arXiv.org/abs/hep-th/0201001}{\texttt{hep-th/0201001}}.
%
\bibitem{Berenstein:2002jq}
D.~Berenstein, J.~M.~Maldacena and H.~Nastase,
\textit{``Strings in flat space and pp waves from {$\mathcal{N}=\mathord{}$4}
  {Super} {Yang Mills}''},
\textsf{JHEP~0204,~013~(2002)},
\href{http://arXiv.org/abs/hep-th/0202021}{\texttt{hep-th/0202021}}.
%
\bibitem{Gubser:2002tv}
S.~S.~Gubser, I.~R.~Klebanov and A.~M.~Polyakov,
\textit{``A semi-classical limit of the gauge/string correspondence''},
\textsf{Nucl.~Phys.~B636,~99~(2002)},
\href{http://arXiv.org/abs/hep-th/0204051}{\texttt{hep-th/0204051}}.
%
\bibitem{Frolov:2003qc}
S.~Frolov and A.~A.~Tseytlin,
\textit{``Multi-spin string solutions in {$AdS_5\times S^5$}''},
\textsf{Nucl.~Phys.~B668,~77~(2003)},
\href{http://arXiv.org/abs/hep-th/0304255}{\texttt{hep-th/0304255}}.
%
\bibitem{Frolov:2003xy}
S.~Frolov and A.~A.~Tseytlin,
\textit{``Rotating string solutions: AdS/CFT duality in non-supersymmetric
  sectors''},
\textsf{Phys.~Lett.~B570,~96~(2003)},
\href{http://arXiv.org/abs/hep-th/0306143}{\texttt{hep-th/0306143}}.
%
\bibitem{Arutyunov:2003uj}
G.~Arutyunov, S.~Frolov, J.~Russo and A.~A.~Tseytlin,
\textit{``Spinning strings in $AdS_5\times S^5$ and integrable systems''},
\textsf{Nucl.~Phys.~B671,~3~(2003)},
\href{http://arXiv.org/abs/hep-th/0307191}{\texttt{hep-th/0307191}}.
%
\bibitem{Beisert:2003xu}
N.~Beisert, J.~A.~Minahan, M.~Staudacher and K.~Zarembo,
\textit{``Stringing Spins and Spinning Strings''},
\textsf{JHEP~0309,~010~(2003)},
\href{http://arXiv.org/abs/hep-th/0306139}{\texttt{hep-th/0306139}}.
%
\bibitem{Beisert:2003ea}
N.~Beisert, S.~Frolov, M.~Staudacher and A.~A.~Tseytlin,
\textit{``Precision Spectroscopy of AdS/CFT''},
\textsf{JHEP~0310,~037~(2003)},
\href{http://arXiv.org/abs/hep-th/0308117}{\texttt{hep-th/0308117}}.
%
\bibitem{Arutyunov:2003za}
G.~Arutyunov, J.~Russo and A.~A.~Tseytlin,
\textit{``Spinning strings in {$AdS_5\times S^5$}: New integrable system
  relations''},
\textsf{Phys.~Rev.~D69,~086009~(2004)},
\href{http://arXiv.org/abs/hep-th/0311004}{\texttt{hep-th/0311004}}.
%
\bibitem{Arutyunov:2003rg}
G.~Arutyunov and M.~Staudacher,
\textit{``Matching Higher Conserved Charges for Strings and Spins''},
\textsf{JHEP~0403,~004~(2004)},
\href{http://arXiv.org/abs/hep-th/0310182}{\texttt{hep-th/0310182}}.
%
\bibitem{Arutyunov:2004xy}
G.~Arutyunov and M.~Staudacher,
\textit{``Two-loop commuting charges and the string/gauge duality''},
\href{http://arXiv.org/abs/hep-th/0403077}{\texttt{hep-th/0403077}}.
%
\bibitem{Kruczenski:2003gt}
M.~Kruczenski,
\textit{``Spin chains and string theory''},
\href{http://arXiv.org/abs/hep-th/0311203}{\texttt{hep-th/0311203}}.
%
\bibitem{Kruczenski:2004kw}
M.~Kruczenski, A.~V.~Ryzhov and A.~A.~Tseytlin,
\textit{``Large spin limit of $AdS_5\times S^5$ string theory and low energy
  expansion of ferromagnetic spin chains''},
\textsf{Nucl.~Phys.~B692,~3~(2004)},
\href{http://arXiv.org/abs/hep-th/0403120}{\texttt{hep-th/0403120}}.
%
\bibitem{Callan:2003xr}
C.~G.~Callan,~Jr., H.~K.~Lee, T.~McLoughlin, J.~H.~Schwarz, I.~Swanson and
  X.~Wu,
\textit{``Quantizing string theory in $AdS_5\times S^5$: Beyond the pp-wave''},
\textsf{Nucl.~Phys.~B673,~3~(2003)},
\href{http://arXiv.org/abs/hep-th/0307032}{\texttt{hep-th/0307032}}.
%
\bibitem{Callan:2004uv}
C.~G.~Callan,~Jr., T.~McLoughlin and I.~Swanson,
\textit{``Holography beyond the Penrose limit''},
\href{http://arXiv.org/abs/hep-th/0404007}{\texttt{hep-th/0404007}}.
%
\bibitem{Kazakov:2004qf}
V.~A.~Kazakov, A.~Marshakov, J.~A.~Minahan and K.~Zarembo,
\textit{``Classical/quantum integrability in AdS/CFT''},
\textsf{JHEP~0405,~024~(2004)},
\href{http://arXiv.org/abs/hep-th/0402207}{\texttt{hep-th/0402207}}.
%
\bibitem{Russo:2002sr}
J.~G.~Russo,
\textit{``Anomalous dimensions in gauge theories from rotating strings in
  {$AdS_5 \times S^5$}''},
\textsf{JHEP~0206,~038~(2002)},
\href{http://arXiv.org/abs/hep-th/0205244}{\texttt{hep-th/0205244}}.
%
\bibitem{Minahan:2002rc}
J.~A.~Minahan,
\textit{``Circular semiclassical string solutions on {$AdS_5\times S^5$}''},
\textsf{Nucl.~Phys.~B648,~203~(2003)},
\href{http://arXiv.org/abs/hep-th/0209047}{\texttt{hep-th/0209047}}.
%
\bibitem{Tseytlin:2002ny}
A.~A.~Tseytlin,
\textit{``Semiclassical quantization of superstrings: {$AdS_5\times S^5$} and
  beyond''},
\textsf{Int.~J.~Mod.~Phys.~A18,~981~(2003)},
\href{http://arXiv.org/abs/hep-th/0209116}{\texttt{hep-th/0209116}}.
%
\bibitem{Frolov:2003tu}
S.~Frolov and A.~A.~Tseytlin,
\textit{``Quantizing three-spin string solution in {$AdS_5 \times S^5$}''},
\textsf{JHEP~0307,~016~(2003)},
\href{http://arXiv.org/abs/hep-th/0306130}{\texttt{hep-th/0306130}}.
%
\bibitem{Dimov:2004qv}
H.~Dimov and R.~C.~Rashkov,
\textit{``A note on spin chain/string duality''},
\href{http://arXiv.org/abs/hep-th/0403121}{\texttt{hep-th/0403121}}.
%
\bibitem{Hernandez:2004uw}
R.~Hernandez and E.~Lopez,
\textit{``The SU(3) spin chain sigma model and string theory''},
\href{http://arXiv.org/abs/hep-th/0403139}{\texttt{hep-th/0403139}}.
%
\bibitem{Stefanskijr.:2004cw}
B.~Stefa{\'n}ski,~Jr. and A.~A.~Tseytlin,
\textit{``Large spin limits of AdS/CFT and generalized Landau- Lifshitz
  equations''},
\textsf{JHEP~0405,~042~(2004)},
\href{http://arXiv.org/abs/hep-th/0404133}{\texttt{hep-th/0404133}}.
%
\bibitem{Mikhailov:2004xw}
A.~Mikhailov,
\textit{``Supersymmetric null-surfaces''},
\href{http://arXiv.org/abs/hep-th/0404173}{\texttt{hep-th/0404173}}.
%
\bibitem{Mikhailov:2004qf}
A.~Mikhailov,
\textit{``Slow evolution of nearly-degenerate extremal surfaces''},
\href{http://arXiv.org/abs/hep-th/0402067}{\texttt{hep-th/0402067}}.
%
\bibitem{Engquist:2003rn}
J.~Engquist, J.~A.~Minahan and K.~Zarembo,
\textit{``Yang-Mills Duals for Semiclassical Strings''},
\textsf{JHEP~0311,~063~(2003)},
\href{http://arXiv.org/abs/hep-th/0310188}{\texttt{hep-th/0310188}}.
%
\bibitem{Kristjansen:2004ei}
C.~Kristjansen,
\textit{``Three-spin strings on $AdS_5\times S^5$ from
  {$\mathcal{N}=\mathord{}$4} SYM''},
\textsf{Phys.~Lett.~B586,~106~(2004)},
\href{http://arXiv.org/abs/hep-th/0402033}{\texttt{hep-th/0402033}}.
%
\bibitem{Engquist:2004bx}
J.~Engquist,
\textit{``Higher conserved charges and integrability for spinning strings in
  $AdS_5\times S^5$''},
\href{http://arXiv.org/abs/hep-th/0402092}{\texttt{hep-th/0402092}}.
%
\bibitem{Stefanski:2003qr}
B.~Stefa{\'n}ski,~Jr.,
\textit{``Open spinning strings''},
\textsf{JHEP~0403,~057~(2004)},
\href{http://arXiv.org/abs/hep-th/0312091}{\texttt{hep-th/0312091}}.
%
\bibitem{Chen:2004yf}
B.~Chen, X.-J.~Wang and Y.-S.~Wu,
\textit{``Open spin chain and open spinning string''},
\textsf{Phys.~Lett.~B591,~170~(2004)},
\href{http://arXiv.org/abs/hep-th/0403004}{\texttt{hep-th/0403004}}.
%
\bibitem{Tseytlin:2003ii}
A.~A.~Tseytlin,
\textit{``Spinning strings and AdS/CFT duality''},
\href{http://arXiv.org/abs/hep-th/0311139}{\texttt{hep-th/0311139}}.
%
\bibitem{Ferretti:2004ba}
G.~Ferretti, R.~Heise and K.~Zarembo,
\textit{``New Integrable Structures in Large-N QCD''},
\href{http://arXiv.org/abs/hep-th/0404187}{\texttt{hep-th/0404187}}.
%
\bibitem{Ryzhov:2004nz}
A.~V.~Ryzhov and A.~A.~Tseytlin,
\textit{``Towards the exact dilatation operator of {$\mathcal{N}=\mathord{}$4}
  super Yang- Mills theory''},
\href{http://arXiv.org/abs/hep-th/0404215}{\texttt{hep-th/0404215}}.
%
\bibitem{Beisert:2004ry}
N.~Beisert,
\textit{``The Dilatation Operator of {$\mathcal{N}=\mathord{}$4} Super
  Yang-Mills Theory and Integrability''},
\href{http://arXiv.org/abs/hep-th/0407277}{\texttt{hep-th/0407277}}.
%
\bibitem{Faddeev:1996iy}
L.~D.~Faddeev,
\textit{``How Algebraic Bethe Ansatz works for integrable model''},
\href{http://arXiv.org/abs/hep-th/9605187}{\texttt{hep-th/9605187}}.
%
\bibitem{Swanson:2004mk}
I.~Swanson,
\textit{``On the Integrability of String Theory in $AdS_5 \times S^5$''},
\href{http://arXiv.org/abs/hep-th/0405172}{\texttt{hep-th/0405172}}.
%
\bibitem{Gross:2002su}
D.~J.~Gross, A.~Mikhailov and R.~Roiban,
\textit{``Operators with large R charge in {$\mathcal{N}=\mathord{}$4}
  Yang-Mills theory''},
\textsf{Annals~Phys.~301,~31~(2002)},
\href{http://arXiv.org/abs/hep-th/0205066}{\texttt{hep-th/0205066}}.
%
\bibitem{Klebanov:2002mp}
I.~R.~Klebanov, M.~Spradlin and A.~Volovich,
\textit{``New effects in gauge theory from pp-wave superstrings''},
\textsf{Phys.~Lett.~B548,~111~(2002)},
\href{http://arXiv.org/abs/hep-th/0206221}{\texttt{hep-th/0206221}}.
%
\bibitem{Beisert:2002tn}
N.~Beisert,
\textit{``BMN Operators and Superconformal Symmetry''},
\textsf{Nucl.~Phys.~B659,~79~(2003)},
\href{http://arXiv.org/abs/hep-th/0211032}{\texttt{hep-th/0211032}}.
%
\bibitem{Kostov:1992pn}
I.~K.~Kostov and M.~Staudacher,
\textit{``Multicritical phases of the O(n) model on a random lattice''},
\textsf{Nucl.~Phys.~B384,~459~(1992)},
\href{http://arXiv.org/abs/hep-th/9203030}{\texttt{hep-th/9203030}}.
%
\end{thebibliography}
\bibliographystyle{nb}

\end{document}